\documentclass[reprint,prd,nofootinbib,superscriptaddress]{revtex4}
\usepackage{axodraw}
\usepackage{graphicx}
\usepackage{dcolumn}
\usepackage{bm}
\usepackage{color}
\usepackage{eurosym}
\usepackage{multirow}
\usepackage{amsfonts}
\usepackage{amsmath}
\usepackage{hyperref}
\usepackage{amssymb}
\usepackage[english]{babel}
\usepackage{graphicx}
\usepackage{epsfig}
\usepackage{subfigure}
\usepackage{unicode}
\begin{document}
\newcommand{\hs}{\hspace*{0.2cm}}
\newcommand{\hsp}{\hspace*{0.5cm}}
\newcommand{\vs}{\vspace*{0.5cm}}
\newcommand{\be}{\begin{equation}}
\newcommand{\ee}{\end{equation}}
\newcommand{\bea}{\begin{eqnarray}}
\newcommand{\eea}{\end{eqnarray}}
\newcommand{\ben}{\begin{enumerate}}
\newcommand{\een}{\end{enumerate}}
\newcommand{\bde}{\begin{widetext}}
\newcommand{\ede}{\end{widetext}}
\newcommand{\nn}{\nonumber}
\newcommand{\crn}{\nonumber \\}
\newcommand{\Tr}{\mathrm{Tr}}
\newcommand{\non}{\nonumber}
\newcommand{\noi}{\noindent}
\newcommand{\al}{\alpha}
\newcommand{\la}{\lambda}
\newcommand{\bet}{\beta}
\newcommand{\ga}{\gamma}
\newcommand{\va}{\varphi}
\newcommand{\om}{\omega}
\newcommand{\pa}{\partial}
\newcommand{\+}{\dagger}
\newcommand{\fr}{\frac}
\newcommand{\sq}{\sqrt}
\newcommand{\bc}{\begin{center}}
\newcommand{\ec}{\end{center}}
\newcommand{\Ga}{\Gamma}
\newcommand{\de}{\delta}
\newcommand{\De}{\Delta}
\newcommand{\ep}{\epsilon}
\newcommand{\varep}{\varepsilon}
\newcommand{\ka}{\kappa}
\newcommand{\La}{\Lambda}
\newcommand{\si}{\sigma}
\newcommand{\Si}{\Sigma}
\newcommand{\ta}{\tau}
\newcommand{\up}{\upsilon}
\newcommand{\Up}{\Upsilon}
\newcommand{\ze}{\zeta}
\newcommand{\ps}{\psi}
\newcommand{\Ps}{\Psi}
\newcommand{\ph}{\phi}
\newcommand{\vph}{\varphi}
\newcommand{\Ph}{\Phi}
\newcommand{\Om}{\Omega}
\newcommand{\lh}[1]{{#1}}
\newcommand{\Long}[1]{{#1}}
\newcommand{\Vien}[1]{{#1}}
\newcommand{\Antonio}[1]{{#1}}
\newcommand{\Green}[1]{{#1}}
\newcommand{\Red}[1]{{#1}}
\newcommand{\Revised}[1]{{#1}}
\title{Lepton masses and mixings, and muon anomalous magnetic moment \\ in an extended $B-L$ model with type I seesaw mechanism}
\author{V. V. Vien\footnote{Corresponding author}}
\email{vovanvien@tdmu.edu.vn}
\affiliation{Institute of Applied Technology, Thu Dau Mot University, Binh Duong Province, Vietnam, \\
Department of Physics, Tay Nguyen University, 
DakLak province, Vietnam.}
\author{Hoang Ngoc Long}
	\email{hnlong@iop.vast.vn}
	\affiliation{Institute of \Revised{Physics}, Vietnam Academy of Science and Technology, 10 Dao Tan, Ba Dinh, 10000 Hanoi, Vietnam.}
\author{A. E. C\'arcamo Hern\'andez}
\email{antonio.carcamo@usm.cl}
\affiliation{Department of Physics, Universidad T\'{e}cnica Federico Santa Mar\'{\i}a,\\
	Casilla 110-V, Valpara\'{\i}so, Chile, \\
	Centro Cient\'{\i}fico-Tecnol\'ogico de Valpara\'{\i}so, Casilla 110-V, Valpara\'{\i}so, Chile,\\
	Millennium Institute for Subatomic physics at high energy frontier - SAPHIR, Fernandez Concha 700, Santiago, Chile.}

\begin{abstract}
We propose a $B-L$ model combined with the $S_4\times Z_3\times Z_4$ discrete symmetry which successfully explains the recent $3+1$ sterile - active neutrino data. The smallness of neutrino mass is obtained through the type-I seesaw mechanism. The active-active and sterile-active
neutrino mixing angles are predicted to be consistent with the recent constraints in which $0.3401\, (0.3402) \leq \sin^2\theta_{12}\leq 0.3415\, (0.3416), \, 0.456\, (0.433) \leq \sin^2\theta_{23}\leq 0.544\, (0.545), \, 2.00\, (2.018) \leq 10^2\times \sin^2\theta_{13}\leq 2.405\, (2.424),\,\, 156 \, (140.8) \leq \delta^{(\circ)}_{CP}\leq 172\, (167.2)$ for normal (inverted) ordering of the three neutrino scenario, and $0.015 \,(0.022) \leq s^2_{14}\leq 0.045 \,(0.029), \, 0.005 (0.0095)\leq s^2_{24}\leq  0.012\, (0.012), \, 0.003 \,(0.009)\leq s^2_{34} \leq 0.011$ for normal (inverted) ordering of the $3+1$ neutrino scenario. Our model predicts flavour conserving leptonic neutral scalar interactions and successfully explains the muon $g-2$ anomaly.
\end{abstract}
\date{\today}
\maketitle
\section{\label{intro} Introduction}
Recently, there have been experimental observations that cannot be explained by the three-neutrino oscillation framework \cite{Aguilar2001, Acero2008,Aguilar2010,Mention2011,Aguilar2013,An2014,Abe2015,Aguilar2018,Behera2019,Adamson2019,Adamson2020,Aartsen2020,Behera2020}. However, these observations could be explained by adding at least one additional neutrino (called sterile neutrino) with mass in the eV range having non-trivial mixing with active neutrinos. The mentioned sterile neutrinos are singlets under $SU(2)_L$ which do not take part in the weak interaction but mix with the active ones that can be verified in the oscillation experiments.
 	Nowadays, there are a number of schemes favouring the existence of sterile neutrinos, including the (3+1) scheme \cite{Kang2013,Girardi2014,Rivera2015,Gariazzo2017,Coloma2018,Liu2018,Gariazzo2018,Dentler2018,Gupta2018,Thakore2018,Dev2019,Miranda2019,Giunti2019,Boser2020,Giunti2020,Behera2020,Diaz2020} in which one sterile neutrino with mass in eV scale is heavier than the three active ones; The (3+1+1) scheme \cite{Nelson2011,Fan2012,Kuflik2012,Huang2013,Giunti2013} in which one sterile neutrino with mass in eV scale and the other is much heavier than 1 eV; The (1+3+1) scheme \cite{Kopp2011, Kopp2013,Goswami2017} in which one of the sterile neutrinos is lighter than the three active ones and the other is heavier; The (3+2) scheme \cite{Sorel2004,Karagiorgi2007,Maltoni2007,Karagiorgi2009,Giunti2011,Donini2012,Archidiacono2012} in which the two sterile neutrinos are lighter than the three active ones are added to the standard three-neutrino oscillation framework. Among the schemes with sterile neutrinos, the one with one additional sterile neutrino with mass in the eV range (called four neutrino scheme) is the simplest extension of standard three neutrino mixing that can accommodate the anomalous results of short-baseline neutrino oscillations. Among four neutrino schemes, the 3+1 scheme is preferred because the 1+3 scheme whose one sterile neutrino is lighter than the active ones and the three active neutrinos are in eV scale is ruled out by Cosmology while the 2+2 scheme is not suitable with the atmospheric and the solar neutrino oscillation data.
 	Currently, the neutrino mass squared differences and the mixing angles in the three-neutrino scheme\cite{Salas2020} and 3+1 neutrino mixing angles\cite{Deepthi2020}, at the best-fit points and $3\sigma$ range, are shown in Table \ref{experconstrain}.
\begin{table}[ht]
\begin{center}
\caption{The neutrino oscillation data for three-neutrino scheme at $3\sigma$ range and the best fit points
taken from \cite{Salas2020}, and 3+1 neutrino mixing constraints
taken from Ref. \cite{Deepthi2020}} \label{experconstrain}
\vspace{0.15 cm}
\begin{tabular}{|c|c|c|}
    \hline
    \multicolumn{1}{|c|}{\multirow{2}{*}{Parameters}}&\multicolumn{2}{c|}{3$\sigma$ range (best fit)}\\
    \cline{2-3} 
    \multicolumn{1}{|c|}{}&\hspace{0.5 cm} Normal hierarchy (NH)&\hspace{0.5 cm} Inverted hierarchy (IH)\\    \hline
    $\Delta m^{2}_{21}(10^{-5}\, \mathrm{eV}^2)$&$6.94\rightarrow 8.14\, (7.50)$&$6.94\rightarrow 8.14\, (7.50)$\\ \hline
$| \Delta m^{2}_{31}|(10^{-3}\, \mathrm{eV}^2)$&$2.47\rightarrow2.63\, (2.55)$&$2.37\rightarrow2.53\, (2.45)$\\ \hline
$s^{2}_{12}/10^{-1}$&$2.71\rightarrow3.69 \,(3.18)$&$2.71\rightarrow3.69\, (3.18)$ \\ \hline
$s^{2}_{23}/10^{-1}$&$4.34\rightarrow 6.10 \,(5.74)$&$4.33\rightarrow 6.08\,(5.78)$\\ \hline
$s^{2}_{13}/10^{-2}$&$2.000\rightarrow2.405\, (2.200)$&$2.018\rightarrow2.424 \,(2.225)$\\ \hline
$\delta/\pi$&$0.71\rightarrow 1.99 \,(1.08)$&$1.11\rightarrow1.96 \,(1.58)$\\ \hline
$s_{14}^{2}$&$0.0098\rightarrow 0.0310$&$0.0098\rightarrow 0.0310$\\ \hline
$s_{24}^{2}$&$0.0059\rightarrow0.0262$&$0.0059\rightarrow0.0262$\\ \hline
$s_{34}^{2}$&$0 \rightarrow 0.0369$&$0 \rightarrow 0.0369$\\ \hline
\end{tabular}
\vspace{-0.5 cm}
\end{center}
\end{table}
Besides, the $3\sigma$ CL ranges on the magnitude of the elements of the leptonic mixing matrix, for three neutrino scheme, are \cite{Esteban2020}:
\bea
\left|U^{3\sigma}_\text{with \, SK-atm}\right| =\left(
\begin{array}{ccc}
0.801 \to 0.845 &\hs\hs 0.513 \to 0.579 &\hs\hs  0.143 \to 0.155 \\
0.234 \to 0.500 &\hs\hs  0.417 \to 0.689 &\hs\hs 0.637 \to 0.776 \\
0.271 \to 0.535 &\hs\hs 0.477 \to 0.694 &\hs\hs  0.613 \to 0.756
\end{array}\right). \label{2020lepmix}
\eea
One notable attribute of discrete symmetries is that they provide an explanation of
the neutrino oscillation data. For the $3+1$ scheme, the $U(1)_{B-L}$ extension with $S_3$ symmetry was presented
in Ref. \cite{Machado2013S3} without mentioning
the sterile-active neutrino mass and mixing which has been addressed
in Ref. \cite{VienS3EPJC21} with the results achieved
at the first-order approximation,
and only the normal mass spectrum was mentioned. The sterile neutrino issue has also been considered in Refs. \cite{Barry2011, Barry2012, Zhang2012, Borah2017fqj, Das2019ea, Sarma2019, Krishnan2020, VienS4JPG22} with the $A_4$ symmetry and a very large amount of scalar fields, which is substantially different than the model considered in this paper whose scalar sector has a moderate amount of particle content.
Namely, in Refs. \cite{Barry2011, Barry2012} the Standard Model (SM) symmetry is enlarged by the
inclusion of the $A_4\times Z_3\times U(1)_R$ discrete group.
In Ref. \cite{Zhang2012}, the SM symmetry is enlarged by the $A_4\times Z_4$ symmetry in which only $|U_{e4}|$ has been predicted without $|U_{\mu 4}|$ and $|U_{\tau4}|$; in Ref. \cite{Borah2017fqj}, the SM symmetry is enlarged
 by the symmetry $A_4\times Z_3\times Z^{'}_3\times U(1)_R$ where
two doublets and seventeen singlets are used;
 in Ref. \cite{Das2019ea}, the symmetry $A_4\times Z_3\times Z_4$  is added to the SM
 in which up to three Higgs doublets and twelve singlets are considered in the scalar sector;
 in Ref. \cite{Sarma2019} the symmetry $A_4\times Z_4$  is added to the SM
in which one doublets and up to twelve singlet scalars are introduced\Revised{;} in Ref.\cite{Krishnan2020}, the symmetry $A_4\times C_4\times C_6 \times C_2\times U(1)_s$ is added to the SM
 in which one doublets and up to twelve singlets are used, and only the normal mass spectrum is satisfied\Revised{, and \Revised{in Ref. \cite{VienS4JPG22}, the symmetry $A_4\times Z_3\times Z_4$  is added to the $B-L$ in which one Higgs doublet and up to eleven singlets are considered in the scalar sector and without mention in the muon anomalous magnetic moment}}. Thus, it would be useful to build a $S_{4}$ flavoured model with a much more economical scalar content. To our knowledge, $A_4$ has not been considered before in the 3+1 scheme with $B-L$ extension.

This paper is organized as follows. The description of the model is presented in section \ref{model}. Its implications in lepton masses and mixings
as well as the corresponding numerical analysis is presented in section \ref{neutrinomixing}. Section \ref{gminus2s} is devoted to the muon anomalous magnetic
moment. Finally, some conclusions are given in section \ref{conclusion}.
Appendix \ref{Higgspotential} provides the scalar potential of the model.
\section{The model \label{model}}
The $B-L$ gauge model \cite{U1X3, U1X4} is supplemented by one discrete symmetry $S_4$ along with two Abelian symmetries $Z_3$ and $Z_4$. Further,
three right-handed neutrinos \Revised{$\nu_R$}, one sterile neutrino \Revised{$\nu_s$},
one $SU(2)_L$ doublet ($H^'$) and four singlet scalars ($\phi, \varphi, \rho, \phi_s$) are additional introduction
to the $B-L$ model. Three right-handed neutrinos
and three left-handed leptons \Revised{$\psi_L$}
are grouped
 in $\mathbf{3}$ under $S_4$ symmetry whereas the first right-handed charged leptons $l_{1R}$ is assigned as
 $\mathbf{1}$ and the two others are grouped  in $\mathbf{2}$ under the $S_4$ symmetry. The sterile neutrino $\nu_{s}$ is assigned as
 $\mathbf{1}$ under the $S_4$ symmetry. The particle content of the model under consideration\footnote{Under $SU(3)_C$ symmetry, all leptons and scalars are singlets.} and its \Revised{corresponding} assignments under the symmetry $SU(2)_L\times U(1)_Y\times U(1)_{B-L} \times S_4\times Z_3\times Z_4\equiv \mathbf{\Gamma}$ are shown in Table \ref{lepcont}.
\begin{table}[h]
\caption{\label{lepcont} The particle and scalar contents of the model.}
\begin{center}
\begin{tabular}{|c|cc|ccccc|cc|}
\hline
Fields  &\hspace{0.25cm}$\psi_{L}$  &$l_{1R} (l_{\al R})$ &\hspace{0.25cm}$H \, (H^')$&$\phi\, (\varphi)$&$\rho$&$\chi$&$\phi_s$
&\hspace{0.2cm}$\nu_{R}$&\hspace{0.15cm}$\nu_s$ \\ \hline
$[\mathrm{SU}(2)_L, \mathrm{U}(1)_Y]$  & \hspace{0.2cm}$[\textbf{2}, -1/2]$ &$[\textbf{1}, -1]$&$[\textbf{2}, 1/2]$&$[\textbf{1}, 0]$ &$[\textbf{1}, 0]$& $[\textbf{1}, 0]$& $[\textbf{1}, 0]$
&\hspace{0.2cm}$[\textbf{1}, 0]$ &\hspace{0.15cm} $[\textbf{1}, 0]$  \\
$\mathrm{U}(1)_{B-L}$ & $-1$ &$-1$&$0$&$0$ &$0$& $2$& $0$&$-1$& $-1$ \\
$S_4$&\hspace{0.15cm}  $\mathbf{3}$  &$\mathbf{1} (\mathbf{2})$&\hspace{0.15cm}$\mathbf{1}\, (\mathbf{1}^')$ &$\mathbf{3}$ & $\mathbf{1}$& $\mathbf{1}$&$\mathbf{3}$
&\hspace{0.2cm}$\mathbf{3}$&\hspace{0.15cm} $\mathbf{1}$  \\
$Z_3$&\hspace{0.15cm}  $\om$  &$1$&$1$&$\om \, (1)$\hspace{0.1cm}&$1$& $\om$  &\hspace{0.1cm} $\om^2$
&\hspace{0.15 cm}$\om$&\hspace{0.1cm} $\om^2$ \\
$Z_4$&\hspace{0.15cm}  $i$  &$1$&$1$&$i$ &$i$&$1$& $i$&\hspace{0.15cm}$1$& $-i$ \\ \hline
\end{tabular}
\end{center}
\vspace*{-0.3cm}
\end{table}\\
With the above specified particle content, the following five dimensional Yukawa interactions invariant under the symmetries of the model arise:
\bea -\mathcal{L}_{clep}&=&\fr{x_{1}}{\La} (\bar{\psi}_{L} l_{1R})_{\Revised{\mathbf{3}}} (H \phi)_{\Revised{\mathbf{3}}}
+ \fr{x_{2}}{\La } (\bar{\psi}_{L} l_{\al R})_{\Revised{\mathbf{3}}} (H\phi)_{\Revised{\mathbf{3}}}
+\fr{x_{3}}{\La } (\bar{\psi}_{L} l_{\al R})_{\Revised{\mathbf{3}^'}} (H^'\phi)_{\Revised{\mathbf{3}^'}}\crn
&+& \frac{x_{1\nu}}{\Lambda} (\bar{\psi}_{L} \nu_{R} )_{\Revised{\mathbf{1}}} (\widetilde{H}\rho)_{\Revised{\mathbf{1}}}+
\frac{x_{2\nu}}{\Lambda} (\bar{\psi}_{L} \nu_{R} )_{\textbf{3}}(\widetilde{H}\varphi)_{\textbf{3}}
+ \frac{x_{3\nu}}{\Lambda} (\bar{\psi}_{L} \nu_{R} )_{\textbf{3}^'}(\widetilde{H^'}\varphi)_{\textbf{3}^'} \crn
&+&\frac{y_{\nu}}{2} (\bar{\nu}^c_{R}\nu_{R})_{\Revised{\mathbf{1}}} \chi
 + \frac{y_s}{\Lambda} (\bar{\nu}^c_{s}\nu_{R})_{\Revised{\mathbf{3}}}(\chi\phi_s)_{\Revised{\mathbf{3}}} +H.c. \label{Lyukawa}
\eea
In Eq. (\ref{Lyukawa}), $\Lambda$ is the cut-off scale of the theory, $y_{\nu}, y_s$, $x_{k}$ and $x_{k\nu}\, (k=1,2,3)$ are dimensionless constants. The VEVs of scalar fields  satisfying the scalar potential minimum condition, 
presented in section \ref{Higgspotential}, are given as follows
 \bea
&&\langle H \rangle =\left(  0 \hspace{0.3 cm}   v\right)^T,\hspace{0.2cm} \langle H^' \rangle =\left(  0 \hspace{0.3 cm}   v^'\right)^T,
\hspace{0.2cm} \langle \phi \rangle = (\langle \phi_{1} \rangle,\hspace{0.2 cm} \langle \phi_{2}
\rangle, \hspace{0.1 cm}\langle \phi_{3} \rangle),\hspace{0.1 cm} \langle \phi_{1} \rangle =\langle \phi_{2} \rangle =\langle \phi_{3} \rangle= v_{\phi}, \label{scalarvev}\\
&&\langle \varphi \rangle = (0 \hspace{0.15 cm} \langle \varphi_{2} \rangle \hspace{0.15 cm} 0), \hspace{0.05 cm} \langle \varphi_{2} \rangle = v_{\varphi}, \hspace{0.05 cm} \langle \rho \rangle =v_{\rho},\hspace{0.05 cm} \langle \chi \rangle =v_{\chi}, \hspace{0.05 cm}
\langle \phi_s \rangle =(\langle \phi_{1s} \rangle \hspace{0.15 cm} 0 \hspace{0.15 cm} \langle \phi_{3s} \rangle), \hspace{0.05 cm}
\langle \phi_{1s} \rangle=\langle \phi_{3s} \rangle =v_s. \hspace{0.5 cm} \nonumber \eea
In fact the electroweak symmetry breaking scale is one order of magnitude lower than TeV scale and the $B-L$ scale is assumed to be at the TeV scale, i.e.,
\bea
v^'\sim v =1.23\times 10^{2}\, \mathrm{GeV},\hs v_\chi=10^3\, \mathrm{GeV}. \label{SMBLscale} \eea
Furthermore, in order to have very heavy right handed Majorana neutrino masses, thus allowing the implementation of the type I seesaw mechanism that produces the tiny masses of the active neutrinos, the VEVs of flavons $\phi,\varphi,\rho$ and $\phi_s$ are assumed to be at a very high scale:
\bea
v_\phi \sim v_\varphi \sim v_\rho \sim v_{\phi_s} = 10^{11} \, \mathrm{GeV}. \label{flavonscale}
\eea
\Revised{It is interesting to note that each of additional symmetry $U(1)_{B-L}, S_4, Z_3$ and $Z_4$ plays a crucial role in forbidding the unwanted terms, which are listed in Table \ref{forbidtable} of Appendix \ref{forbidappen}, to obtain the lepton mass matrices in Eqs. (\ref{Mltq}) and (\ref{MDRS}). 
Furthermore, there exist invariant terms via Weinberg operators whose dimensions greater than or equal to six, $\fr{1}{\La^{2(k+l+1)}}(\overline{\psi}_L \psi^c_L) ({\widetilde{\mathbf{H}}}^2 \chi^*) (\mathbf{H}^\+\mathbf{H})^k (\mathbf{P}^*\mathbf{P})^l$ with $k, l= 0, 1, 2, ...$; $\mathbf{H}=H, H^'$, and $\mathbf{P}=\phi, \varphi, \rho, \chi, \phi_s$. The fact that $v_H\ll v_{\mathbf{P}} \ll \La$. Hence, the 
neutrino mass obtained by those Weinberg operators, 
 $v_{\mathbf{H}} \left(\fr{ v_{\mathbf{H}} }{\La}\right)^{2k+1} \left(\fr{v_P}{\La}\right)^{2l+1}$, is very small
 compared to the one obtained by the type-I seesaw mechanism as
 in Eq.(\ref{MDRS}) and thus have been ignored.}
\section{\label{neutrinomixing} Neutrino mass and mixing}
\subsection{Lepton mass and mixing in the three neutrino framework}
From the Yukawa interactions in Eq. (\ref{Lyukawa}), and using the tensor product rules of the $S_4$ discrete group \cite{ishi},
together with the VEVs of scalar fields in Eq. (\ref{scalarvev}), the charged lepton mass 
Lagrangian is written in the form:
 \bea
\mathcal{L}^{\mathrm{mass}}_{cl}
&=& - (\bar{l}_{1L} \hs \bar{l}_{2L}\hs \bar{l}_{3L})
M_l (l_{1R}\hs l_{2R}\hs l_{3R})^T+H.c, \eea
where
\bea M_l=\left(%
\begin{array}{ccc}
x_1 v \frac{ v_{\phi}}{\Lambda} & \frac{v_\phi}{\Lambda}\left(x_2 v+x_3 v^'\right) &\frac{v_\phi}{\Lambda}\left(x_2 v-x_3 v^'\right)   \\
x_1 v \frac{ v_{\phi}}{\Lambda} &  \om^2 \frac{v_\phi}{\Lambda}\left(x_2 v+x_3 v^'\right) &  \om \frac{v_\phi}{\Lambda}\left(x_2 v-x_3 v^'\right)\\
x_1 v \frac{ v_{\phi}}{\Lambda} &  \om \frac{v_\phi}{\Lambda}\left(x_2 v+x_3 v^'\right) & \om^2 \frac{v_\phi}{\Lambda}\left(x_2 v - x_3 v^'\right)\\
\end{array}%
\right), \label{Mltq}\eea
which is diagonalized as, $ U^\dagger_l M_l U_r= \mathrm{diag}(m_e, \, m_\mu , \,m_\tau)$, with
\bea U^\dagger_l&=&\fr{1}{\sqrt{3}}\left(%
\begin{array}{ccc}
  1 &\,\,\, 1 &\,\,\, 1 \\
  1 &\,\,\, \om &\,\,\, \om^2 \\
  1 &\,\,\, \om^2 &\,\,\, \om \\
\end{array}%
\right),\hs U_r=\mathbf{I}_{3\times 3} \hs\,\, \left(\om=e^{\frac{i2\pi}{3}}\right),  \label{Uclep}\\
m_e &=&\sqrt{3} x_1 v \frac{ v_{\phi}}{\Lambda},\hs  m_{\mu,\tau}= \sqrt{3} \frac{v_\phi}{\Lambda}\left(x_2 v \pm x_3 v^'\right). \label{memtv}\eea
Equation (\ref{Uclep}) tells us that $U_l$ is non trivial
and hence it will affect on
the lepton mixing matrix. Furthermore, Eq. (\ref{memtv}) shows that $m_{\mu}$ and $m_{\tau}$ are distinguished by $v^'$. This is the reason why $H^'$ is additionally introduced to $H$.

Furthermore, assuming the gauge singlet scalar fields adjure VEVs much larger than the electroweak symmetry breaking scale, the scalar spectrum at low energies is the same as in the 2HDM theory. Then, the physical CP even and CP odd neutral scalar fields are given by:
\begin{eqnarray}
H_{R}^{0} &=&\sin \alpha h-\cos \alpha H,\hspace{0.6cm}H_{R}^{\prime 0}=-\cos
\alpha h-\sin \alpha H, \\
H_{I}^{0} &=&\cos \beta G_{Z}+\sin \beta A^{0},\hspace{0.2 cm}H_{I}^{\prime
0}=\sin \beta G_{Z}-\cos \beta A^{0},
\end{eqnarray}
where $G_{Z}$ is the Goldstone boson associated with the longitudinal component of the gauge boson $Z$, $h$ is the $126$ GeV SM like Higgs, $H$ is the heavy CP even Higgs, $A$ is the heavy CP odd Higgs boson, $\tan\beta=\frac{v}{v^'}$ and $\alpha$ is the mixing angle in the neutral scalar sector. Then, the leptonic Yukawa interactions involving neutral scalar fields are given by:
\bea
&&\mathcal{L}_{Y}^{h}=\sqrt{\frac{3}{2}}y_{1}\sin \alpha \overline{e}%
_{L}he_{R}+\sqrt{\frac{3}{2}}\left( y_{2}\sin \alpha -y_{3}\cos \alpha
\right) \overline{\mu }_{L}h\mu _{R}\crn
&&\hspace{0.6 cm}+\, \sqrt{\frac{3}{2}}\left( y_{2}\sin
\alpha +y_{3}\cos \alpha \right) \overline{\tau }_{L}h\tau _{R}+h.c,\label{Lyh}\\
&&\mathcal{L}_{Y}^{H^{0}}=-\sqrt{\frac{3}{2}}y_{1}\cos \alpha \overline{e}%
_{L}H^{0}e_{R}+\sqrt{\frac{3}{2}}\left( y_{2}\cos \alpha +y_{3}\sin \alpha
\right) \overline{\mu }_{L}H^{0}\mu _{R}\crn
&&\hspace{0.8 cm}+\,\sqrt{\frac{3}{2}}\left( y_{2}\cos
\alpha -y_{3}\sin \alpha \right) \overline{\tau }_{L}H^{0}\tau _{R}+h.c,\label{LyH}\\
&&\mathcal{L}_{Y}^{A^{0}}=i\sqrt{\frac{3}{2}}y_{1}\sin \beta \overline{e}%
_{L}A^{0}e_{R}+i\sqrt{\frac{3}{2}}\left( y_{2}\sin \beta -y_{3}\cos \beta
\right) \overline{\mu }_{L}A^{0}\mu _{R}\crn
&&\hspace{0.8 cm}+\, i\sqrt{\frac{3}{2}}\left( y_{2}\sin
\beta +y_{3}\cos \beta \right) \overline{\tau }_{L}A^{0}\tau _{R}+h.c,\label{LyA}
\eea
where
\begin{equation}
y_k=x_k\frac{v_{\phi}}{\Lambda}\hspace{0.3 cm} (k=1,2,3).
\end{equation}
Equations (\ref{Lyh}), (\ref{LyH}) and (\ref{LyA}) imply that the flavour changing leptonic neutral scalar interactions are absent in our model. Consequently, there are no scalar contributions to the charged lepton flavor violating decays $\tau\to\mu\gamma$, $\tau\to e\gamma$ and $\mu\to e\gamma$. The charged lepton flavor violating decays $\tau\to\mu\gamma$, $\tau\to e\gamma$ and $\mu\to e\gamma$ will receive one loop level contributions due to the electrically charged scalars and right handed Majorana neutrinos, however these contributions will scale with the inverse of fourth power of the model cutoff as well as with the fourth power of the very small neutrino Yukawa coupling, thus making their branching ratios very tiny. Another contributions to the charged lepton flavor violating decays will arise from the virtual exchange of a $W$ gauge boson and sterile neutrinos however those contributions will have very tiny branching ratios as in the SM.

As we will see in the following, our model can successfully accommodate the SM charged lepton masses. Now, comparing the obtained result in Eq. (\ref{memtv}) with the experimental values of $m_{e, \mu, \tau}$ \cite{PDG2020},  $m_e=0.51099 \,\mathrm{MeV},  m_\mu = 105.65837\,\mathrm{MeV}, m_\tau = 1776.86 \,\mathrm{MeV}$, and considering the benchmark point $v=v^' =123\, \mathrm{GeV},\, v_{\phi}=10^{11}\, \mathrm{GeV}$ and $\Lambda =10^{13}\, \mathrm{GeV}$, we get:
\bea
|x_{1}| \sim 10^{-4},\hs |x_{2}| \sim |x_{3}| \sim 10^{-1}.
\eea
Furthermore, from the above given leptonic Yukawa interactions and using the benchmark point given above, it follows that the coupling of the $126$ GeV SM like Higgs boson with the SM lepton-antilepton pair is about $0.7$ the SM expectation.\\
From the particle content given in Tab. \ref{lepcont}, we get the following neutrino Yukawa interactions:
\bea &&-\mathcal{L}^{l}_{Y} =\frac{x_{1\nu}}{\Lambda} (\bar{\psi}_{1L} \nu_{1R} +\bar{\psi}_{2L} \nu_{2R}+\bar{\psi}_{3L} \nu_{3R})\widetilde{H} \rho \crn
&&+\frac{x_{2\nu}}{\Lambda} \left[(\bar{\psi}_{2L}\nu_{3R}+\bar{\psi}_{3L}\nu_{2R})(\widetilde{H} \varphi_1)
+(\bar{\psi}_{3L}\nu_{1R}+\bar{\psi}_{1L}\nu_{3R})(\widetilde{H} \varphi_2)
+(\bar{\psi}_{1L}\nu_{2R}+\bar{\psi}_{2L}\nu_{1R})(\widetilde{H} \varphi_3)\right]\crn
&&+\frac{x_{3\nu}}{\Lambda} \left[(\bar{\psi}_{2L}\nu_{3R}-\bar{\psi}_{3L}\nu_{2R})(\widetilde{H^'} \varphi_1)
+(\bar{\psi}_{3L}\nu_{1R}-\bar{\psi}_{1L}\nu_{3R})(\widetilde{H^'} \varphi_2)
+(\bar{\psi}_{1L}\nu_{2R}-\bar{\psi}_{2L}\nu_{1R})(\widetilde{H^'} \varphi_3)\right]  \crn
&&+ \frac{y_{\nu}}{2} (\bar{\nu}^c_{1R}\nu_{1R}+\bar{\nu}^c_{2R}\nu_{2R}+\bar{\nu}^c_{3R}\nu_{3R}) \chi
 + \frac{y_s}{\Lambda} \left(\bar{\nu}^c_{s}\nu_{1R}  \chi \phi_{1s}+ \bar{\nu}^c_{s}\nu_{2R}   \chi \phi_{2s}
+\bar{\nu}^c_{s}\nu_{3R}   \chi \phi_{3s}\right)
 + \mathrm{H.c}. \label{Lylep0}\eea
With the VEV configuration given in Eq. (\ref{scalarvev}), after symmetry breaking,
the $7\times 7$ neutrino mass matrix in the $(\nu^c_{L},\nu_{R}, \nu_s)$ basis takes the form
\bea
M_{\nu}^{7\times7} = \begin{pmatrix}
0     & M_{D} &0 \\
M^T_{D}  & M_{R} &M^T_S \\
0 & M_S   & 0
\end{pmatrix}, \label{Mnu77}
\eea
where $M_{D}, M_{R}$ and $M_S$ are the Dirac, Majorana and sterile neutrino mass matrices, respectively. They are given by:
\bea M_D&=&
\left(%
\begin{array}{ccc}
\frac{v_\rho}{\Lambda} v x_{1\nu}  & 0   & \frac{v_{\varphi}}{\Lambda}\left( v x_{2\nu} -v^' x_{3\nu} \right)\\
 0                                          &\frac{v_\rho}{\Lambda} v x_{1\nu} & 0 \\
\frac{v_{\varphi}}{\Lambda}\left( v x_{2\nu} + v^' x_{3\nu} \right)  & 0   &\frac{v_\rho}{\Lambda} v x_{1\nu} \\
\end{array}%
\right),  \crn
 M_R&=&y_\nu v_\chi \mathbf{I}, \hs\hs M_{\Revised{S}}= y_s v_\chi \frac{v_s}{\Lambda} (1 \hspace{0.35 cm} 0\hspace{0.35 cm} 1). \label{MDRS}\eea
 In the case of $M_D < M_S \ll M_R $,
 the effective $4\times4$ light neutrino
mass matrix 
is determined by\cite{Barry2011, Zhang2012, Das2019ea, Krishnan2020}:
\begin{equation}
M_{\nu}= -\left( \begin{array}{cc}
              M_{D} M^{-1}_R M^T_D &  M_D M^{-1}_R M^T_S\\
              M_S (M^{-1}_R)^TM^T_{D} & M_SM^{-1}_{R}M^T_S
                      \end{array} \right).
\label{Mnu44}
\end{equation}
The expression (\ref{MDRS}) shows that, in the neutrino sector, there are five complex parameters, thus implying the existence of ten real parameters. Considering the case of real VEVs for the scalars $H, H^', \phi, \chi$ and $\eta$ while taking the VEV of $\varphi$ to be complex ($v_{\varphi}=v_{0} e^{i\alpha}$), the phase redefinition of
$\psi_L$ and $\nu_R$ allows to rotate away the phases of three Yukawa couplings $x_{1\nu}$, $y_{\nu}$ and $y_{s}$, reducing to seven parameters. Furthermore, two parameters are absorbed during the formation of the neutrino mass matrix.
Thus, there are left five real and dimensionless parameters
$k_{2, 3},  m, m_s$ and $\alpha$, where $k_{2, 3},  m$ and $m_s$ are defined as follows
\bea
k_{2}=\frac{v_0}{v_\rho}\frac{x_{2\nu}}{x_{1\nu}},\hs k_3=\frac{v_0}{v_\rho}\frac{v^'}{v}\frac{x_{3\nu}}{x_{1\nu}}, \hs m=\frac{v^2 v^2_\rho  x_{1\nu}^2}{\Lambda^2 v_\chi y_{\nu}}, \hs m_s=\frac{2v_\chi v_s^2 y_s^2}{\Lambda^2 y_{\nu}}. \label{k2k3mms}
\eea
Combining Eqs.(\ref{MDRS}) and (\ref{Mnu44}) yields the $4\times 4$ active-sterile mass matrix in the explicit form
\bea
 &&M^{4\times4}_{\nu} =-\left(
\begin{array}{cccc}
\hspace{-0.1 cm}m\left(\begin{array}{ccc}
1 + (k_2 - k_3)^2 e^{2 i \al}  \hs  & 0 & 2 k_2 e^{i \al } \\
                       0 & 1 & 0\\
2 k_2 e^{i \al } & 0 &\hs 1 + (k_2 + k_3)^2 e^{2 i \al}  \\
                      \end{array}\right) & \sqrt{\frac{m m_s}{2}}\left(\begin{array}{c}
                    1+ (k_{2}-k_{3})e^{i \al} \\
                    0 \\
                 1+ (k_{2}+k_{3})e^{i \al}\\
                      \end{array}\right)\\
\hspace{-0.6 cm}\sqrt{\frac{m m_s}{2}}\left(\begin{array}{ccc}
1+ (k_{2}-k_{3})e^{i \al}\hspace{0.15 cm}&0\hspace{0.5 cm}& 1+ (k_{2}+k_{3})e^{i \al} \\
                      \end{array}\right) &\hspace{1.25 cm}  m_{s}\\
\end{array}\right). \crn
\label{M44separate}
\eea
In the case of $M_D < M_{\Revised{S}}$, one can apply the type-I seesaw mechanism on Eq.(\ref {Mnu44}) to get the active neutrino mass matrix as \cite{Barry2011, Zhang2012, Das2019ea, Krishnan2020}
\bea
 M_\nu &=&  M_D M_R^{-1}M_{\Revised{S}}^T (M_{\Revised{S}} M_R^{-1} M_{\Revised{S}}^T)^{-1} M_{\Revised{S}} M_R^{-1} M_D^T - M_D M_R^{-1} M_D^T \crn
 &=&\frac{m}{2}\left(
\begin{array}{ccc}
 -\left[ (k_{2}-k_{3})e^{i \al}-1\right]^2 & 0 & 1+ (k_{2}^2-k_{3}^2)e^{2 i \al} -2 e^{i \al} k_{2}\\
 0 & -2 & 0 \\
 1+ (k_{2}^2-k_{3}^2)e^{2 i \al} -2 e^{i \al} k_{2} & 0 & -\left[ (k_{2}+k_{3}) e^{i \al} -1\right]^2 \\
\end{array}
\right).\label{Mnu33}
\eea
We note that $M_\nu$ is complex because $m, k_{2,3}$ and $\alpha$ are real parameters.
Hence, to determine the active neutrino masses, we define a Hermitian matrix $\mathbf{m}^{2}_{\nu}$ given by
\bea
\mathbf{m}^{2}_{\nu}= M_\nu M^\+_{\nu}
=\frac{m^2 k_0}{2}  \left(%
\begin{array}{ccc}
k_{-} &0 &\hs 2 k_{3}^2- k_{0}-i 2 k_{3} \sin \al  \\
 0          &\frac{2}{k_0} & 0  \\
2 k_{3}^2- k_{0}+ i 2 k_{3} \sin \al \hs  & 0 & k_{+} \\
\end{array}%
\right),\label{Mnu33}
\eea
where
\bea
k_0=1+k_2^2+k_3^2-2 k_2 \cos \al, \hs k_\mp =1+(k_2 \mp k_3)^2-2 (k_2 \mp k_3)\cos \al. \label{k0kpm}
\eea
The squared matrix $\mathbf{m}^{2}_{\nu}$ in Eq. (\ref{Mnu33}) is diagonalized by the rotation matrix $U_{\nu}$ satisfying
\be
U_{\nu }^\+ \mathbf{m}^{2}_{\nu} U_{\nu }=\left\{
\begin{array}{l}
\left(%
\begin{array}{ccc}
0\quad & 0 &0 \\
0\quad  & m^2 & 0 \\
0\quad & 0 & m^2 k_0^2\\
\end{array}%
\right),\hspace{0.1cm} U_{\nu }=\left(%
\begin{array}{ccc}
g_0 (g_1 + ig_2)&0&-r_0 (g_1 + ig_2) \\
0                          &1& 0  \\
\frac{1}{\sqrt{2} g_0} &0&\frac{1}{\sqrt{2} r_0} \\
\end{array}%
\right) \hspace{0.2cm}\mbox{for NH,}\ \  \\
\left(%
\begin{array}{ccc}
m^2 k_0^2& 0 &\hspace{0.1 cm} 0 \\
0 & m^2  &\hspace{0.1 cm}  0 \\
0& 0 & \hspace{0.1 cm} 0\\
\end{array}%
\right),\hspace{0.1cm} U_{\nu }=\left(%
\begin{array}{ccc}
-r_0 (g_1 + ig_2)&0&  g_0 (g_1 + ig_2)\\
0                          &1& 0  \\
\frac{1}{\sqrt{2} r_0} &0&\frac{1}{\sqrt{2} g_0}  \\
\end{array}%
\right) \hspace{0.2cm}\mbox{for IH,}
\end{array}%
\right.  \label{Unu}
\ee
where
\bea
&&g_{0}=\sqrt{\frac{k_0}{k_{-}}}, \hs r_{0}=\sqrt{\frac{k_0}{k_{+}}}, \hs g_1=\frac{1}{\sqrt{2}}-\frac{\sqrt{2} k_{3}^2}{k_0}, \hs g_2=\frac{\sqrt{2} k_{3} \sin \al}{k_0}. \label{g0g1g2r0}
\eea
The corresponding leptonic mixing matrix is
\bea
U_{\mathrm{lep}}=U_{L}^{\dag} U_{\nu }=\left\{
\begin{array}{l}
\fr{1}{\sqrt{3}}\left(
\begin{array}{ccc}
\frac{1}{\sqrt{2} g_{0}}+ g_{0} (g_{1}+i g_{2}) & 1 & \frac{1}{\sqrt{2} r_{0}}-(g_{1}+ig_{2}) r_{0} \\
 \frac{\om^2}{\sqrt{2} g_{0}}+g_{0} (g_{1}+ig_{2}) & \om & \frac{\om^2}{\sqrt{2} r_{0}}-(g_{1}+i g_{2} ) r_{0} \\
\frac{\om}{\sqrt{2} g_{0}}+ g_{0} (g_{1}+ig_{2}) & \om^2 & \frac{\om}{\sqrt{2} r_{0}}-(g_{1}+ig_{2}) r_{0} \\
\end{array}
\right) \hspace{0.1cm}\mbox{for  NH},  \label{Ulep}  \\
\fr{1}{\sqrt{3}}\left(
\begin{array}{ccc}
\frac{1}{\sqrt{2} r_{0}}-(g_{1}+ig_{2}) r_{0}& 1 &\frac{1}{\sqrt{2} g_{0}}+ g_{0} (g_{1}+i g_{2})  \\
\frac{\om^2}{\sqrt{2} r_{0}}-(g_{1}+i g_{2} ) r_{0} & \om &\frac{\om^2}{\sqrt{2} g_{0}}+g_{0} (g_{1}+ig_{2})  \\
\frac{\om}{\sqrt{2} r_{0}}-(g_{1}+ig_{2}) r_{0}& \om^2 &\frac{\om}{\sqrt{2} g_{0}}+ g_{0} (g_{1}+ig_{2})  \\
\end{array}
\right) \hspace{0.1cm}\mbox{for IH}.
\end{array}%
\right.
\eea
Three light neutrino masses are given by
\bea
\left\{
\begin{array}{l}
m^2_1=0, \hspace{1.0 cm} m^2_2 =m^2 , \hspace{0.4 cm} m^2_3=k_0^2 m^2\hspace{0.4cm}\mbox{for \, NH},    \\
m^2_1= k_0^2 m^2, \hspace{0.3 cm} m^2_2 =m^2, \hspace{0.4 cm} m^2_3=0\hspace{1.1cm}\,\mbox{for\, IH},
\end{array}%
\right. \label{m1m2m3}
\eea
which implies the neutrino mass ordering should be either $(0, m, m k_0)$ or $(m k_0, m, 0)$. It is important
to note that \Revised{in the considered model} both NH ($m_1<m_2 \Revised{<m_3}$) and IH ($\Revised{m_3<} m_1<m_2$) are satisfied \Revised{due to the fact that $k_0$ in Eq. (\ref{k0kpm}) can be greater or less than the one} in the case of
\Revised{$k_2$ and $k_3$} are real parameters. Hence, the model can predict both NH and IH spectrum
which are consistent with the recent experimental data \cite{Salas2020} and different from that of Ref. \cite{Krishnan2020} in which only the NH is allowed.

  In the three-neutrino scheme \cite{PDG2020}, the lepton mixing angles are determined from Eq. (\ref{Ulep}),
\bea &&s_{13}^2=\left| \mathrm{U}_{e 3}\right|^2=\left\{
\begin{array}{l}
\fr{2}{3} \frac{k_3^2}{k_0}\hspace{0.3cm}\mbox{for \, NH},    \\
\frac{2}{3}-\frac{2}{3} \frac{k_{3}^2}{k_0}\hspace{0.25cm}\,\mbox{for \, IH},
\end{array}%
\right.  \label{s13sq}\\
&& s_{12}^2 =\frac{\left| \mathrm{U}_{e 2}\right|^2}{1-\left| \mathrm{U}_{e 3}\right|^2}=
\frac{1}{3c^2_{13}}\hspace{0.25cm}\,\mbox{for \, both \,NH\, and \, IH}, \label{s12s13relation}\\
&& s_{23}^2=\frac{\left| \mathrm{U}_{\mu 3}\right|^2}{1-\left| \mathrm{U}_{e 3}\right|^2}=
\left\{
\begin{array}{l}
\fr{1}{2}+\frac{\sqrt{3} k_3 \sin \al}{3k_0-2k_3^2}\hspace{0.3cm}\mbox{for \, NH},    \\
\frac{1}{2}-\frac{\sqrt{3} k_{3} \sin \alpha}{k_0+2 k_{3}^2} \hspace{0.25cm}\,\mbox{for \, IH},
\end{array}%
\right. \label{s23sq}\eea
where $s_{ij}=\sin \theta_{ij}$, $c_{ij}=\cos \theta_{ij}$, $t_{12}=\frac{s_{12}}{c_{12}}$ and $t_{23}=\frac{s_{23}}{c_{23}}$ with
$\theta_{ij}$ are neutrino mixing angles.

Furthermore, from Eqs. (\ref{m1m2m3}), we can express $m$ and $k_0$ in terms of two observables $\Delta m^2_{21}$ and $\Delta m^2_{31}$ as follows:
\bea
&&m=\left\{
\begin{array}{l}
\sqrt{\Delta m^2_{21}}\hspace{1.6cm}\mbox{for NH},    \\
\sqrt{\Delta m^2_{21}-\Delta m^2_{31}} \hspace{0.1cm}\,\mbox{for\,  IH},
\end{array}%
\right. \label{mv}\\
&&k_{0}=\left\{
\begin{array}{l}
\frac{\sqrt{\Delta m^2_{31}}}{m}\hspace{0.4cm}\mbox{for \, NH},    \\
\frac{\sqrt{-\Delta m^2_{31}}}{m} \hspace{0.1cm}\,\mbox{for \, IH}{\Revised{.}}
\end{array}
\right. \label{k0v}\eea
The fact that the neutrino mass ordering can be normal ($m_1< m_2 < m_3$) or inverted ($m_3< m_1< m_2$) depending on the sign of $\Delta m^2_{31}$ \cite{Salas2020, Kelly2021}. We will show that the considered model can provide the satisfied explanation on neutrino masses and mixings data, for both three-neutrino scheme and $3+1$ scheme, given in Table \ref{experconstrain}.
\subsection{3+1 sterile-active neutrino mixing \label{neutrino4}}
In the case of $M_D < M_{\Revised{S}}$, the mass of the $4^\text{th}$ mass eigenstate is given by\cite{Barry2011, Zhang2012, Das2019ea, Krishnan2020},
\bea
 m_4= M_{\Revised{S}} M_R^{-1}M_{\Revised{S}}^T=m_s. \label{m4}
\eea
Combining Eqs. (\ref{m1m2m3}) and (\ref{m4}) yields:
\bea
m_{s}=m_4=\left\{
\begin{array}{l}
\sqrt{\Delta m^2_{41}}\hspace{1.5cm}\mbox{for  NH},    \\
\sqrt{m^2 k^2_0 +\Delta m^2_{41}} \hspace{0.1cm}\,\mbox{for IH}.
\end{array}%
\right. \label{msm4}
\eea
The $4\times 4$ neutrino mixing matrix is given by \cite{Barry2011, Zhang2012, Das2019ea, Krishnan2020}:
\bea
U =U^\dagger_L U_\nu= \left(\begin{array}{cc}
       U^\dagger_L(1-\frac{1}{2}R R^\dagger) U_\nu & U^\dagger_L R \\
       -R^\dagger U_\nu & 1-\frac{1}{2}R^\dagger R
       \end{array}\right), \label{U44}
\eea
where
$R$ is given by \cite{Barry2011, Zhang2012, Das2019ea, Krishnan2020}
\bea
  R = M_D M_R^{-1}M_{\Revised{S}}^T (M_{\Revised{S}} M_R^{-1}M_{\Revised{S}}^T)^{-1}=\sqrt{\frac{m}{2m_{s}}}\left(
\begin{array}{c}
 1+ (k_2-k_3) e^{i \al} \\
 0 \\
 1+ (k_2+k_3) e^{i \al} \\
\end{array}
\right). \label{Rmatrix}
\eea
Combining Eqs. (\ref{Uclep}) and (\ref{Rmatrix}), we find that the strength of the active-sterile mixing is given by
\bea
U^\dagger_L R&=&\sqrt{\frac{m}{6m_s}}\left(
\begin{array}{c}
2+2 k_2 e^{i \al} \\
\left[\om^2 (k_2+k_3)+k_2-k_3\right] e^{i \al} -\om\\
\left[\om (k_2+k_3)+k_2-k_3\right] e^{i \al} -\om^2 \\
\end{array}
\right)\equiv \left(
\begin{array}{c}
U_{e4}  \\
U_{\mu 4} \\
U_{\tau 4} \\
\end{array}
\right), \label{Uemutau4}
\eea
which leads to
\bea
&&s^2_{14}= |U_{e4}|^2=\frac{2}{3}\frac{m}{m_s}\left(1+k_2^2+2 k_2 \cos \al\right), \crn
&&s^2_{24}=\frac{ |U_{\mu 4}|^2}{1- |U_{e 4}|^2}=\frac{1+k_{2}^2+3 k_{3}^2+2 k_2 \cos\alpha +2 \sqrt{3} k_{3} \sin\alpha }{ \frac{6m_s}{m}-4 \left(1+k_{2}^2+2 k_{2} \cos\alpha \right)},  \crn
&&s^2_{34}=\frac{ |U_{\tau 4}|^2}{1- |U_{e 4}|^2- |U_{\mu 4}|^2}=\frac{1+k_{2}^2+3 k_{3}^2+2k_{2} \cos \alpha  -2 \sqrt{3} k_{3} \sin \alpha}{\frac{6 m_s}{m}-\left(5+ 5k^2_{2}+3 k_{3}^2+10 k_{2} \cos \alpha +2 \sqrt{3} k_{3} \sin \alpha\right)}{\Revised{.}} \label{s14s24s34sq}
\eea

\subsection{\label{Meff}Effective neutrino mass parameter and Jarlskog invariant}
The Jarlskog invariant, obtained from Eq. (\ref{Ulep}), has the form \cite{Jarlskog1, Jarlskog2, Jarlskog3, PDG2020}
\bea
&&J_{CP}=\mathrm{Im} (U_{12} U_{23} U^*_{13} U^*_{22})=\left\{
\begin{array}{l}
\hspace{0.3 cm}\frac{k_3 (k_2-\cos\al)}{3\sqrt{3}k_0}\hspace{0.25cm}\mbox{for NH},    \\
-\frac{k_3 (k_2-\cos\al)}{3\sqrt{3}k_0} \hspace{0.175cm}\,\mbox{for  IH}.
\end{array}%
\right. \label{Jm}
\eea
The effective neutrino masses get the following forms \cite{betdecay2, betdecay4, betdecay5},
\bea
&&m_{\beta}^{(3)} = \sqrt{\sum^3_{i=1} \left|U_{ei}\right|^2 m_i^2 }=
\sqrt{\frac{m^2}{3} + \frac{ m^2 k_0 \left[\left(2 k_3^2-k_0+k_{+}\right)^2+4 k^2_3 \sin^2 \alpha\right]}{6 k_{+}}}, \label{mb3v}\\
&& \langle m_{ee}^{(3)}\rangle = \left| \sum^3_{i=1} U_{ei}^2 m_i \right|\crn
&&\hspace{1.1 cm}=\frac{m}{6} \sqrt{\frac{16 k_{3}^2 \sin^2\alpha \left(2 k_{3}^2-k_0+k_{+}\right)^2+\left[\left(k_{+}-k_0+2 k_3^2\right)^2-4 k_3^2 \sin^2\alpha+2 k_{+}\right]^2}{k_{+}^2}},\\
&& m_{\beta } = \sqrt{\sum^4_{i=1} \left|U_{ei}\right|^2 m_i^2 } 
=\left\{\frac{m^2}{3} + \frac{m^2 k_0 \left[\left(2 k_3^2-k_0+k_{+}\right)^2+4 k^2_3 \sin^2 \alpha\right]}{6 k_{+}} \right. \crn
&&\left.\hspace{3.65 cm}+\,\frac{2 m m_s}{3} \left[(k_2 \cos\alpha +1)^2+k_2^2 \sin^2\alpha\right]\right\}^{\frac{1}{2}}, \hspace{0.2 cm}\\
&& \langle m_{ee}\rangle = \left| \sum^4_{i=1} U_{ei}^2 m_i \right|
=\frac{m}{3} \left\{16 \sin^2 \alpha\left[2 k_2  k_{+} (k_2 \cos\alpha +1)+k_0 k_3-2 k_3^3-k_3 k_{+}\right]^2 \right. \crn
&&\hspace{3.35 cm} \left.+ \left[ k_0^2+6 k_{+} + 4 k_2 k_{+} (k_2\cos\alpha+2)\cos\alpha  -2 k_0\left(2 k_3^2+k_{+}\right) \right.\right. \crn
&&\hspace{3.3 cm} \left. \left. -\, 4 \left(k_2^2 k_{+}+k_3^2\right) \sin^2\alpha +\left(2 k_3^2+k_{+}\right)^2\right]^2\right\}^{\frac{1}{2}}{\Revised{.}} \label{meev}
\eea
\subsection{\label{NR} Numerical analysis}
Firstly, provided that $s_{13}$ is located inside the $3\sigma$ experimentally allowed range\footnote{In present work, numbers are displayed with 4 significant digits to the right of the decimal point.} taken from Ref. \cite{Salas2020}, i.e., $s^2_{13}\in (2.000, 2.405) 10^{-2}$ for NH and $s^2_{13}\in (2.018, 2.424) 10^{-2}$ for IH, from Eq. (\ref{s12s13relation}), we predict solar mixing angle
\bea
s^2_{12}\in \left\{
\begin{array}{l}
(0.3401, 0.3415) \,\hspace{0.3 cm} \mbox{\Revised{f}or \, NH,} \\
(0.3402, 0.3416)\,  \hspace{0.3 cm}  \mbox{\Revised{f}or \, IH,}%
\end{array}%
\right.  \label{s12constraint}
\eea
which is consistent with the $2\sigma$ experimental range given in Ref. \cite{Salas2020}.\\
Furthermore, Eqs. (\ref{mv}) and (\ref{k0v}) imply that
\bea
&&m\in\left\{
\begin{array}{l}
(8.331, 9.022) \, \mbox{meV} \hspace{0.15 cm} \mbox{for  NH,} \\
(49.39, 51.10)\, \mbox{meV} \hspace{0.15 cm}  \mbox{for  IH,}%
\end{array}%
\right.   
\hs k_0\in\left\{
\begin{array}{l}
(5.51, 6.16) \hspace{0.3 cm} \mbox{for \, NH,} \\
(0.9833, 0.9866) \hspace{0.15 cm}  \mbox{for  IH,}%
\end{array}%
\right.   \label{k0range}\\
&&\left\{
\begin{array}{l}
m_1= 0, \hspace{0.25 cm} m_2\in (8.331, 9.022)\, \mbox{\Revised{meV}}, \hspace{0.25 cm} m_3\in (49.70, 51.28)\, \mbox{\Revised{meV}} \hspace{0.3 cm} \mbox{for  NH,} \\
m_1\in (48.68, 50.30)\, \mbox{\Revised{meV}}, \hspace{0.25 cm} m_2\in (49.39, 50.10)\, \mbox{\Revised{meV}}, \hspace{0.25 cm} m_3=0 \hspace{0.35 cm}  \mbox{for  IH,}%
\end{array}%
\right.   \label{m3range}\\
&&{\sum}_{i=1}^{3} m_i \in\left\{
\begin{array}{l}
(58.03, 60.31) \, \mbox{\Revised{meV}}\hspace{0.325 cm} \mbox{for \hspace{0.01 cm} NH,} \\
(98.07, 101.30) \, \mbox{\Revised{meV}} \hspace{0.15 cm}  \mbox{for \hspace{0.07 cm}IH,}%
\end{array}%
\right.   \label{sumrange}
\eea
provided that the light neutrino mass-squared differences $\Delta m^2_{21}$ and $\Delta m^2_{31}$ to be varied in $3\sigma$ range of Ref. \cite{Salas2020} as shown in Table \ref{experconstrain}.

To find the allowed ranges of
$k_2$, $k_3$, $m, m_s$ and $\al$, and predictive ranges of the experimental parameters $\sin^2 \theta_{23}$, $\sin \delta$, $\langle m_{ee} \rangle$ and $s^2_{k4}\, (k=1, 2, 3)$, we utilize the observables $\Delta m^2_{21}$, $\Delta m^2_{31}, \Delta m^2_{41}$, $\sin^2 \theta_{13}$ and $\sin^2 \theta_{23}$ with values given in Table \ref{experconstrain}\Revised{\footnote{\Revised{As will be mentioned below, at present, there are various experimental bounds on $\Delta m^2_{41}$. In this work, $\Delta m^2_{41}$ is assumed in the range of $\Delta m^2_{41}\in (5.0, 10)\, \mathrm{eV}^2$ for NH while $\Delta m^2_{41}\in (30.0, 50.0)\, \mathrm{eV}^2$ for IH.}}}.

From Eqs. (\ref{k0kpm}), (\ref{s13sq})-(\ref{s23sq}) and (\ref{Jm}), we get:
\bea
&&k_2=\left\{
\begin{array}{l}
-\frac{1}{s_{13}} \sqrt{\frac{t_N k_{\Phi}}{2}+\Phi_N+s^2_{13}} \hspace{1.2 cm}  \mbox{for  NH,} \\
\sqrt{1+\frac{1}{2 - 3 s^2_{13}}\left(\frac{3 t_I k_\Phi }{2}+\sqrt{3} \Phi_I\right)}\hspace{0.3 cm}  \mbox{for IH,}%
\end{array}%
\right.   \label{k2expres}\\
&&k_3=\left\{
\begin{array}{l}
\sqrt{\frac{3t_N}{2}} s_{13} \hspace{1.3 cm}  \mbox{for NH,} \\
\sqrt{\left(1 - \frac{3}{2}s^2_{13}\right) t_I} \hspace{0.25 cm} \mbox{for  IH,}
\end{array}%
\right.   \label{k3expres}\\
&&\cos\alpha=\left\{
\begin{array}{l}
\frac{\sqrt{\frac{t_N k_{\Phi}}{2}+\Phi_N+s^2_{13}} \left(t_N c^4_{13} \cos^2 2\theta_{23}+\Phi_N-2 s^2_{13}\right)}{ \left[t_N  (3 s^2_{13}-2)+2\right] s^3_{13} } \hspace{1.9 cm} \mbox{for NH,} \\
\frac{\sqrt{2-3 s^2_{13}+\frac{3 t_I k_\Phi}{2}+\sqrt{3} \Phi_I} \left(3 t_I c^4_{13} \cos^2 2\theta_{23}+\sqrt{3} \Phi_I+6 s^2_{13}-4\right)}{\sqrt{2-3 s^2_{13}} \left(3 t_I s^2_{13}-2\right)}\hspace{0.3 cm}  \mbox{for IH, }
\end{array}%
\right.   \label{c1expres}\\
&&\sin\delta_{CP}=\left\{
\begin{array}{l}
-\frac{\sqrt{t_N k_\Phi+2 (\Phi_N+s^2_{13})} \left(s_{13} + \frac{t_N c^4_{13} \cos^2 2\theta_{23}+\Phi_N-2 s^2_{13}}{s_{13} \left[t_N (3 s^2_{13}-2)+2\right]}\right)}
{\sqrt{t_N} s^2_{13} \sin 2\theta_{23}\sqrt{2-3 s^2_{13}}} \hspace{1.6 cm} \mbox{for NH,} \\
-\frac{s_{13} \sqrt{2-3 s^2_{13}+\frac{3 t_I k_\Phi}{2}+\sqrt{3} \Phi_I} \left(\frac{3 t_I c^4_{13} \cos^2 2\theta_{23}+\sqrt{3}\Phi_I+6 s^2_{13}-4}{(3 s^2_{13}-2) \left(3 t_N s^2_{13}-2\right)}+1\right)}{\sqrt{6} s^2_{13} s_{23}c_{23} \sqrt{(2-3 s^2_{13}) t_N}} \hspace{0.3 cm} \mbox{for IH,}%
\end{array}%
\right.   \label{sdexpres}
\eea
with
\bea
&&\Phi_N =\sqrt{\left(\cos^2 2\theta_{13}-c^4_{13} \sin^2 2\theta_{23}\right)\left(t_N c^4_{13} \cos^2 2\theta_{23}-2 s^2_{13}\right) t_N},\\
&&\Phi_I =\sqrt{\left(\cos^2 2\theta_{13}-c^4_{13} \sin^2 2\theta_{23}\right) \left(3t_I  c^4_{13} \cos^2 2\theta_{23}+6 s^2_{13}-4\right) t_I },\\
&&k_{\Phi}=(6 -5 s_{13}^2)s^2_{13}+ 2(c^4_{13} \sin^2 2\theta_{23} -1),  \\
&&t_N=\sqrt{\frac{\Delta m^2_{31}}{\Delta m^2_{21}}}, \hs t_I=\sqrt{\frac{-\Delta m^2_{31}}{\Delta m^2_{21}-\Delta m^2_{31}}}. \label{tNI}
\eea
Equations (\ref{k2expres})-(\ref{tNI}) imply that $k_3$ depends on three parameters $\Delta m^2_{21}, \Delta m^2_{31}$ and $s_{13}$ while $k_2$ and $\cos\alpha$ depend on four parameters $\Delta m^2_{21}, \Delta m^2_{31}, s_{13}$ and $s_{23}$. As a consequence, Eqs. (\ref{s14s24s34sq})$-$ (\ref{tNI}) imply that $J_{CP}, \sin\delta_{CP}, \langle m^{(3)}_{ee}\rangle$, $\langle m_{ee}\rangle$ and $m^{(3)}_\beta$ as well as $|U_{ij}| \, (i=1,2,3; j=1,3)$ depend on four parameters $\Delta m^2_{21}, \Delta m^2_{31}, s_{13}$ and $s_{23}$ while $s^2_{k4}\, (k=1,2,3)$ and $m_\beta$ depend on five parameters $\Delta m^2_{21}, \Delta m^2_{31}, s_{13}, s_{23}$ and $m_s\equiv m_4\equiv \sqrt{\Delta m^2_{41}}$ in which three parameters $\Delta m^2_{21}, \Delta m^2_{31}$ and $s_{13}$ are measured with more accuracy
that can be used to constrain the others.

\begin{center}
\begin{figure}[ht]
\begin{center}
\vspace{-0.5 cm}
\hspace*{-2.7 cm}
\includegraphics[width=0.80\textwidth]{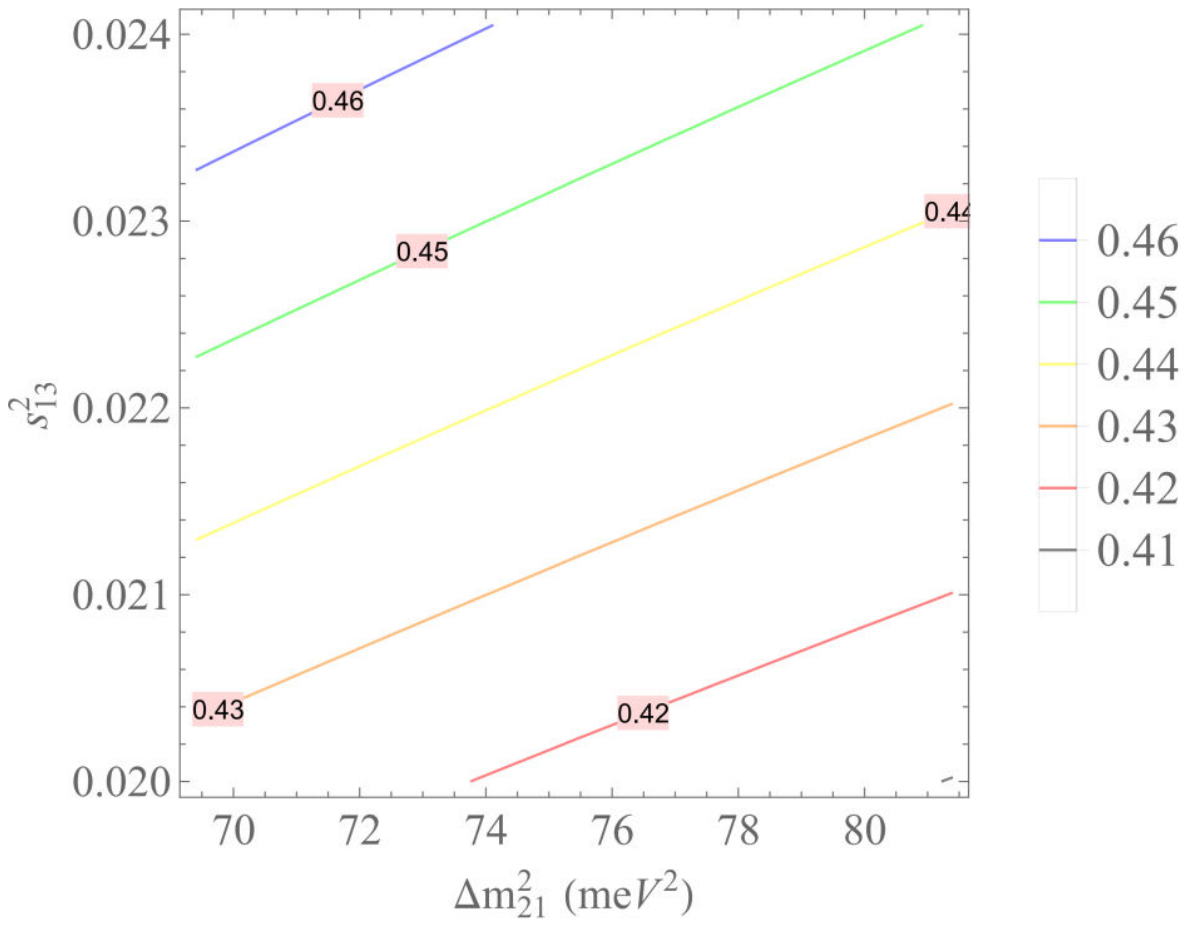}\hspace*{-5.2 cm}
\includegraphics[width=0.80\textwidth]{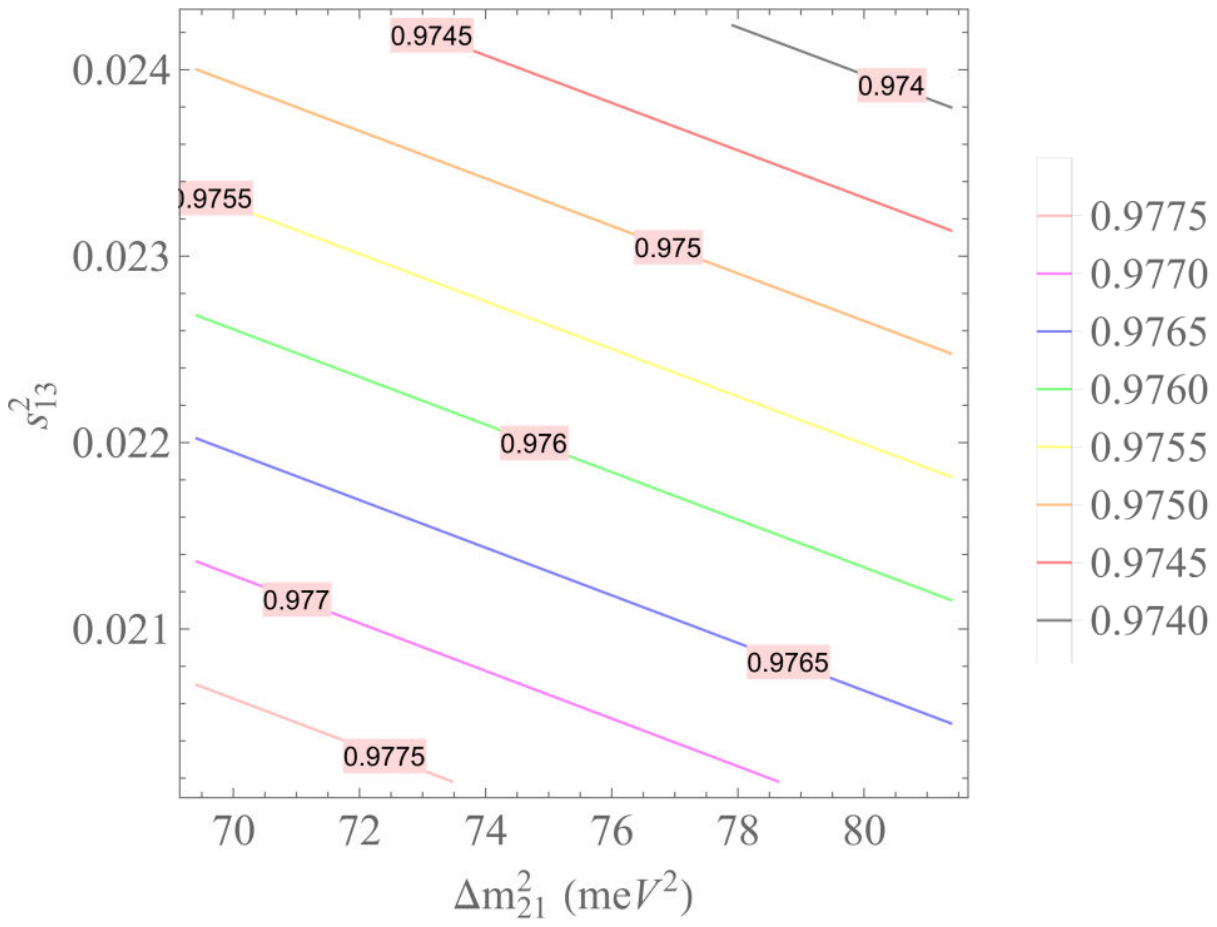}\hspace*{-2.4 cm}
\end{center}
\vspace{-11.65 cm}
\caption{$k_{3}$ as a function of $s^2_{13}$  and $\Delta m^2_{21}$ with $\Delta m^2_{21}\in (69.4, 81.4)\, \mathrm{meV}^2$ and $s^2_{13}\in (2.000, 2.405)10^{-2}$ for NH (in the left panel) while $s^2_{13}\in (2.018, 2.424)10^{-2}$ for IH (in the right panel).}
\label{k3F}
\end{figure}
\end{center}

At the best-fit points of $\Delta m^2_{31}$ \cite{Salas2020}, $\Delta m^2_{31}=2.55\times 10^3\, \mathrm{meV}^2$ for NH and $\Delta m^2_{31}=-2.45\times 10^3\, \mathrm{meV}^2$ for IH, the parameter $k_{3}$ depends on $s^2_{13}$ and $\Delta m^2_{21}$ which is depicted in Fig. \ref{k3F}. This figure implies
\bea
&&k_3\in \left\{
\begin{array}{l}
 (0.41, 0.46)\,  \hspace{0.95cm}\mbox{for  NH},  \\
(0.9740, 0.9775) \hspace{0.25cm}\mbox{for IH},
\end{array}%
\right. \label{k3constraint}\eea

\begin{center}
\begin{figure}[h]
\begin{center}
\vspace{-0.5 cm}
\hspace*{-2.4 cm}
\includegraphics[width=0.8\textwidth]{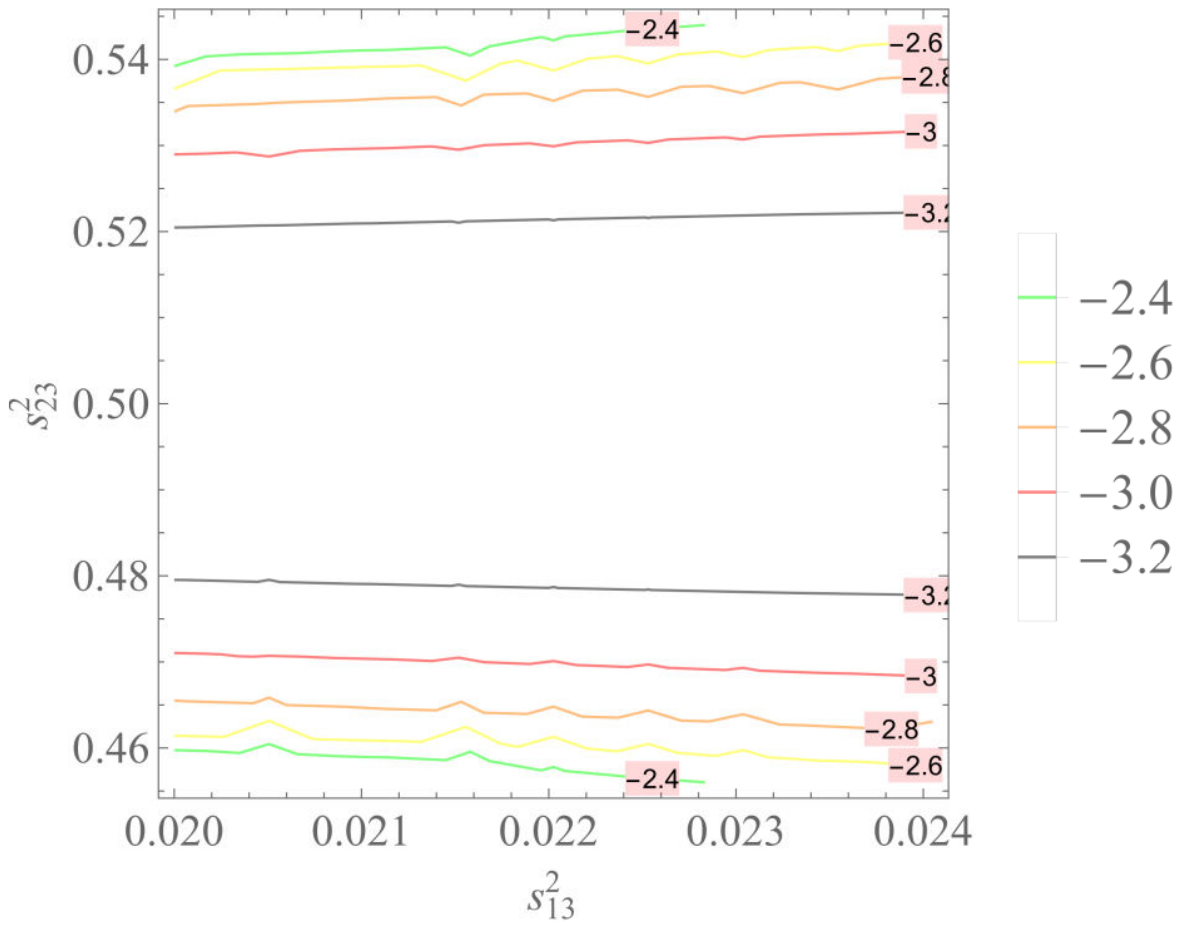}\hspace*{-5.4 cm}
\includegraphics[width=0.8\textwidth]{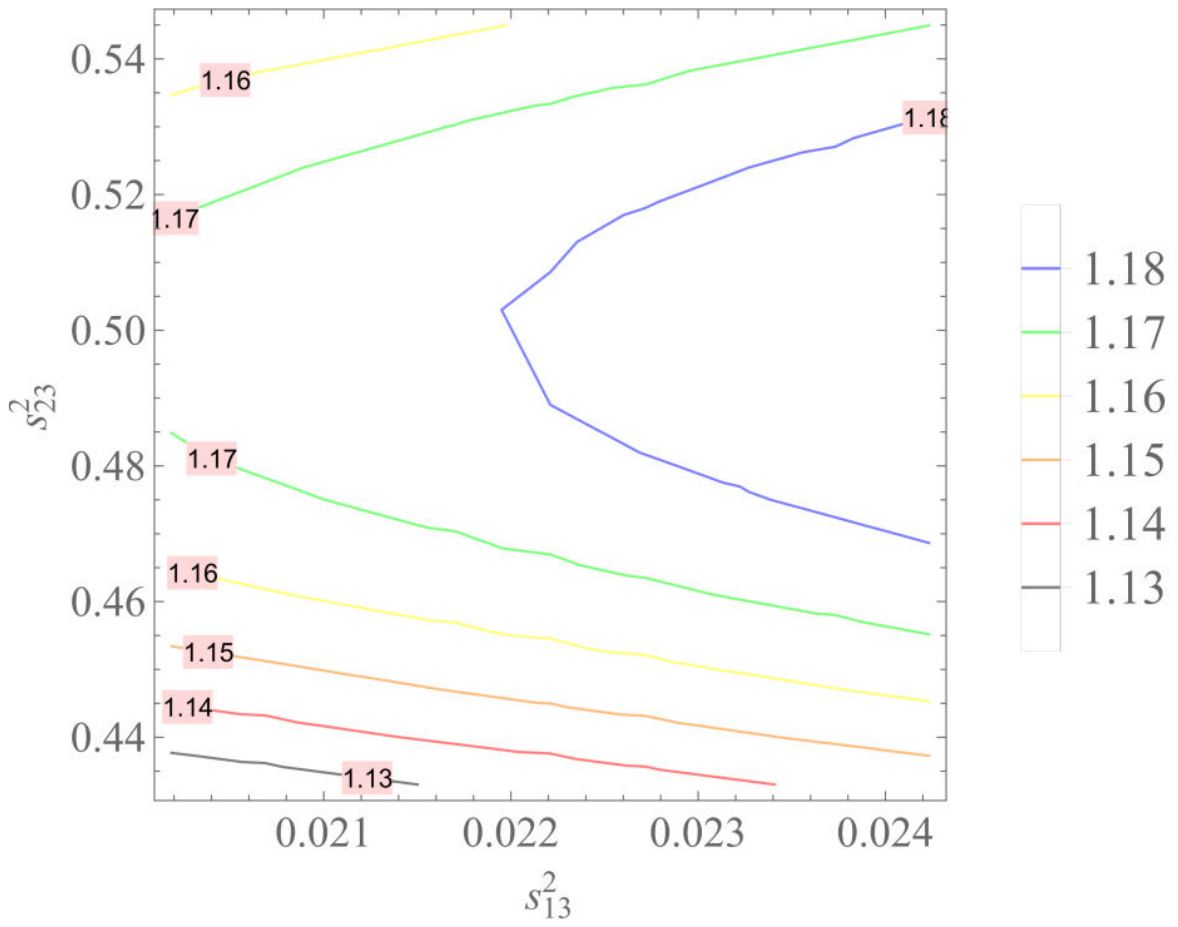}\hspace*{-2.6 cm}
\end{center}
\vspace{-11.85 cm}
\caption{$k_{2}$ as a function of $s^2_{23}$  and $s^2_{13}$ with $s^2_{23}\in (0.456, 0.544)$ and $s^2_{13}\in (2.00, 2.405)10^{-2}$ for NH (in the left panel) while $s^2_{23}\in (0.433,0.545)$ and $s^2_{13}\in (2.018, 2.424)10^{-2}$ for IH (in the right panel).}
\label{k2F}
\end{figure}
\end{center}

\begin{center}
\begin{figure}[h]
\begin{center}
\vspace{-0.4 cm}
\hspace*{-2.5 cm}
\includegraphics[width=0.80\textwidth]{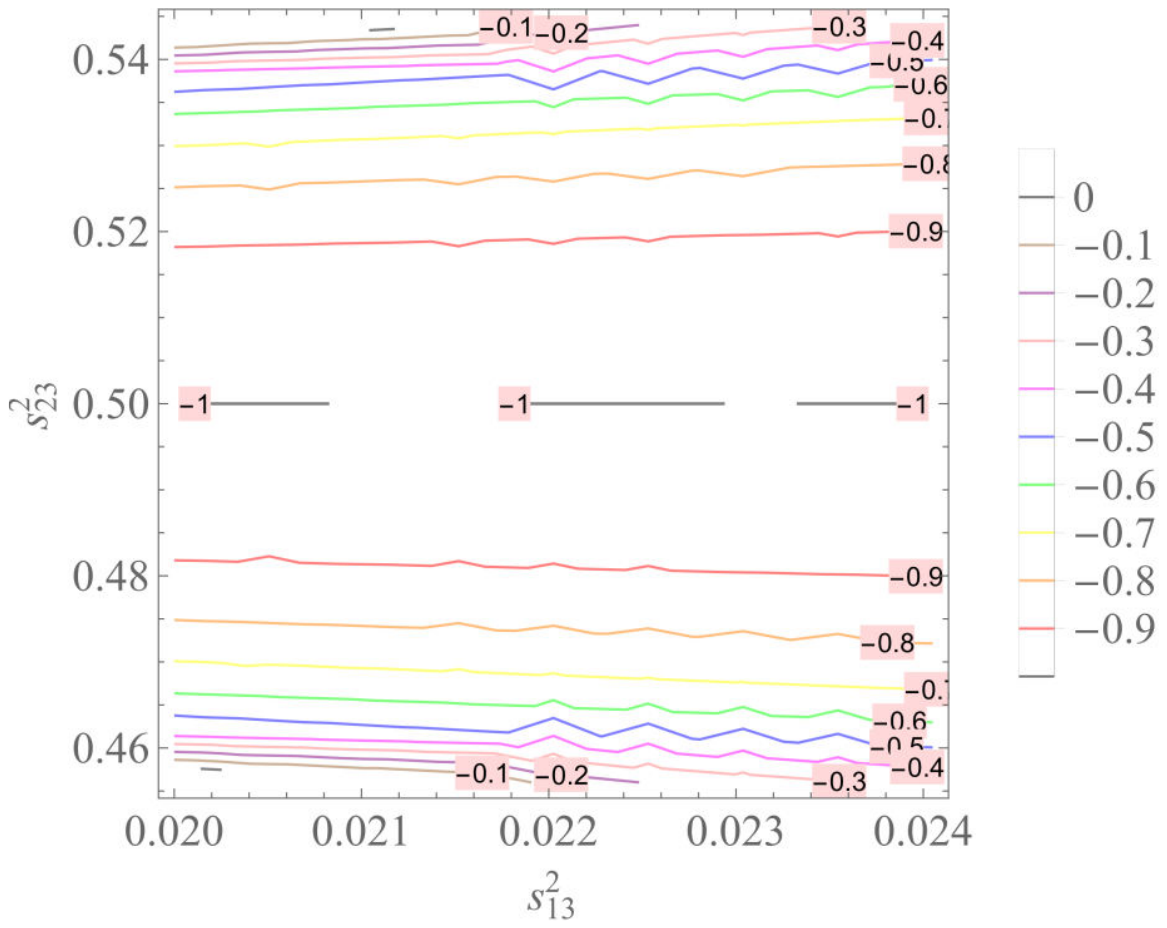}\hspace*{-5.4 cm}
\includegraphics[width=0.80\textwidth]{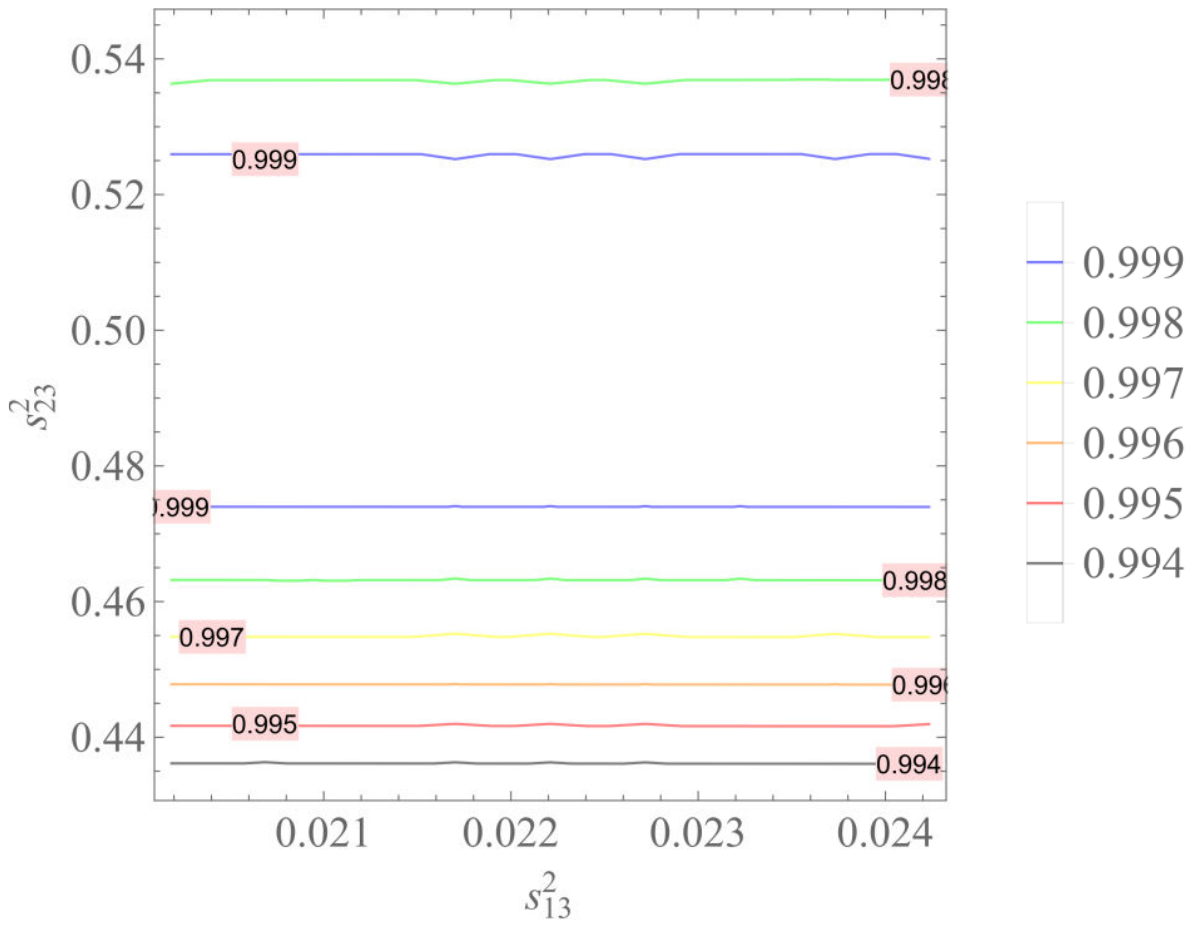}\hspace*{-2.5 cm}
\end{center}
\vspace{-11.75 cm}
\caption{$\cos\alpha$ as a function of $s^2_{23}$  and $s^2_{13}$ with $s^2_{23}\in (0.456, 0.544)$ and $s^2_{13}\in (2.00, 2.405)10^{-2}$ for NH (in the left panel) while $s^2_{23}\in (0.433,0.545)$ and $s^2_{13}\in (2.018, 2.424)10^{-2}$ for IH (in the right panel).}
\label{c1F}
\end{figure}
\end{center}

\begin{center}
\begin{figure}[h]
\begin{center}
\vspace{-0.5 cm}
\hspace*{-2.5 cm}
\includegraphics[width=0.8\textwidth]{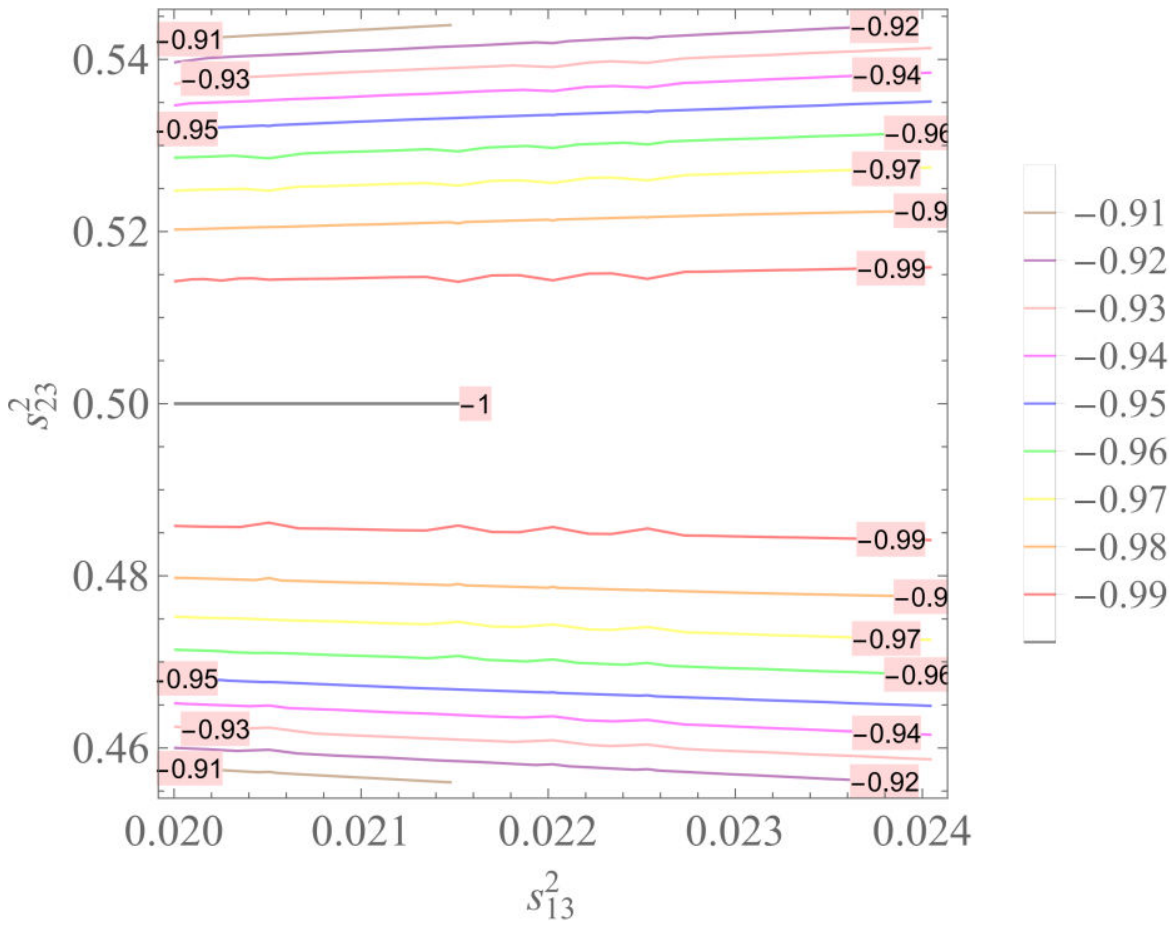}\hspace*{-5.3 cm}
\includegraphics[width=0.8\textwidth]{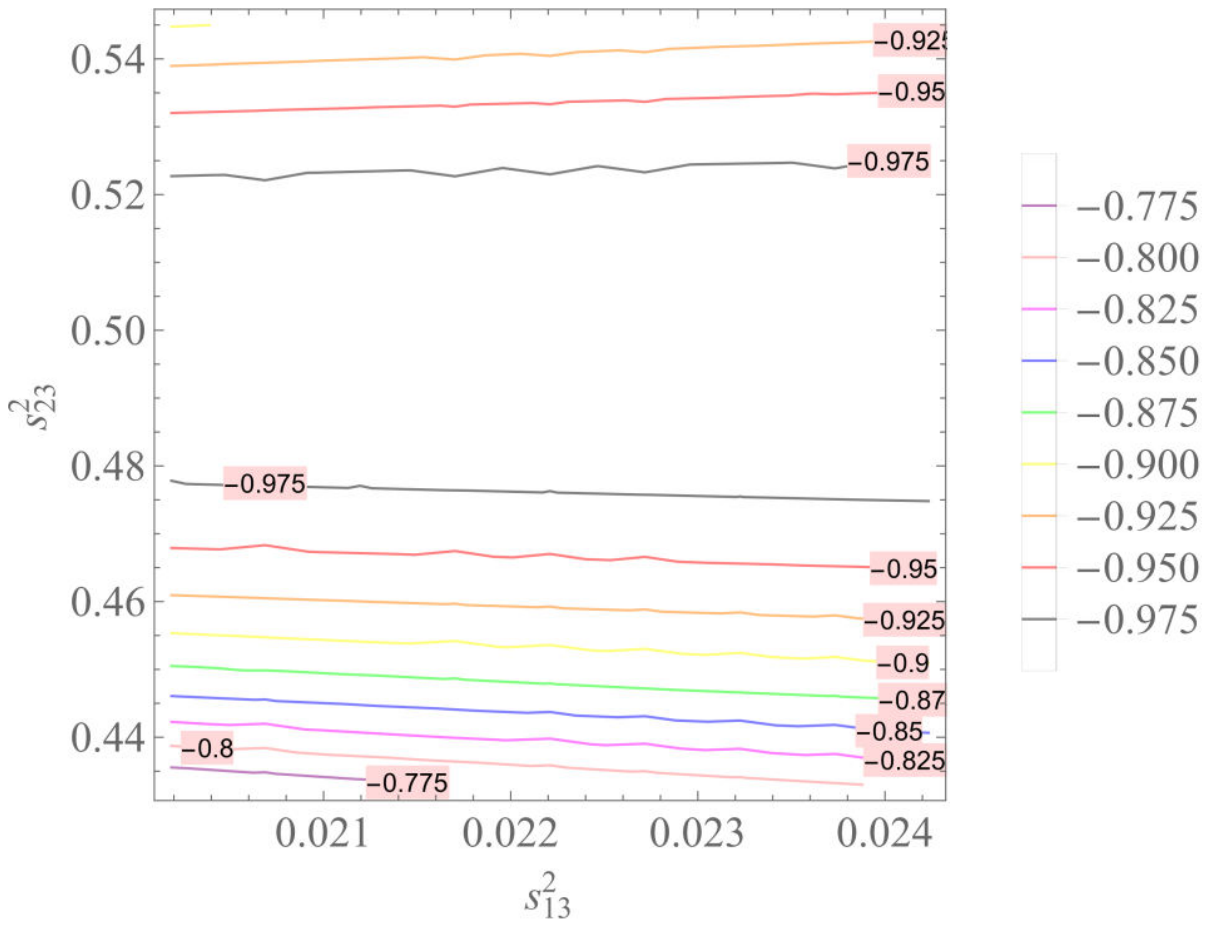}\hspace*{-2.5 cm}
\end{center}
\vspace{-11.85 cm}
\caption{$\sin \delta_{CP}$ as a function of $s^2_{23}$  and $s^2_{13}$ with $s^2_{23}\in (0.456, 0.545)$ and $s^2_{13}\in (2.00, 2.405)10^{-2}$ for NH (in the left panel) while $s^2_{23}\in (0.433,0.545)$ and $s^2_{13}\in (2.018, 2.424)10^{-2}$ for IH (in the right panel).}
\label{sdF}
\end{figure}
\end{center}

\begin{center}
\begin{figure}[h]
\begin{center}
\vspace{-0.5 cm}
\hspace*{-2.5 cm}
\includegraphics[width=0.8\textwidth]{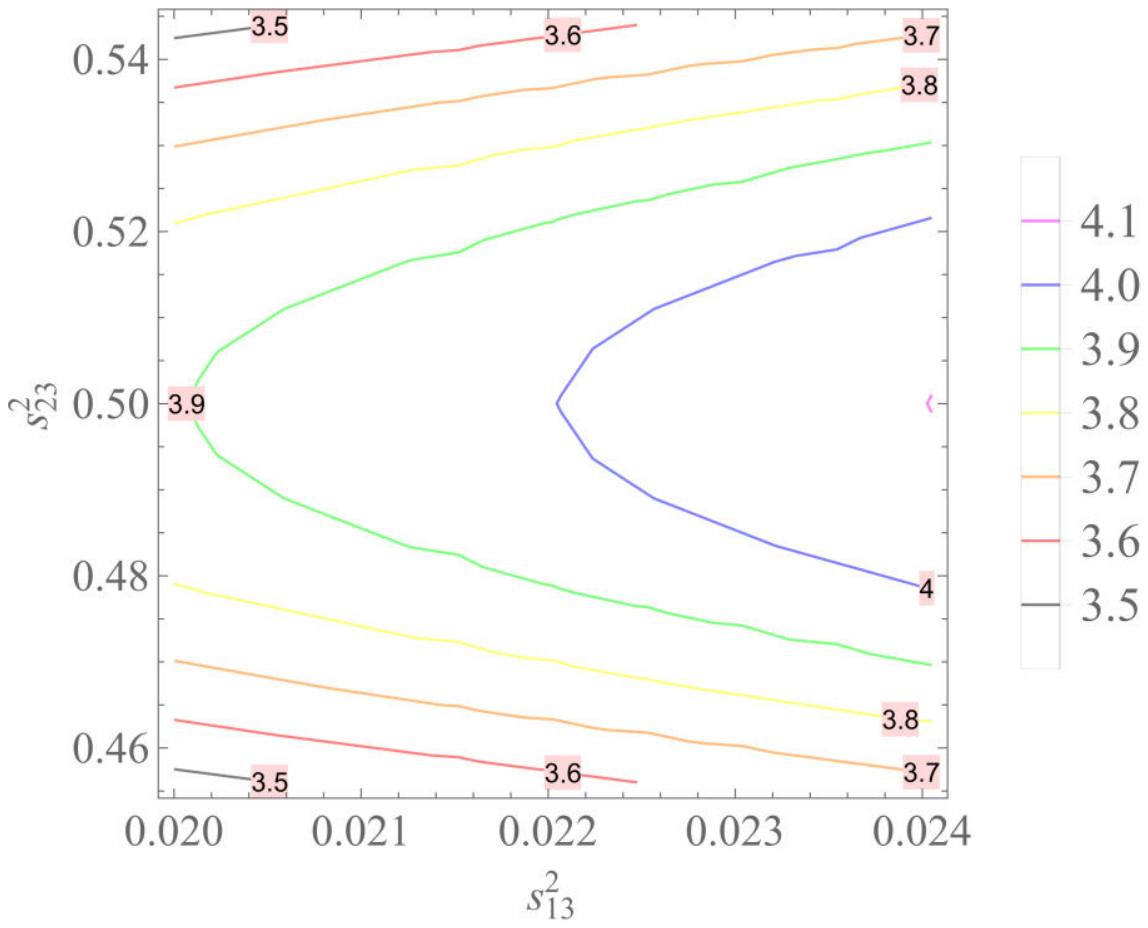}\hspace*{-5.5 cm}
\includegraphics[width=0.8\textwidth]{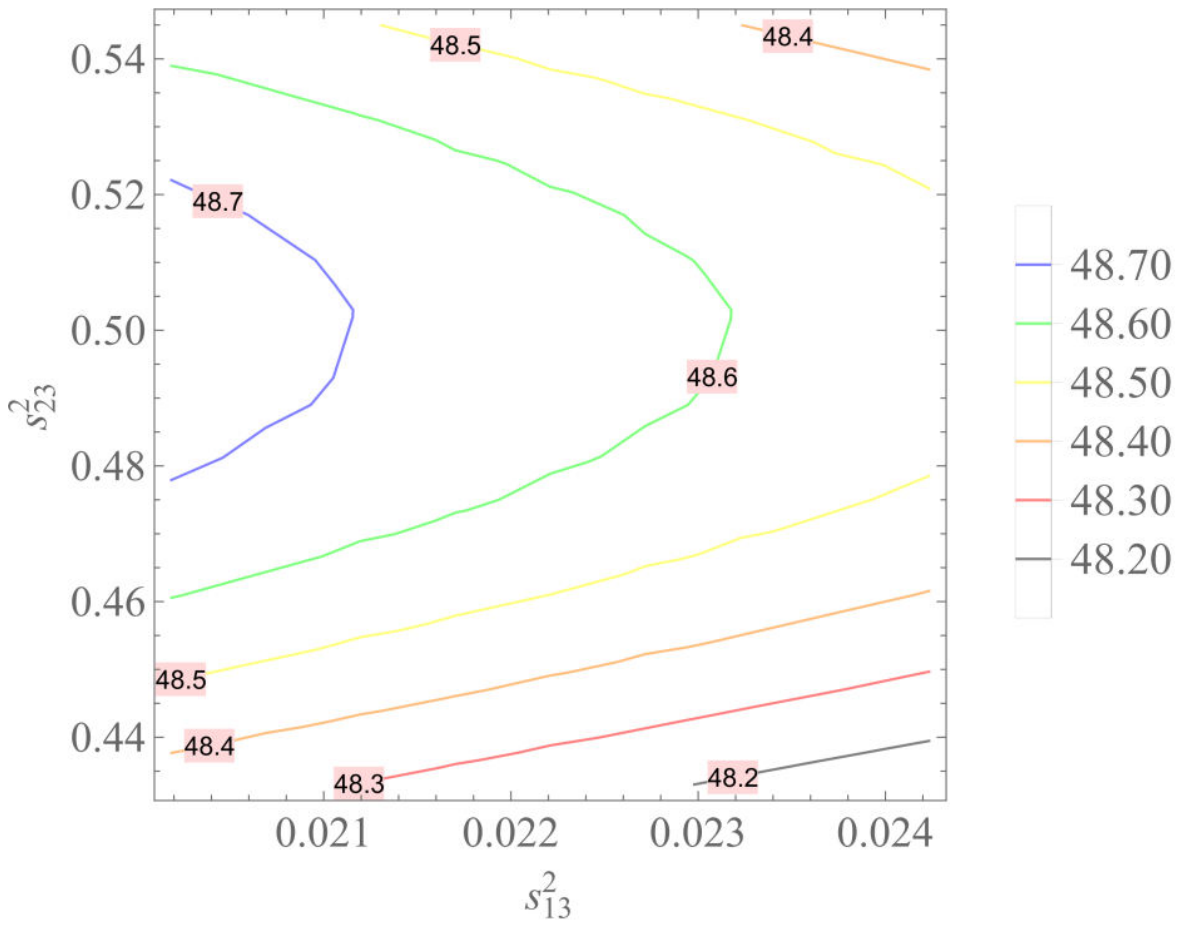}\hspace*{-2.5 cm}
\end{center}
\vspace{-11.85 cm}
\caption{$\langle m^{(3)}_{ee}\rangle\, (\mathrm{meV})$ as a function of $s^2_{23}$  and $s^2_{13}$ with $s^2_{23}\in (0.456, 0.544)$ and $s^2_{13}\in (2.00, 2.405)10^{-2}$ for NH (in the left panel) while $s^2_{23}\in (0.433,0.545)$ and $s^2_{13}\in (2.018, 2.424)10^{-2}$ for IH (in the right panel).}
\label{mee3F}
\end{figure}
\end{center}

\begin{center}
\begin{figure}[h]
\begin{center}
\hspace*{-2.5 cm}
\includegraphics[width=0.8\textwidth]{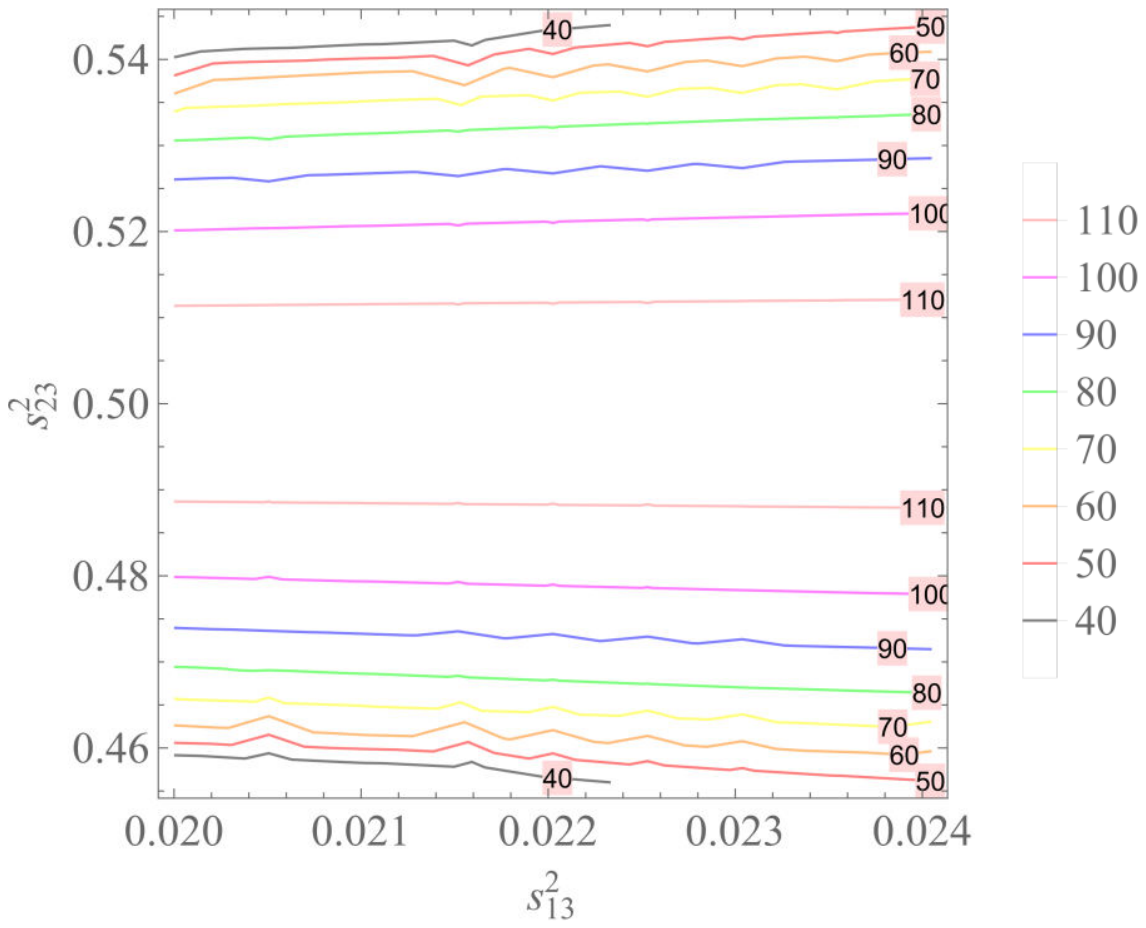}\hspace*{-5.5 cm}
\includegraphics[width=0.8\textwidth]{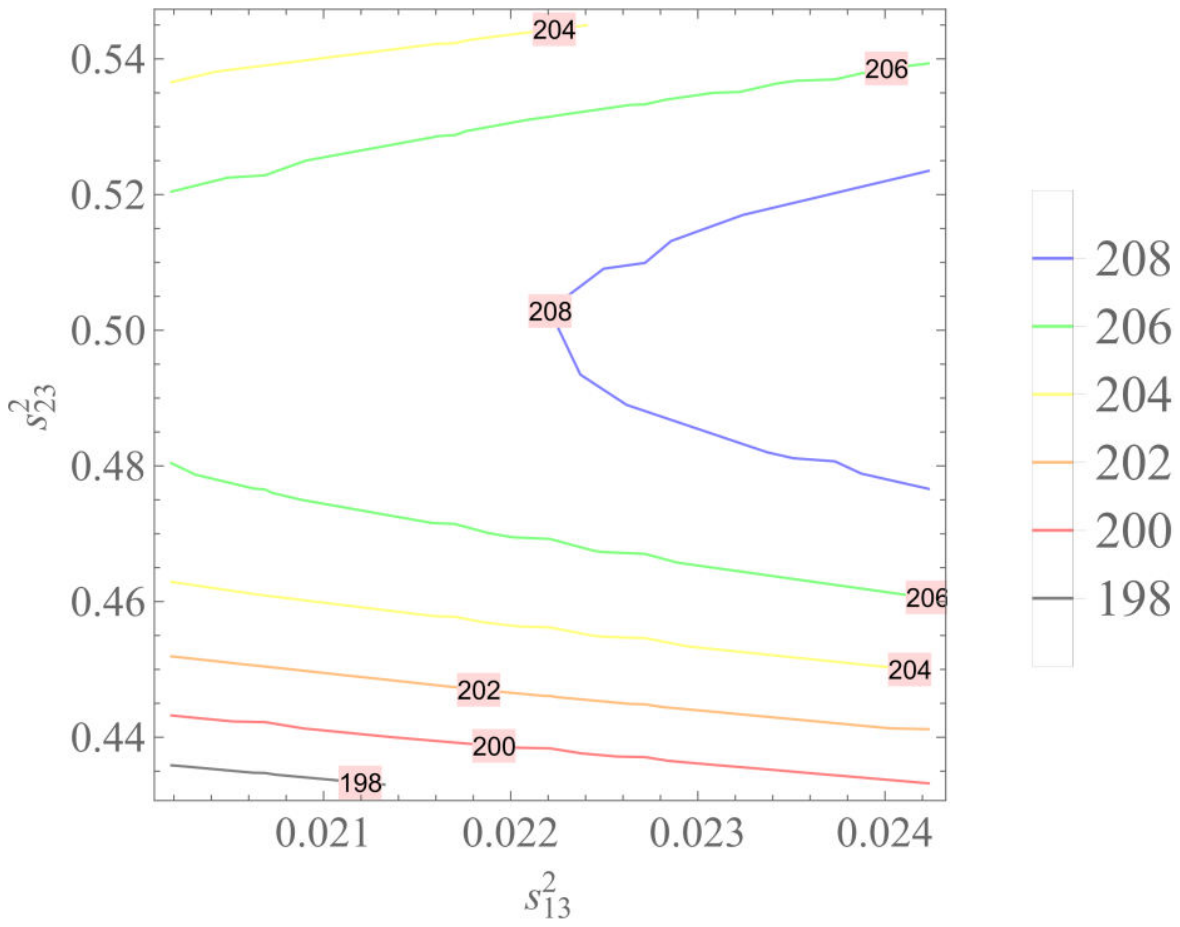}\hspace*{-2.5 cm}
\end{center}
\vspace{-11.85 cm}
\caption{$\langle m_{ee}\rangle\, (\mathrm{meV})$ as a function of $s^2_{23}$  and $s^2_{13}$ with $s^2_{23}\in (0.456, 0.544)$ and $s^2_{13}\in (2.00, 2.405)10^{-2}$ for NH (in the left panel) while $s^2_{23}\in (0.541,0.598)$ and $s^2_{13}\in (2.018, 2.424)10^{-2}$ for IH (in the right panel).}
\label{meeF}
\end{figure}
\end{center}

\begin{center}
\begin{figure}[h]
\begin{center}
\hspace*{-2.5 cm}
\includegraphics[width=0.8\textwidth]{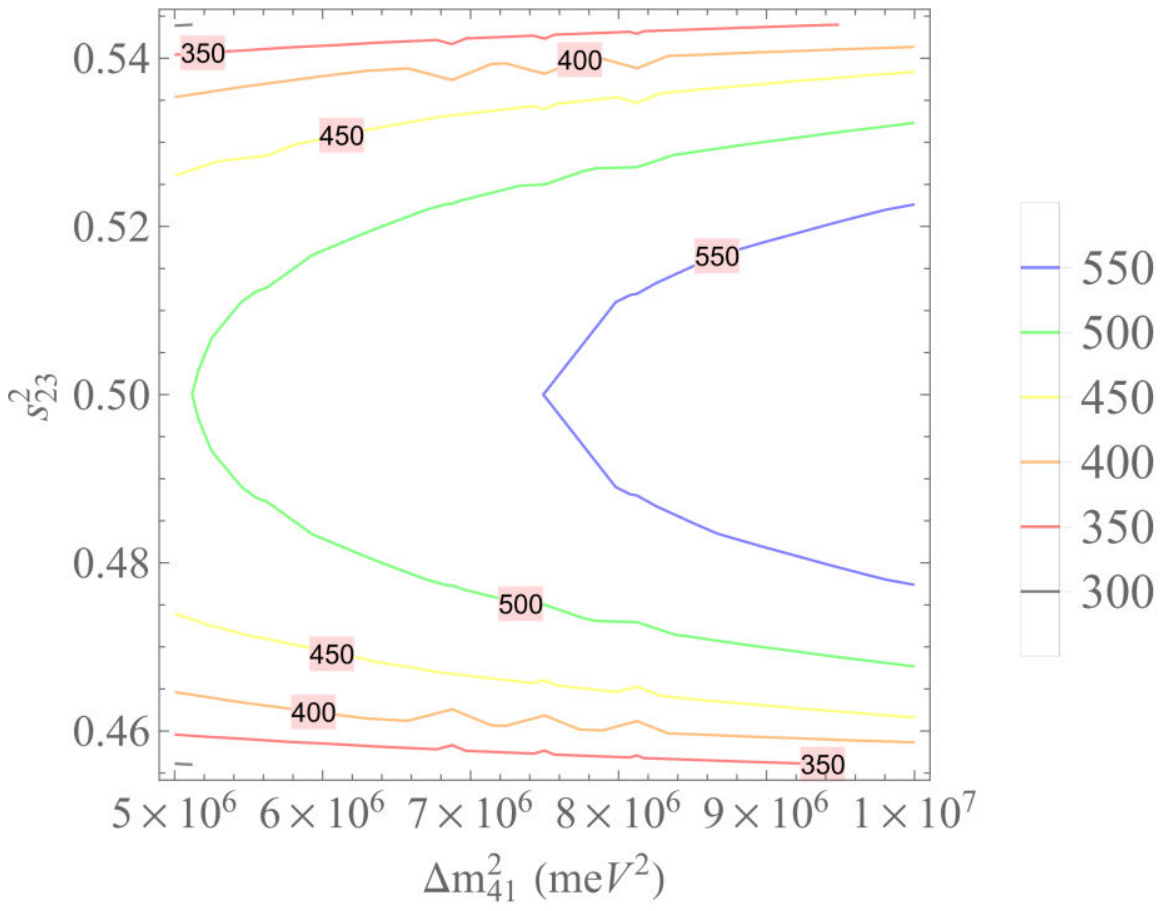}\hspace*{-5.5 cm}
\includegraphics[width=0.8\textwidth]{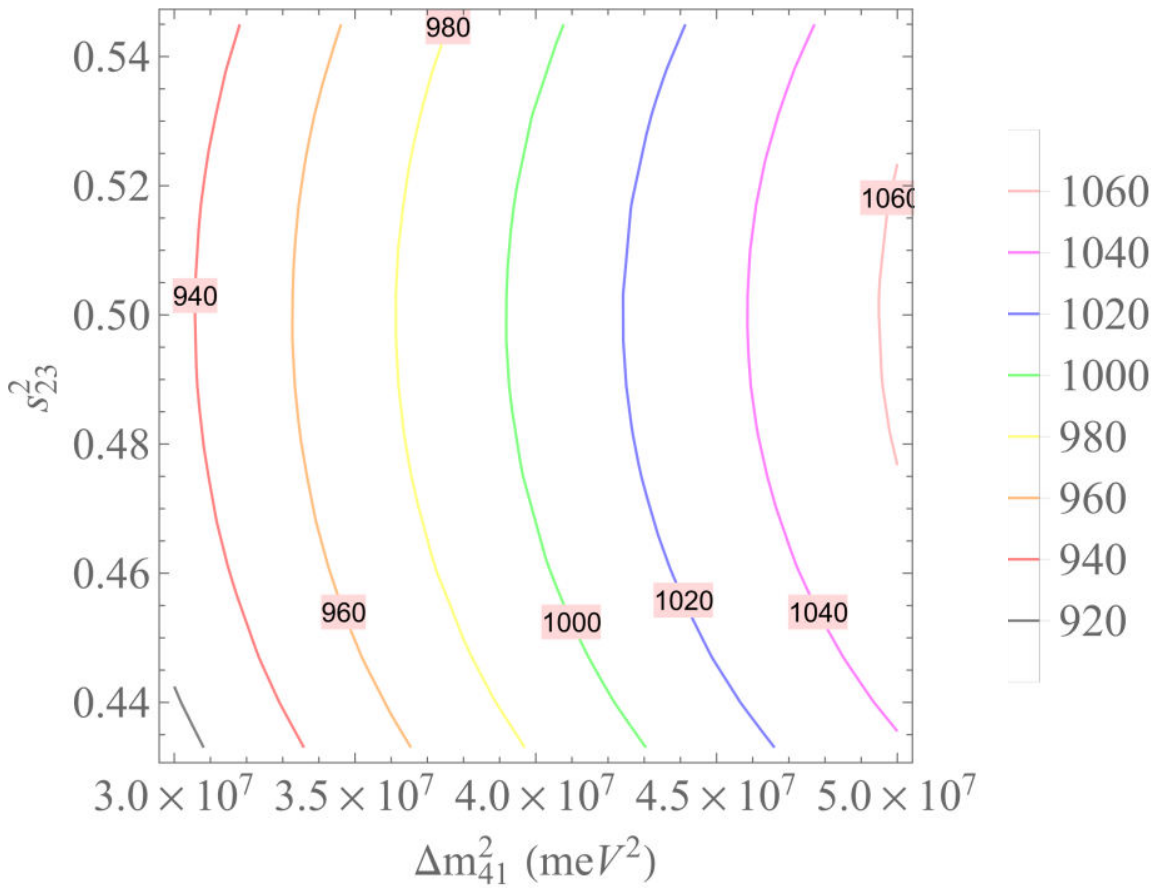}\hspace*{-2.5 cm}
\end{center}
\vspace{-11.85 cm}
\caption{$m_\beta\, (\mathrm{meV})$ as a function of $s^2_{23}$  and $s^2_{13}$ with $s^2_{23}\in (0.456, 0.544)$ and $\Delta m^2_{41} \in (5.0, 10.0) 10^6\, \mathrm{meV}^2$ for NH (in the left panel) while $s^2_{23}\in (0.433, 0.545)$ and $\Delta m^2_{41} \in (30.0, 50.0) 10^6\, \mathrm{meV}^2$ for IH (in the right panel).}
\label{mbF}
\end{figure}
\end{center}

At the best-fit points of two light neutrino mass squared differences taken from Ref. \cite{Salas2020}, $\Delta m^2_{31}=2.55\times 10^3\, \mathrm{meV}^2$ for NH while $\Delta m^2_{31}=-2.45\times 10^3\, \mathrm{meV}^2$ for IH and $\Delta m^2_{31}=75.0 \, \mathrm{meV}^2$, we get
\bea
&&k_0= \left\{
\begin{array}{l}
5.831 \,  \hspace{0.25cm}\mbox{for  NH},  \\
0.985 \,\hspace{0.25cm}\mbox{for IH},
\end{array}%
\right. \label{k0valuest}\eea
and the parameters $k_2, \cos\alpha$, $\sin \delta_{CP}, \langle m^{(3)}_{ee}\rangle$, $\langle m_{ee}\rangle$ and $m^{(3)}_\beta$ depend on $s_{13}$ and $s_{23}$ which are, respectively, depicted in Figs. \ref{k2F}, \ref{c1F}, \ref{sdF}, \ref{mee3F}, \ref{meeF} and \ref{mbF}.

Figure \ref{k2F} implies that
\bea
&&k_2\in \left\{
\begin{array}{l}
 (-3.20, -2.40) \hspace{0.2cm}\mbox{for  NH},  \\
(1.13, 1.18) \hspace{0.8cm}\mbox{for IH}.
\end{array}%
\right. \label{k2constraint}\eea

From figure \ref{c1F}, it follows that:
\bea
&&\cos \alpha \in \left\{
\begin{array}{l}
 (-0.90, 0.0)\,  \hspace{0.4cm}\mbox{for  NH},  \\
(0.994, 0.999) \hspace{0.2cm}\mbox{for IH},
\end{array}%
\right. \mathrm{\hs i.e.,\,} \alpha^{(\circ)} \in \left\{
\begin{array}{l}
 (90.0, 154.0)\,  \hspace{0.25cm}\mbox{for  NH},  \\
(2.563, 6.28) \,\hspace{0.25cm}\mbox{for IH}{\Revised{.}}
\end{array}%
\right. \label{c1constraint}\eea

\begin{center}
\begin{figure}[h]
\begin{center}
\vspace{-0.85 cm}
\hspace*{-2.5 cm}
\includegraphics[width=0.8\textwidth]{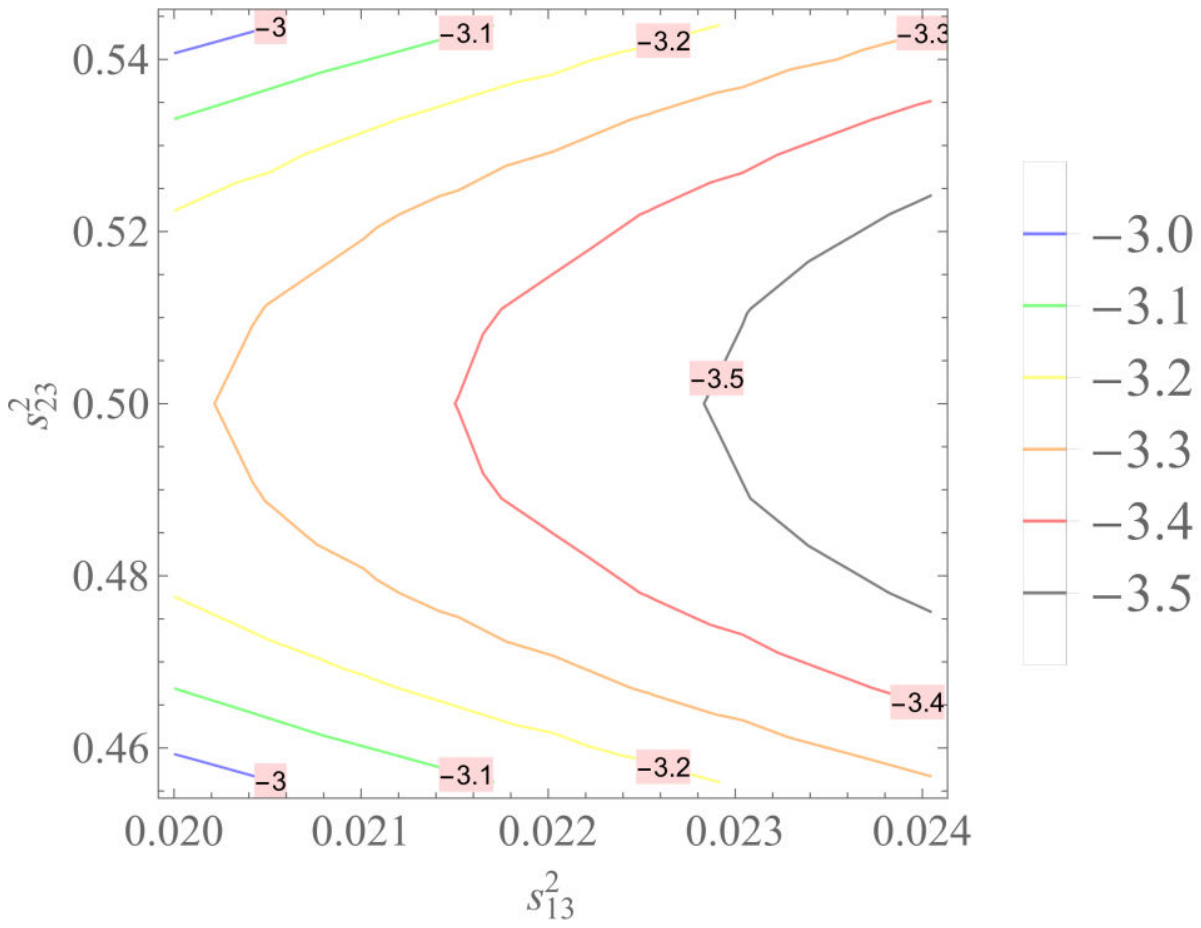}\hspace*{-5.3 cm}
\includegraphics[width=0.8\textwidth]{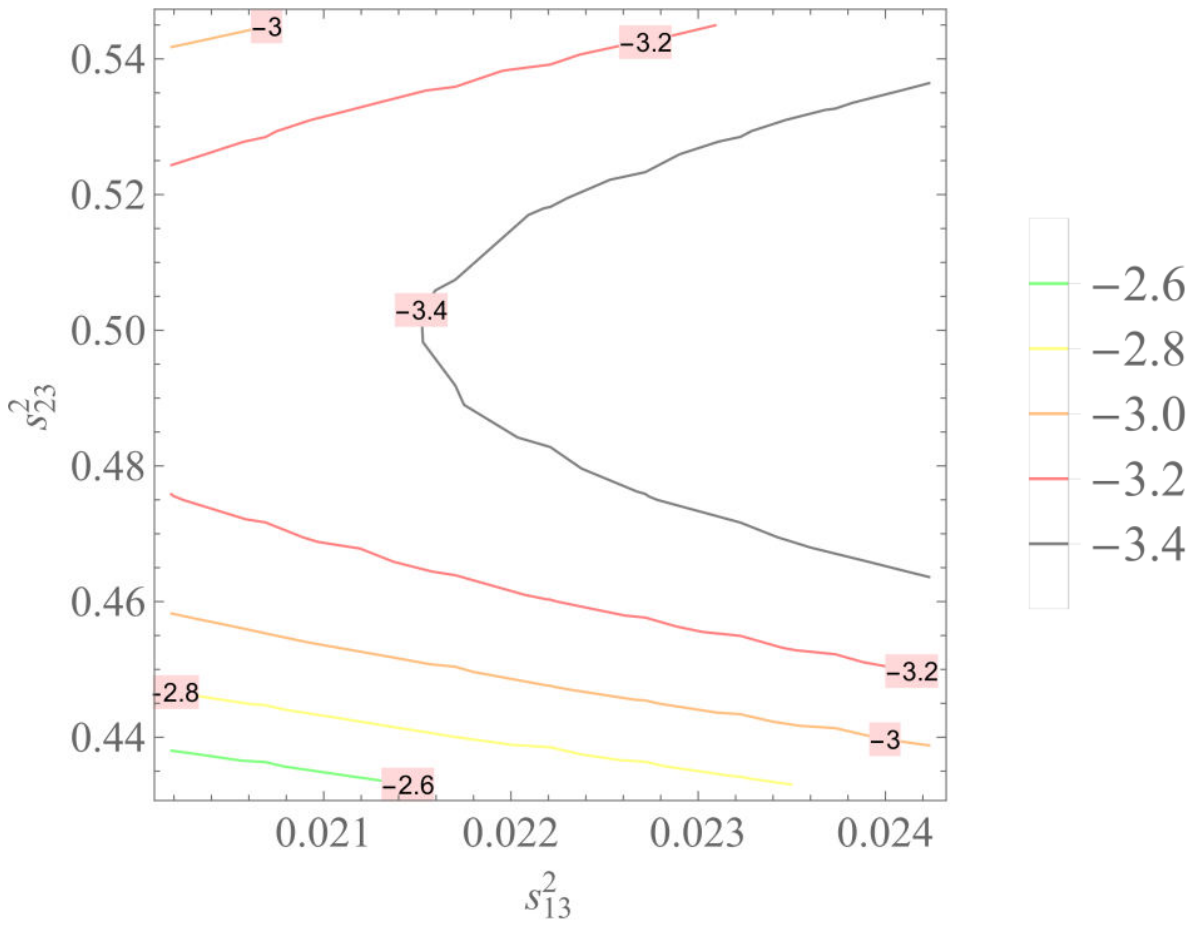}\hspace*{-2.5 cm}
\end{center}
\vspace{-11.85 cm}
\caption{$J_{CP}\times 10^2$ as a function of $s^2_{23}$  and $s^2_{13}$ with $s^2_{23}\in (0.456, 0.544)$ and $s^2_{13}\in (2.00, 2.405)10^{-2}$ for NH (in the left panel) while $s^2_{23}\in (0.433,0.545)$ and $s^2_{13}\in (2.018, 2.424)10^{-2}$ for IH (in the right panel).}
\label{JF}
\end{figure}
\end{center}

The dependence of $J_{CP}$ on $s^2_{23}$  and $s^2_{13}$ is depicted in Fig. \ref{JF}.

Figure \ref{JF} implies that the Jarlskog invariant takes the values \cite{PDG2020}:
\bea
J_{CP} \in \left\{
\begin{array}{l}
(-3.5, -3.0)\times 10^{-2} \hspace{0.2 cm} \mbox{for  NH,} \\
(-3.4, -2.6)\times 10^{-2} \hspace{0.2 cm} \mbox{for  IH.}%
\end{array}%
\right.   \label{JCPvalues}
\eea

Figure \ref{sdF} implies that
\bea
&&\sin \delta_{CP} \in \left\{
\begin{array}{l}
 (-0.99, -0.91)\,  \hspace{0.5cm}\mbox{for  NH},  \\
(-0.975, -0.775) \hspace{0.2cm}\mbox{for IH},
\end{array}%
\right. \mathrm{\hs i.e.,\,\, \,} \delta^{(\circ)}_{CP} \in \left\{
\begin{array}{l}
 (156.0, 172.0)\,  \hspace{0.25cm}\mbox{for  NH},  \\
(140.8, 167.2) \,\hspace{0.25cm}\mbox{for IH}.
\end{array}%
\right. \label{sdconstraint}\eea
\begin{center}
\begin{figure}[h]
\begin{center}
\vspace{-0.5 cm}
\hspace*{-2.5 cm}
\includegraphics[width=0.8\textwidth]{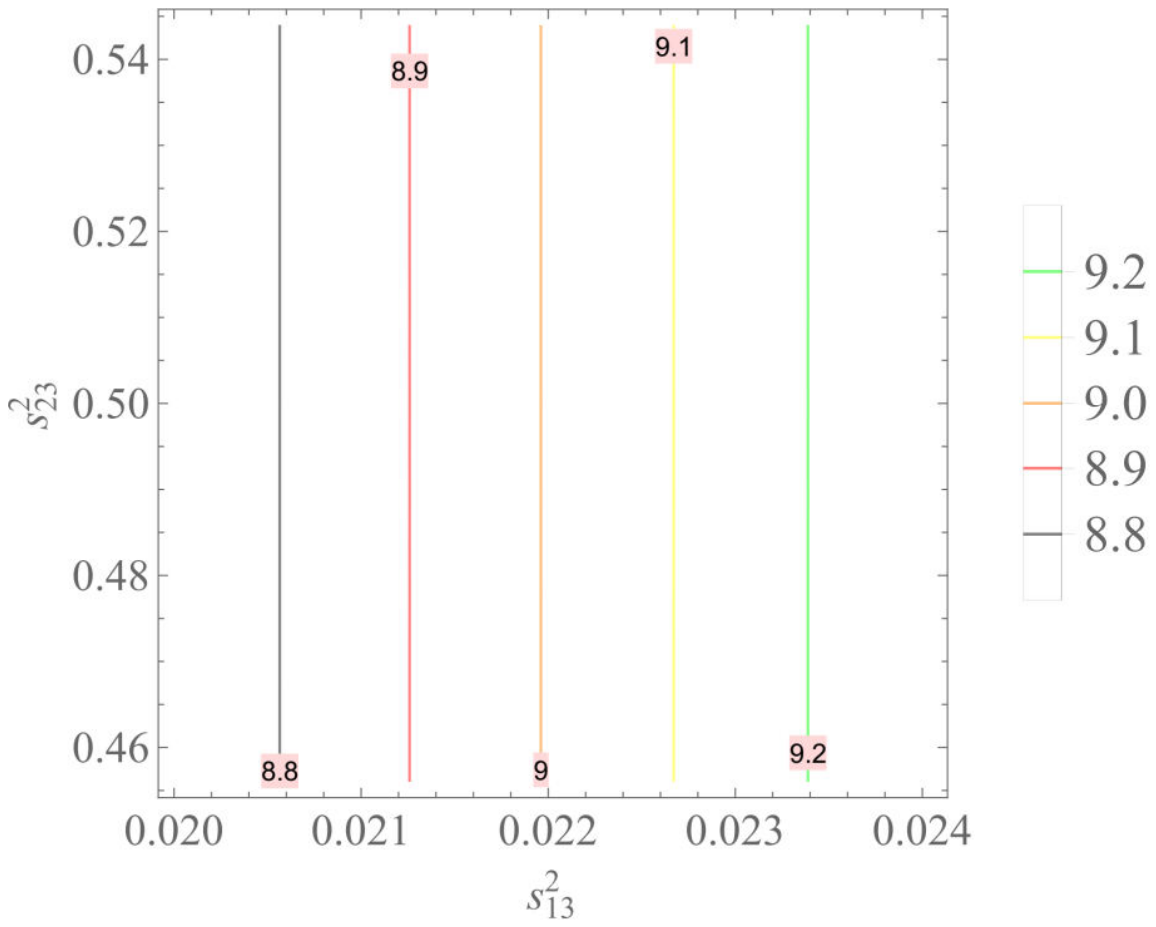}\hspace*{-5.5 cm}
\includegraphics[width=0.8\textwidth]{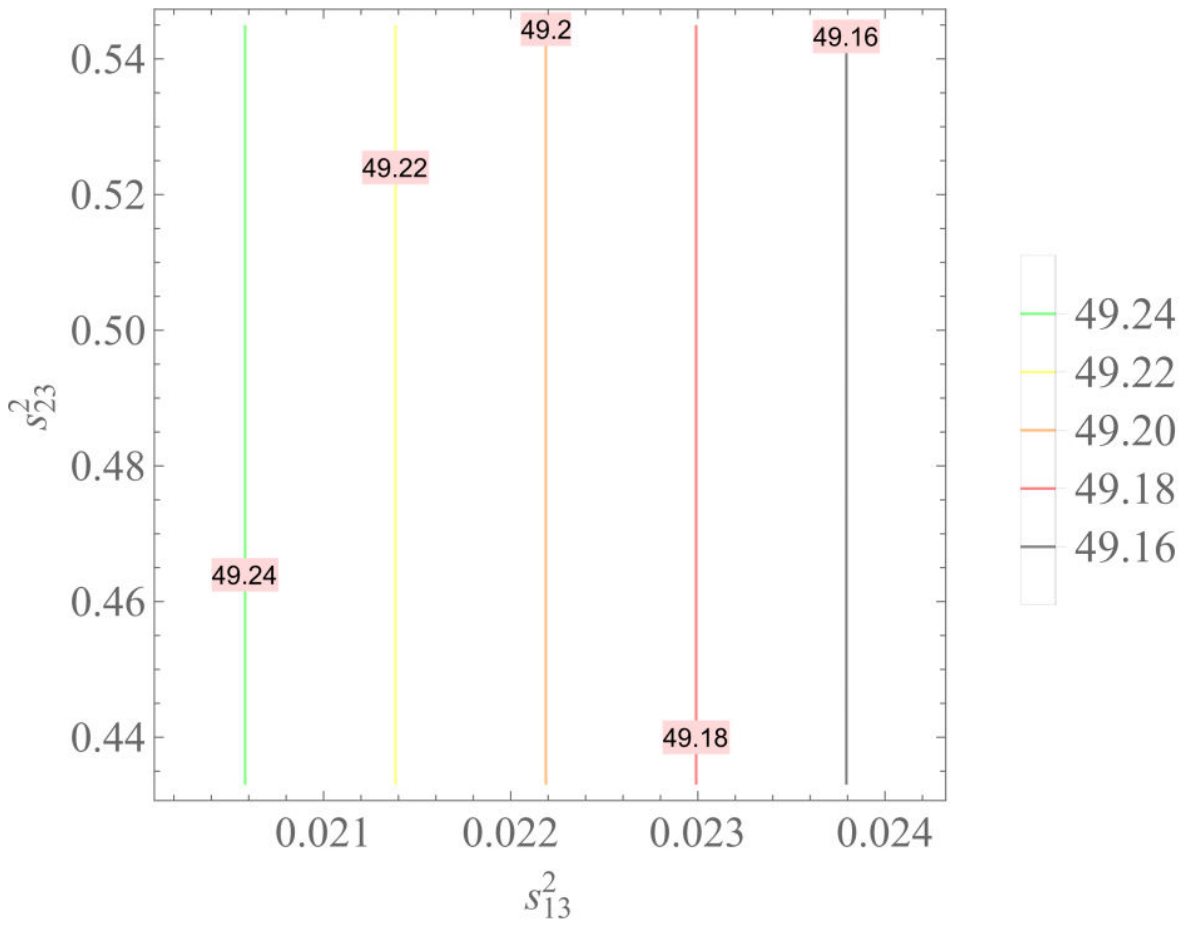}\hspace*{-2.5 cm}
\end{center}
\vspace{-11.85 cm}
\caption{$m^{(3)}_\beta\, (\mathrm{meV})$ as a function of $s^2_{23}$  and $s^2_{13}$ with $s^2_{23}\in (0.456, 0.544)$ and $s^2_{13}\in (2.00, 2.405)10^{-2}$ for NH (in the left panel) while $s^2_{23}\in (0.433,0.545)$ and $s^2_{13}\in (2.018, 2.424)10^{-2}$ for IH (in the right panel).}
\label{mb3F}
\end{figure}
\end{center}

Figures \ref{mee3F} and \ref{mb3F} imply that
\bea
&&\langle m^{(3)}_{ee}\rangle \in \left\{
\begin{array}{l}
 (3.50, 4.10)\, \mathrm{meV}  \hspace{0.575cm}\mbox{for  NH},  \\
(48.20,48.70)\, \mathrm{meV} \hspace{0.2cm}\mbox{for IH},
\end{array}%
\right. \label{mee3constraint}\\
&&m^{(3)}_{\beta} \in \left\{
\begin{array}{l}
 (8.80, 9.20)\, \mathrm{meV}  \hspace{0.575cm}\mbox{for  NH},  \\
(49.16, 49.24)\, \mathrm{meV} \hspace{0.2cm}\mbox{for IH},
\end{array}%
\right. \label{mb3constraint}\\
&&\langle m_{ee}\rangle \in \left\{
\begin{array}{l}
 (40.0, 110.0)\, \mathrm{meV}  \hspace{0.4cm}\mbox{for  NH},  \\
(198.0, 208.0)\, \mathrm{meV} \hspace{0.2cm}\mbox{for IH}.
\end{array}%
\right. \label{meeconstraint}\eea

We see that the resulting effective neutrino mass for three neutrino scheme in Eqs. (\ref{mee3constraint})-(\ref{meeconstraint}), for both NH and IH, are below all the upper limits taken from GERDA \cite{Agostini18}$\langle m_{ee} \rangle < (120 \div 260) \,\mathrm{meV}$, MAJORANA \cite{MAJO} $\langle m_{ee} \rangle < (24 \div 53) \,\mathrm{meV}$, CUORE \cite{CUORE18} $\langle m_{ee} \rangle < (110 \div 500) \,\mathrm{meV}$, KamLAND-Zen \cite{KamLAND16} $\langle m_{ee} \rangle < (61\div 165) \,\mathrm{meV}$, GERDA \cite{GERDA19}
 $\langle m_{ee} \rangle < (104 \div 228) \,\mathrm{meV}$, CUORE \cite{CUORE20} $\langle m_{ee} \rangle < (75 \div 350) \,\mathrm{meV}$ and CUPID-Mo Collaboration \cite{CUPID2021} $\langle m_{ee} \rangle < (310 \div 540) \,\mathrm{meV}$.

\begin{center}
\begin{figure}[h]
\begin{center}
\vspace{-1.0 cm}
\hspace*{-2.2 cm}
\includegraphics[width=0.8\textwidth]{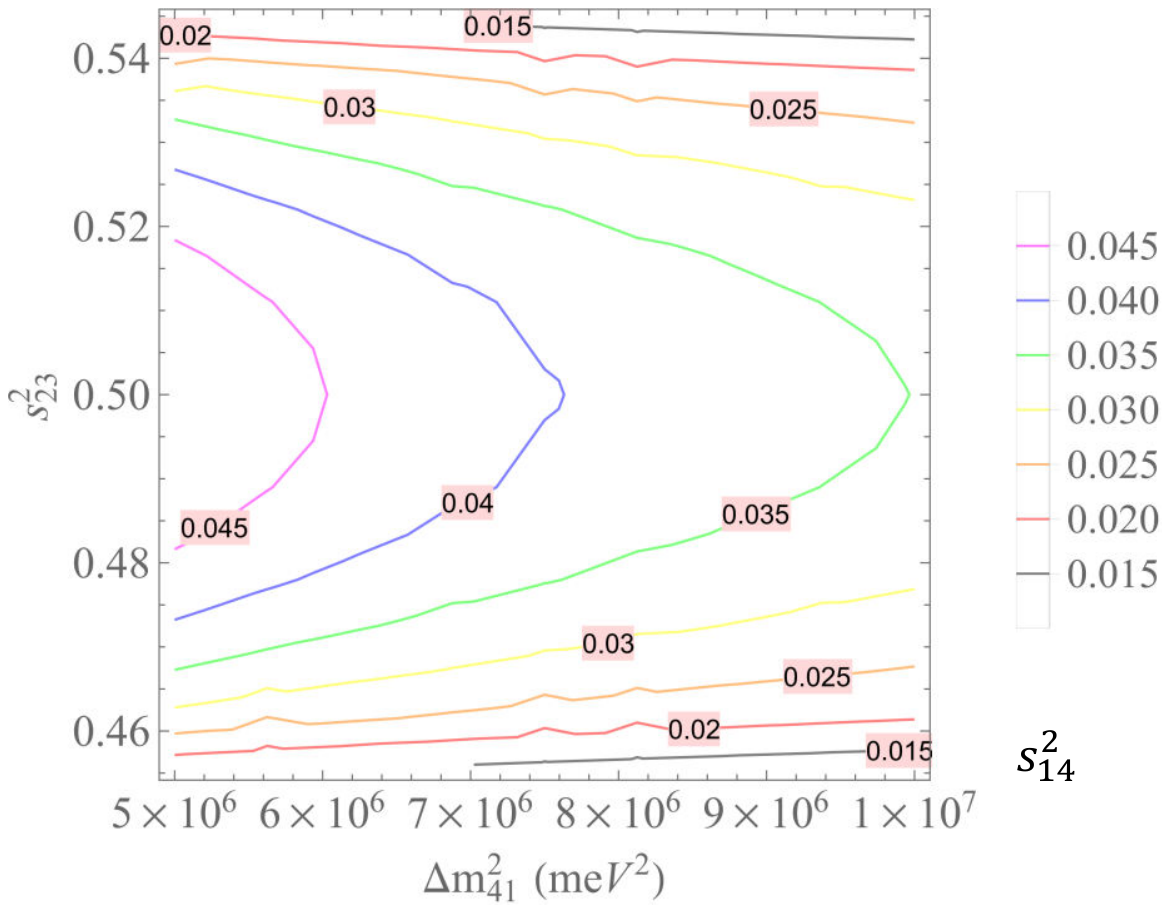}\hspace*{-5.5 cm}
\includegraphics[width=0.8\textwidth]{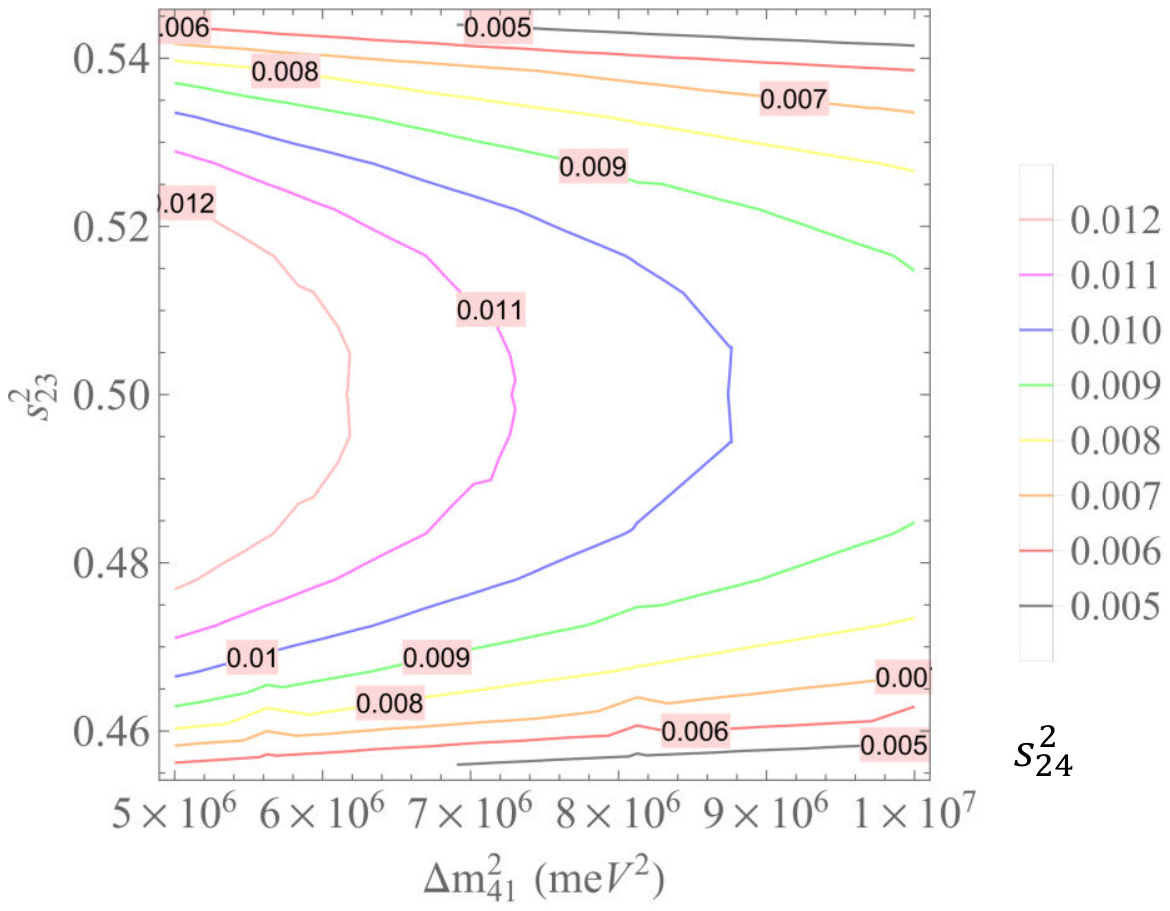}\hspace*{-2.6 cm}\\
\vspace{-12.2 cm}
\hspace*{-11.8 cm}\includegraphics[width=0.8\textwidth]{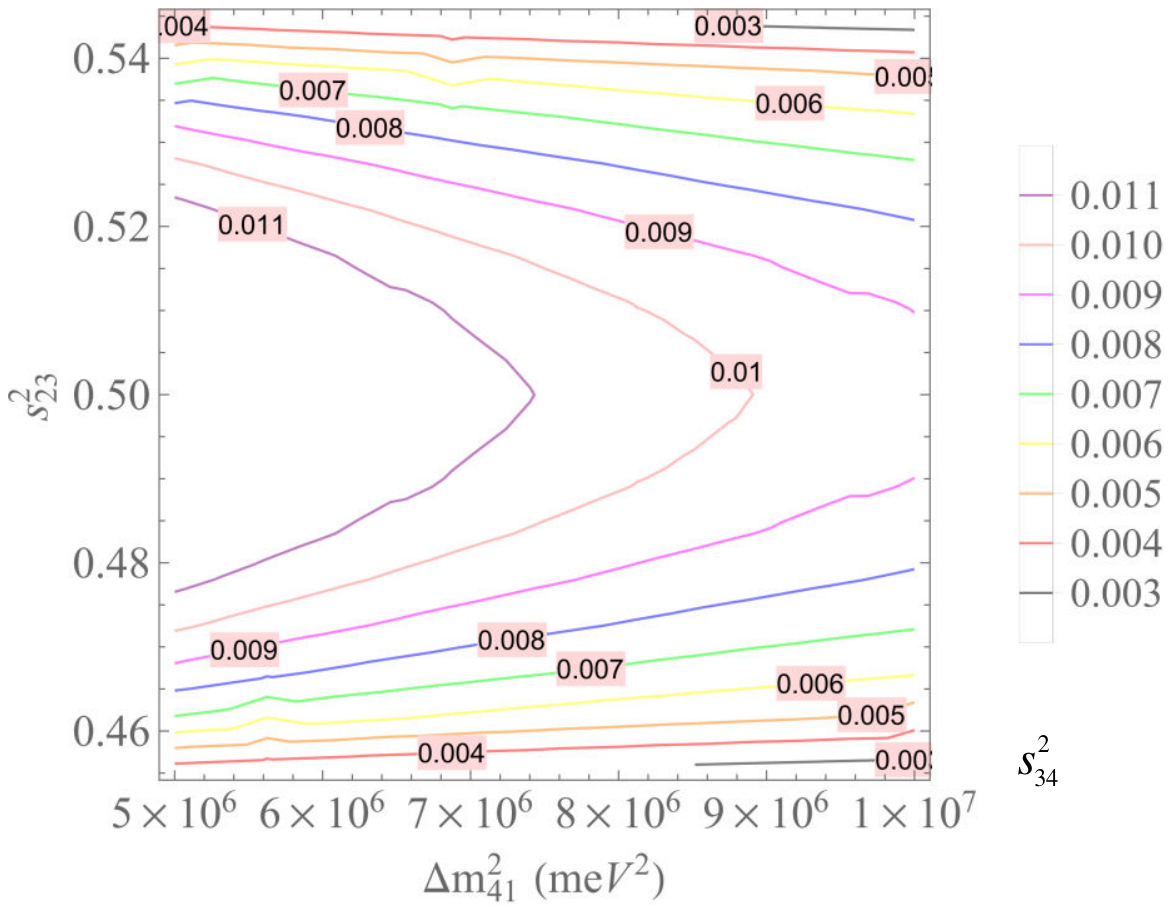}\hspace*{-4.25 cm}
\end{center}
\vspace{-11.85 cm}
\caption{$s^2_{14}, s^2_{24}$ and $s^2_{34}$ as functions of $s^2_{23}$ and $\Delta m^2_{41}$ with $s^2_{23}\in (0.456, 0.544)$ and $\Delta m^2_{41} \in (5.0, 10.0) 10^6\, \mathrm{meV}^2$ for NH.}
\label{si4sqNF}
\end{figure}
\end{center}

\begin{center}
\begin{figure}[h]
\begin{center}
\vspace{-1.0 cm}
\hspace*{-2.0 cm}
\includegraphics[width=0.8\textwidth]{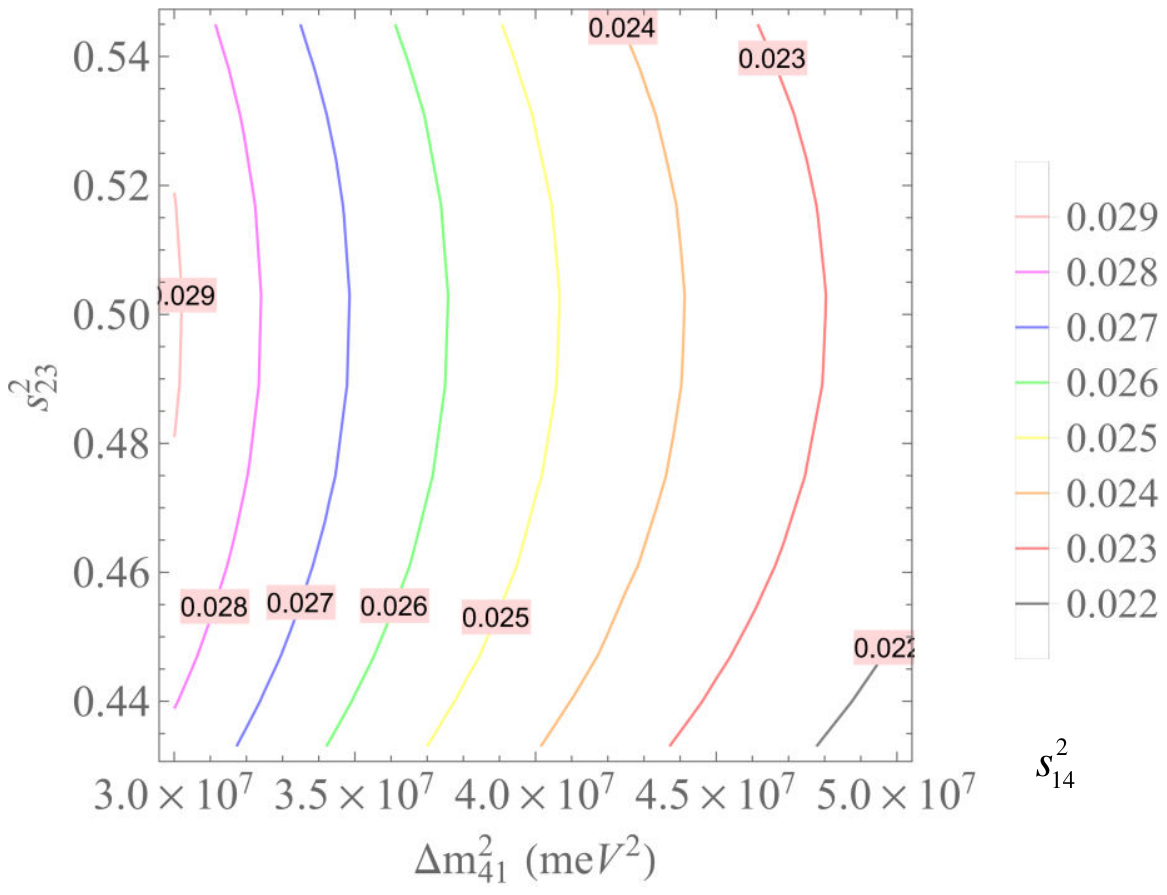}\hspace*{-5.5 cm}
\includegraphics[width=0.8\textwidth]{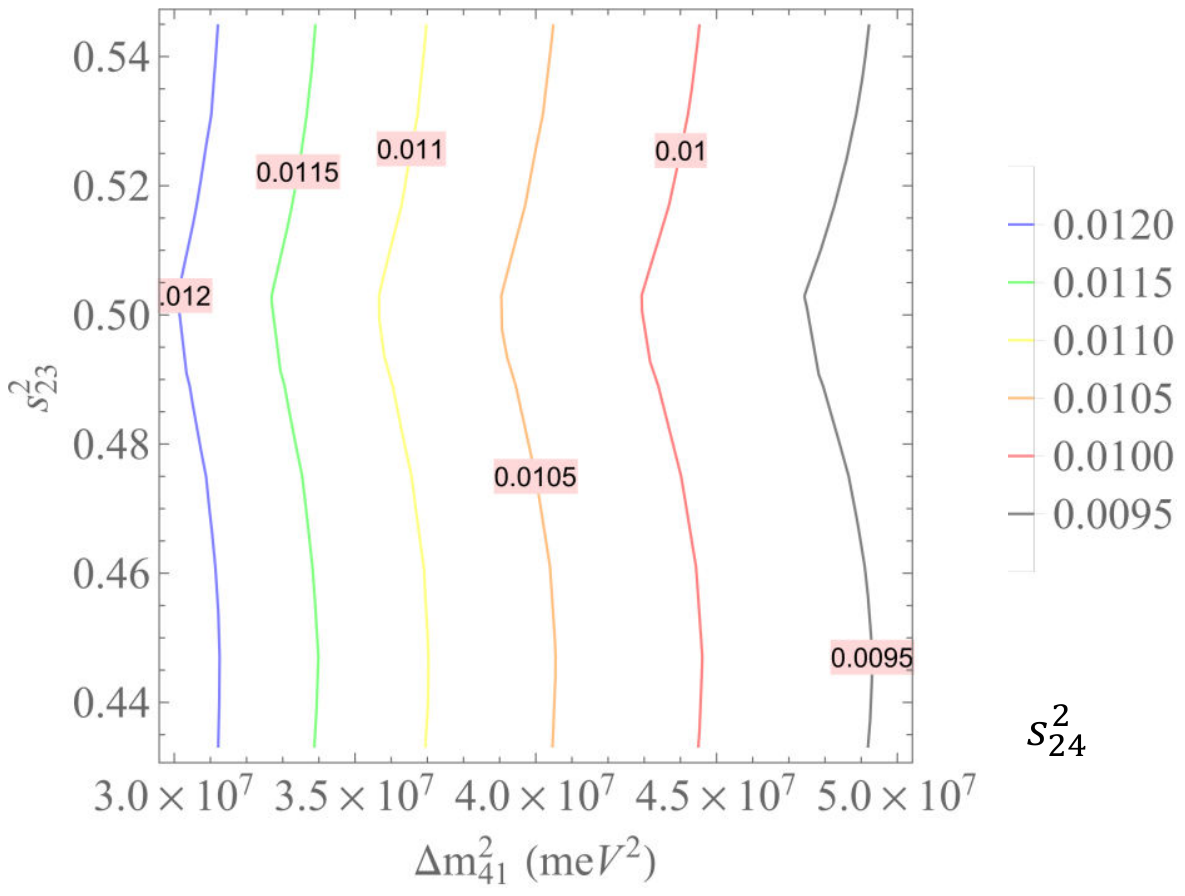}\hspace*{-2.5 cm}\\
\vspace{-12.2 cm}
\hspace*{-11.3 cm}\includegraphics[width=0.8\textwidth]{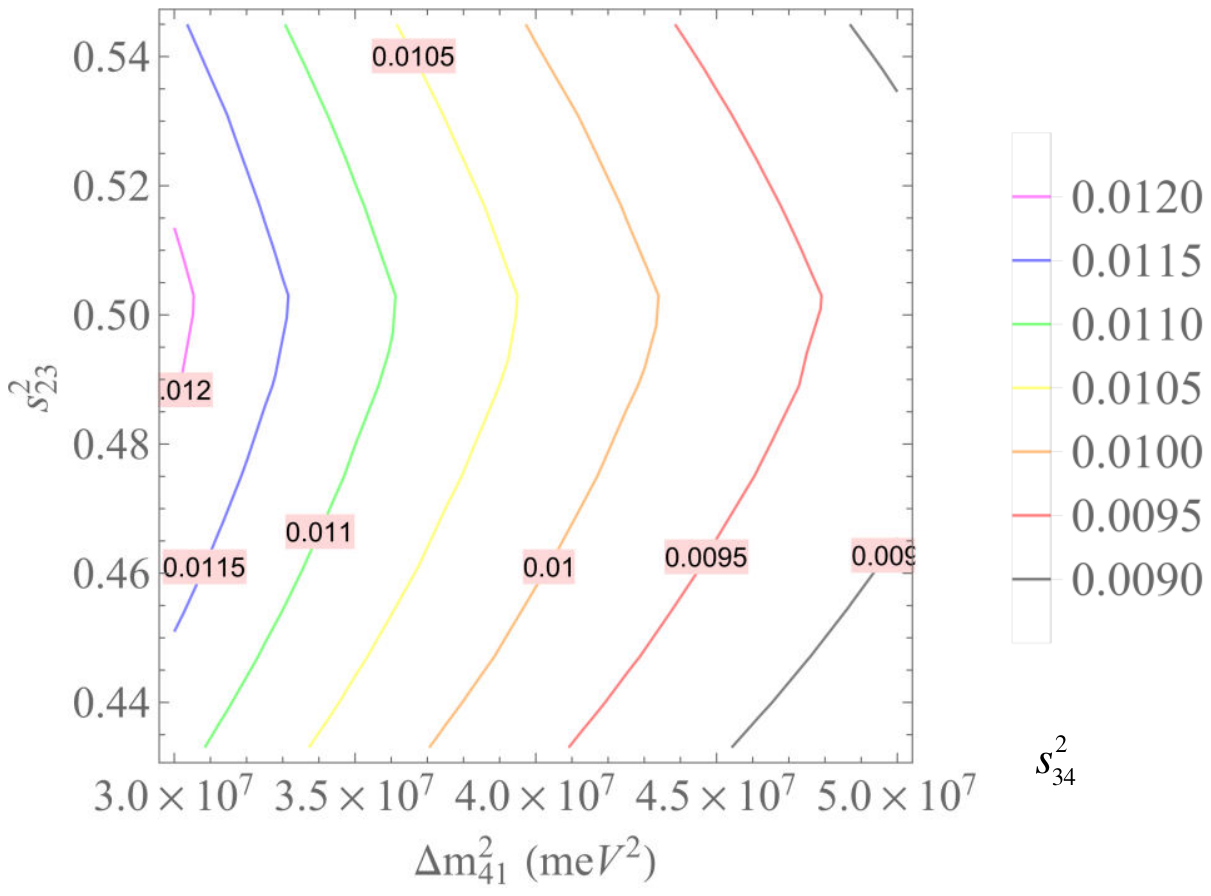}\hspace*{-4.25 cm}
\end{center}
\vspace{-11.95 cm}
\caption{$s^2_{14}, s^2_{24}$ and $s^2_{34}$ as functions of $s^2_{23}$ and $\Delta m^2_{41}$ with $s^2_{23}\in (0.433, 0.545)$ and $\Delta m^2_{41} \in (30.0, 50.0) 10^6\, \mathrm{meV}^2$ for IH.}
\label{si4sqIF}
\end{figure}
\end{center}

The dependence of $\langle m_{ee}\rangle$ on $s^2_{23}$ and $\Delta m^2_{41}$ with $s^2_{23}\in (0.433,0.545)$ and $\Delta m^2_{41} \in (10.0, 30.0) 10^6\, \mathrm{meV}^2$ are depicted in Fig. \ref{meeF}.
The dependences  of the absolute values of the
entries of the lepton mixing matrix in Eq. (\ref{Ulep}) on $s_{13}$ and $s_{23}$ are presented in Figs. \ref{UijNF} and \ref{UijIF} which indicate that
\bea
|U_{\mathrm{lep}}| \, \in\left\{
\begin{array}{l}
\left(
\begin{array}{ccc}
0.8020 \rightarrow 0.8040 & \fr{1}{\sqrt{3}} & 0.142 \rightarrow 0.154 \\
0.370 \rightarrow 0.420 & \fr{1}{\sqrt{3}} & 0.700 \rightarrow 0.730 \\
0.430 \rightarrow 0.470 & \fr{1}{\sqrt{3}} & 0.670 \rightarrow 0.695 \\
\end{array}
\right) \hspace{0.1cm}\mbox{for  NH},  \label{Ulepconst}  \\
\left(
\begin{array}{ccc}
0.8020 \rightarrow 0.8040 & \fr{1}{\sqrt{3}} & 0.144 \rightarrow 0.154 \\
0.440 \rightarrow 0.490 & \fr{1}{\sqrt{3}} & 0.655 \rightarrow 0.690 \\
0.340 \rightarrow 0.400 & \fr{1}{\sqrt{3}} & 0.710 \rightarrow 0.745 \\
\end{array}
\right) \hspace{0.1cm}\mbox{for IH}.
\end{array}%
\right.
\eea
At the best-fit points of the two light neutrino mass squared differences and reactor neutrino mixing angle taken from Ref. \cite{Salas2020}, $\Delta m^2_{31}=75.0 \, \mathrm{meV}^2$ and $\Delta m^2_{31}=2.55\times 10^3\, \mathrm{meV}^2,\, s^2_{13}=2.200\times 10^{-2}$ for NH while $\Delta m^2_{31}=-2.45\times 10^3\, \mathrm{meV}^2,\, s^2_{13}=2.225\times 10^{-2}$for IH, the parameter $m_\beta$ and $s^2_{k 4} \, (k=1, 2, 3)$ depend on $s_{23}$ and $m_{s}=\sqrt{\Delta m^2_{14}}$. At present, there are various experimental bounds on $\Delta m^2_{41}$ \cite{Aguilar2001, Arevalo2013, Acero2008, Arevalo2010, Mention2011, An2014, Arevalo2018, Gariazzo17, Adamson2019, Adamson2020, Aartsen2020, Beheraa2019, BeheraPRD2020}, for example, $\Delta m^2_{41} \in (0.01, 1.0)\, \mathrm{eV}^2$ \cite{Arevalo2013}, $\Delta m^2_{41}>10^{-2} \, \mathrm{eV}^2$ \cite{Adamson2019}, $\Delta m^2_{41} =0.041\, \mathrm{eV}^2$ \cite{Arevalo2018},  $\Delta m^2_{41}\in (0.1, 1.0)\, \mathrm{eV}^2$ \cite{Arevalo2010},  $\Delta m^2_{41} \in (0.2, 10.0)\, \mathrm{eV}^2$ \cite{Aguilar2001},   $\Delta m^2_{41} \geq 0.1\, \mathrm{eV}^2$ \cite{Acero2008}, $\Delta m^2_{41} =1.0\, \mathrm{eV}^2$ \cite{Beheraa2019, BeheraPRD2020}, $\Delta m^2_{41} > 1.5\, \mathrm{eV}^2$ \cite{Mention2011}, $\Delta m^2_{41} =1.7\, \mathrm{eV}^2$ \cite{Gariazzo17}, $\Delta m^2_{41} <10.0\, \mathrm{eV}^2$ \cite{Adamson2020} $\Delta m^2_{41} =1.45\, \mathrm{eV}^2$ \cite{Aartsen2020}. Thus in this work, $\Delta m^2_{41}$ is assumed in the range of $\Delta m^2_{41}\in (5.0, 10)\, \mathrm{eV}^2$ for NH while $\Delta m^2_{41}\in (30.0, 50.0)\, \mathrm{eV}^2$ \Revised{for IH}. Going by this assumption, the dependences of $m_\beta$ and $s^2_{k 4}$ on $s_{23}$ and $\Delta m^2_{41}$, with $s^2_{23}\in (0.456, 0.544)$ and $\Delta m^2_{41} \in (5.0, 10.0) 10^6\, \mathrm{meV}^2$ for NH  while $s^2_{23}\in (0.433, 0.545)$ and $\Delta m^2_{41} \in (30, 50) 10^6\, \mathrm{meV}^2$ for IH, are presented in Figs. \ref{mbF}, \ref{si4sqNF} and \ref{si4sqIF}, respectively.


Figures \ref{si4sqNF}-\ref{si4sqIF} show that the considered model predicts the range of $s_{1 4}^2, s_{2 4}^2$ and $s_{3 4}^2$ as follows
\bea
&&s^2_{14} \in\left\{
\begin{array}{l}
(0.015,\,0.045) \hspace{0.3 cm}    \mbox{for  NH,} \\
(0.022, \, 0.029) \hspace{0.3 cm} \mbox{for  IH,}%
\end{array}%
\right.   \label{s14range}\\
&&s^2_{24}\in\left\{
\begin{array}{l}
(0.005,\,\, 0.012) \hspace{0.3 cm} \mbox{for NH,} \\
(0.0095,\, 0.012) \hspace{0.2 cm} \mbox{for IH,}%
\end{array}%
\right.   \label{s24range} \\
&&s^2_{34}\in\left\{
\begin{array}{l}
 (0.003,\, 0.011) \hspace{0.3 cm}  \mbox{for  NH,} \\
(0.009,\,\, 0.012) \hspace{0.25 cm} \mbox{for  IH,}%
\end{array}%
\right.   \label{s34range}
\eea
which are in consistent with the constraints given in Tab. \ref{experconstrain}.

Finally, from the above analysis, the obtained parameters of the model are summarized in Table \ref{parameterrangeA4}.
\begin{table}[ht]
\caption{The model prediction compared to the experimental range \cite{Salas2020, Deepthi2020}}
{\begin{tabular}{@{}cc|cc|cccccc@{}} \hline
& Parameters & Experimental range(NH)& Prediction(NH)&Experimental range(IH) &Prediction(IH)\\\hline
& \,\,${\sum}_{i=1}^{3} m_i (\mathrm{meV})$ &\, $< 120$&\, $58.03\rightarrow 60.31$&\, $< 150$&\, $98.07\rightarrow 101.30$&\\
& \,\,$\sin^2\theta_{13} \times 10^{2} $ &\, $2.00\rightarrow 2.405$&\, $2.00\rightarrow 2.405$&\, $2.018\rightarrow 2.424$&\, $2.018\rightarrow 2.424$&\\
& \,\,$\sin^2\theta_{12}$ &\, $0.271\rightarrow 0.369$&\, $0.3401\rightarrow0.3415$&\, $0.271\rightarrow 0.369$&\, $0.3402\rightarrow0.3416$&\\
  & \,\,$\sin^2\theta_{23}$ &\, $0.434\rightarrow 0.610$&\, $0.456\rightarrow 0.544$&\, $0.433\rightarrow 0.608$&\, $0.433\rightarrow 0.545$&\\
  & \,\,$\delta_{CP} (^\circ)$ &\, $128 \rightarrow 352$&\, $156\rightarrow 172$&\, $200\rightarrow 353$&\, $140.8 \rightarrow 167.2$&\\
  &\,\, $s^2_{14}$ &\, $0.0098\rightarrow 0.0310$&\, $0.015\rightarrow 0.045$&\, $0.0098\rightarrow 0.0310$&\, $0.022\rightarrow 0.029$&\\
  &\,\, $s^2_{24}$ &\, $0.0059\rightarrow 0.0262$&\, $0.005\rightarrow 0.012$&\, $0.0059\rightarrow 0.0262$&\, $0.0095\rightarrow 0.012$&\\
  &\,\, $s^2_{34}$ &\, $0\rightarrow 0.0369$&\, $0.003 \rightarrow 0.011$&\, $0\rightarrow 0.0369$&\, $0.009 \rightarrow 0.012$&\\
  &\,\, $\langle m^{(3)}_{ee}\rangle\, (\mathrm{meV})$ &\, $\approx 1.0$ \cite{mbet3constraint}&\, $3.50 \rightarrow 4.1$&\, $\approx 1.0$ \cite{mbet3constraint}&\, $47.90 \rightarrow 48.50$&\\
  &\,\, $m^{(3)}_\beta \, (\mathrm{meV})$ &\, $8.9 \rightarrow 12.6$ \cite{mbet3constraint}&\, $8.85 \rightarrow 9.15$&\, $8.9 \rightarrow 12.6$ \cite{mbet3constraint} &\, $47.17 \rightarrow 49.23$&\\
  &\,\, $\langle m_{ee}\rangle\, (\mathrm{meV})$ &\, < $110\rightarrow 520$ \cite{CUORE20}&\, $40.0 \rightarrow 110.0$&\, $< 110\rightarrow 520$ \cite{CUORE20}&\, $185.0 \rightarrow 205.0$&\\
  &\,\, $m_\beta \, (\mathrm{meV})$ &\, $< 1100$ \cite{Karin21}&\, $300.0 \rightarrow 550.0$&\, $< 1100$ \cite{Karin21}&\, $700.0 \rightarrow 900.0$&\\
 \hline
\end{tabular}} \label{parameterrangeA4}
\end{table}
In the case $s^2_{23}=0.544\, (\theta_{23}=47.50^\circ)$ and $\Delta m^2_{41} =5.0\, \mathrm{eV}^2$  for NH while  $s^2_{23}=0.545\, (\theta_{23}=47.58^\circ)$ and $\Delta m^2_{41} =30.0\, \mathrm{eV}^2 \,(m_s=4477.23\, \mathrm{meV})$ for IH, we obtain
the parameters as given in Tab. \ref{parameterS4}
\begin{table}[ht]
\caption{The obtained parameters}
{\begin{tabular}{@{}cccccccccc@{}} \hline
& Parameters & The derived values (NH) &The derived values (IH)\\\hline
  & \,\,$k_2$ &\, $-2.29 $&\, $1.161$&\\
  & \,\,$k_3$ &\, $0.439 $&\, $0.9758$&\\
  & \,\,$\cos\alpha\, \left(\alpha^{(\circ)}\right)$ &\, $-0.136 \, (97.80)$&\, $0.9758\, (4.412)$&\\
  & \,\,$k_0$ &\, $5.83 $&\, $0.985$&\\
  & \,\,$m_1 (\mathrm{meV})$ &\, $0$&\, $49.50$&\\
  & \,\,$m_2 (\mathrm{meV})$ &\, $8.66$&\, $50.25$&\\
  & \,\,$m_1 (\mathrm{meV})$ &\, $50.50$&\, $0$&\\
    & \,\,${\sum}_{i=1}^{3} m_i (\mathrm{meV})$ &\, $59.20$&\, $99.75$&\\
  & \,\,$\sin \delta_{CP} \, \left(\delta_{CP} ^{(\circ)}\right)$ &\, $-0.912\, (294.0)$&\, $-0.9092\, (294.60)$&\\
  & \,\,$m_s \, (\mathrm{eV})$ &\, $2.24 $&\, $4.472$&\\
  &\,\, $s^2_{14}$ &\, $0.0178$&\, $0.02853$&\\
  &\,\, $s^2_{24}$ &\, $0.00589$&\, $0.01225$&\\
  &\,\, $s^2_{34}$ &\, $0.00394$&\, $0.01157$&\\
  &\,\, $\langle m^{(3)}_{ee}\rangle\, (\mathrm{meV})$ &\, $3.575$&\, $48.45$&\\
  &\,\, $\langle m_{ee}\rangle\, (\mathrm{meV})$ &\, $37.27$&\, $203.90$&\\
  &\,\, $m^{(3)}_\beta\, (\mathrm{meV})$ &\, $9.006$&\, $49.20$&\\
  &\,\, $m_\beta\, (\mathrm{meV})$ &\, $294.20$&\, $926.40$&\\
 \hline
\end{tabular}} \label{parameterS4}
\end{table}
and the lepton mixing matrix gets the explicit form:
\bea
U_{lep}=\left\{
\begin{array}{l}
\left(
\begin{array}{ccc}
 0.8012\, +0.05287 i & 0.5774 & -0.1285-0.07406 i \\
 0.09636\, -0.354 i & -0.2887+0.5 i & -0.6317-0.3646 i \\
 0.09636\, +0.4598 i & -0.2887-0.5 i & -0.6317+0.2165 i \\
\end{array}
\right)\hspace{0.1cm}\mbox{for  NH},  \label{Ulepconst}  \\
\left(
\begin{array}{ccc}
 0.8009\, -0.05405 i & 0.5774 & -0.1285+0.07575 i \\
 0.09596\, -0.4611 i & -0.288+0.5 i & -0.6315-0.2147 i \\
 0.09596\, +0.353 i & -0.2888-0.5 i & -0.6315+0.3662 i \\
\end{array}
\right) \hspace{0.1cm}\mbox{for IH}{\Revised{,}}
\end{array}%
\right.
\eea
which are unitary and in good agreement with the constraint on the absolute values of the entries of the lepton mixing matrix given in Eq. (\ref{2020lepmix}).

\section{\label{gminus2s}Muon anomalous magnetic moment}
\begin{figure}[ht]
\vspace*{0.5cm}
\centering
\includegraphics[width=10.0cm, height=9cm]{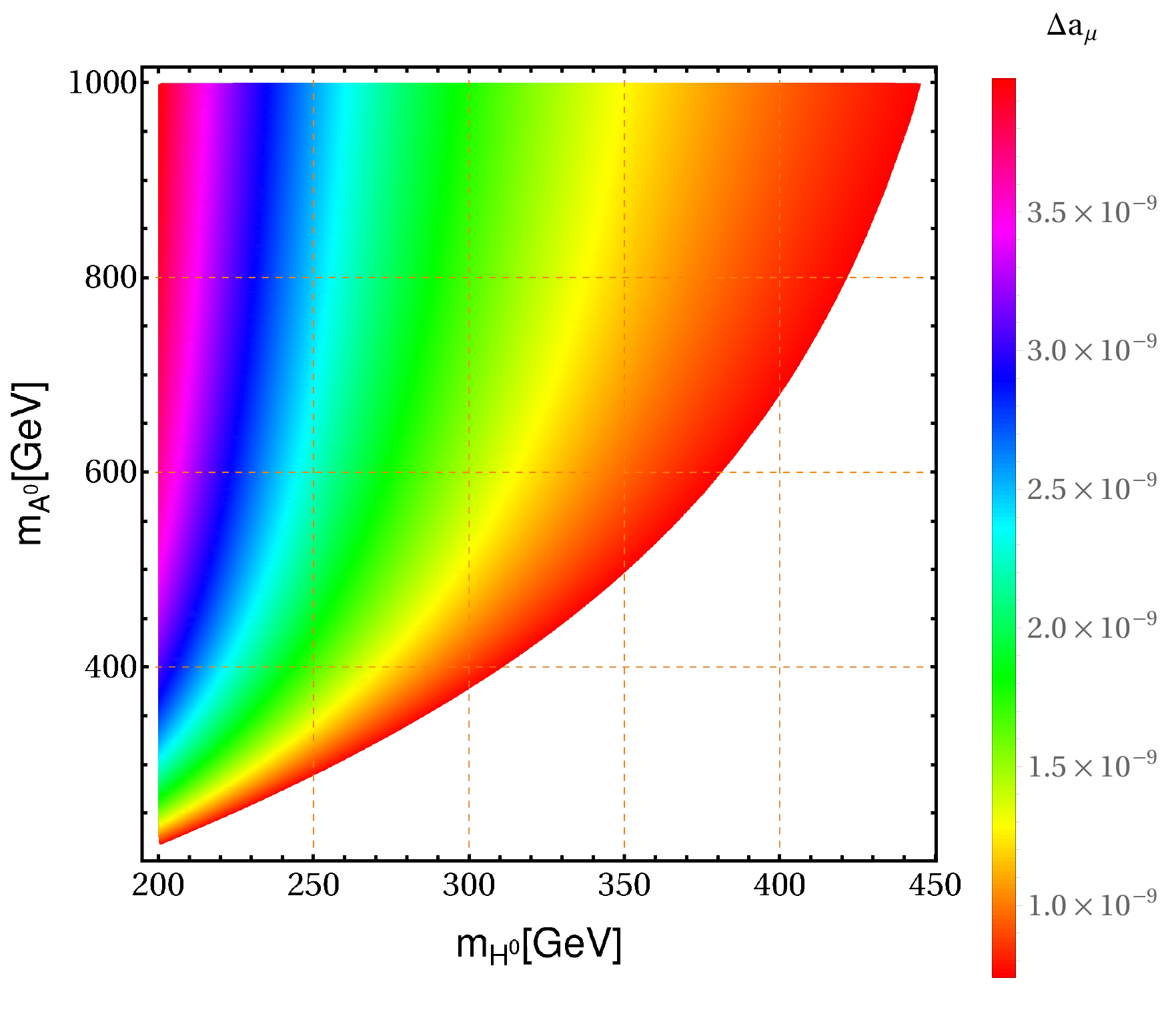}
\vspace*{-0.5cm}
\caption{Allowed parameter space in the $m_{H^{0}}-m_{A^{0}}$ plane consistent with the muon anomalous magnetic moment.}\label{gminus2}
\end{figure}

The experimental data shows a significant deviation of the muon anomalous
magnetic moment from its SM value
\begin{equation}
\Delta a_{\mu }=a_{\mu }^{\mathrm{exp}}-a_{\mu }^{\mathrm{SM}}=\left(
2.51\pm 0.59\right) \times 10^{-9}\hspace{17mm}\mbox{%
\cite{Hagiwara:2011af,Davier:2017zfy,Nomura:2018lsx,Nomura:2018vfz,Blum:2018mom,Keshavarzi:2018mgv,Aoyama:2020ynm,Abi:2021gix}}
\label{eq:a-mu}
\end{equation}%
In this subsection, we will analyze the implications of our model in the
muon anomalous magnetic moment. Muon anomalous magnetic moments mainly
arises from one-loop diagrams involving the exchange of electrically neutral
CP even and CP odd scalars and the muon running in the internal lines of the
loop. It is worth mentioning that due \Revised{to} the symmetries in our model, there
are no tree level flavor changing neutral scalar interaction in the leptonic
Yukawa terms, thus implying that the muon is the only charged lepton
contributing to the muon anomalous magnetic moment. There are also
contributions arising from electrically charged scalar and right handed
Majorana neutrinos but these contributions are strongly \Revised{suppressed.}  Thus, the
leading contributions to the muon anomalous magnetic moment take the form:
\bea
&&\Delta a_{\mu }\simeq \frac{\left[ \left( y_{h\overline{\mu }_{R}\mu
_{L}}\right) ^{2}-\left( y_{h\overline{\mu }_{R}\mu _{L}}^{\left( SM\right)
}\right) ^{2}\right] m_{\mu }^{2}}{8\pi ^{2}}I_{H}^{(\mu )}\left( m_{\mu
},m_{h}\right) \crn
&&\hspace{0.8 cm}+\, \frac{m_{\mu }^{2}}{8\pi ^{2}}\left[ y_{H^{0}\overline{\mu }%
_{R}\mu _{L}}^{2}I_{H}^{(\mu )}\left( m_{\mu },m_{H^{0}}\right) +y_{A^{0}%
\overline{\mu }_{R}\mu _{L}}^{2}I_{A}^{(\mu )}\left( m_{\mu
},m_{A^{0}}\right) \right] ,
\label{deltaamu}
\eea
\Revised{where the Yukawa couplings appearing in Eq. (\ref{deltaamu}) are given by:
\begin{eqnarray}
y_{h\overline{\mu }_{R}\mu _{L}} &=&\sqrt{\frac{3}{2}}\left( y_{2}\sin
\alpha -y_{3}\cos \alpha \right) ,\hspace*{0.2cm}\hspace*{0.2cm}\hspace*{%
	0.2cm}\hspace*{0.2cm}y_{h\overline{\mu }_{R}\mu _{L}}^{\left( SM\right)}=\frac{m_{\mu }}{v}, \\
y_{H^{0}\overline{\mu }_{R}\mu _{L}} &=&\sqrt{\frac{3}{2}}\left( y_{2}\cos
\alpha +y_{3}\sin \alpha \right) ,\hspace*{0.2cm}\hspace*{0.2cm}\hspace*{%
	0.2cm}\hspace*{0.2cm}y_{A^{0}\overline{\mu }_{R}\mu _{L}}=\sqrt{\frac{3}{2}}%
\left( y_{2}\sin \beta -y_{3}\cos \beta \right),
\end{eqnarray}
whereas the} loop function $I_{H\left( A\right) }^{\left( \mu \right) }\left(
m_{f},m_{H,A}\right) $ has the form \cite%
{Diaz:2002uk,Jegerlehner:2009ry,Kelso:2014qka,Lindner:2016bgg,Kowalska:2017iqv}:
\begin{equation}
I_{H\left( A\right) }^{\left( \mu \right) }\left( m_{f},m_{H,A}\right)
=\int_{0}^{1}\frac{x^{2}\left( 1-x\pm \frac{m_{f}}{m_{\mu }}\right) }{m_{\mu
}^{2}x^{2}+\left( m_{f}^{2}-m_{\mu }^{2}\right) x+m_{H,A}^{2}\left(
1-x\right) }dx.
\end{equation}
Figure \ref{gminus2} shows the allowed parameter space in the $m_{H^{0}}-m_{A^{0}}$ plane consistent with the muon anomalous magnetic moment. We find that our model can successfully accommodate the experimental values of the muon anomalous magnetic moment.
\section{\label{conclusion} Conclusions}
We have proposed a $B-L$ model combined with the $S_4\times Z_3\times Z_4$ discrete group which successfully explains the recent $3+1$ sterile-active neutrino data. The tiny masses of the light active neutrinos are obtained through the type-I seesaw mechanism. The active-active and sterile-active neutrino mixing angles are predicted to be consistent with the recent constraints where $0.3401\, (0.3402) \leq \sin^2\theta_{12}\leq 0.3415\, (0.3416), \, 0.456\, (0.433) \leq \sin^2\theta_{23}\leq 0.544\, (0.545), \, 2.00\, (2.018) \leq 10^2\times \sin^2\theta_{13}\leq 2.405\, (2.424),\,\, 156 \, (140.8) \leq \delta^{(\circ)}_{CP}\leq 172\, (167.2)$ for normal (inverted) ordering of the three neutrino scenario, and $0.015 \,(0.022) \leq s^2_{14}\leq 0.045 \,(0.029), \, 0.005 (0.0095)\leq s^2_{24}\leq  0.012\, (0.012), \, 0.003 \,(0.009)$ $\leq s^2_{34} \leq 0.011$ for normal (inverted) ordering of the $3+1$ neutrino scenario. The effective neutrino masses are predicted to be in the ranges $40.0\, (185) \leq \langle m_{ee}\rangle [\mbox{meV}] \leq 110.0\, (205.0)$ and $300 \, (700) \leq \langle m_{\beta}\rangle [\mbox{meV}] \leq 550 \, (900)$, for normal and inverted neutrino mass orderings, respectively, range of values consistent with the recent experimental data. Our model predicts flavour conserving leptonic neutral scalar interactions and successfully explains the muon $g-2$ anomaly.

\section*{Acknowledgments}
A.E.C.H is supported by ANID-Chile FONDECYT 1210378, ANID PIA/APOYO
AFB180002 and \Revised{ANID- Programa Milenio - code} ICN2019\_044.  \Revised{H. N. Long acknowledges the financial support of the International Centre of Physics at the Institute of Physics, VAST under grant No: ICP.2022.02.}
\newpage
\appendix
\Revised{\section{\label{forbidappen}Forbidden terms}
\begin{table}[h]
\begin{center}
\vspace{-0.5 cm}
\caption{\label{forbidtable}\Revised{Forbidden terms}}
\vspace{0.15 cm}
 \begin{tabular}{|c|c|c|c|c|} \hline
Couplings &Forbidden by \\ \hline
$(\overline{\psi}_{L} \psi_{L}^c)_{\mathbf{1}} \widetilde{H}^2, (\overline{\psi}_{L} \psi_{L}^c)_{\mathbf{1}} \widetilde{H^'}^2, (\overline{\nu}^C_{R} \nu_{R})_{\textbf{1}} (\chi^{*2})_{\textbf{1}}, (\overline{\nu}^C_{R} \nu_{R})_{\textbf{1}} (\phi \varphi^*)_{\textbf{1}},
(\overline{\nu}^C_{R} \nu_{R})_{\textbf{2}} (\phi \varphi^*)_{\textbf{2}}, $&\multirow{5}{2.2 cm}{\hspace{0.55 cm}$U(1)_{B-L}$}  \\
$(\overline{\nu}^C_{R} \nu_{R})_{\textbf{3}} (\phi \varphi^*)_{\textbf{3}},
(\overline{\nu}^C_{R} \nu_{R})_{\textbf{3}^'} (\phi \varphi^*)_{\textbf{3}^'}, (\overline{\nu}^C_{R} \nu_{R})_{\textbf{3}} (\phi \rho^*)_{\textbf{3}}, (\overline{\nu}^C_{R} \nu_{R})_{\textbf{1}} (\phi^* \phi_s)_{\textbf{1}}, $ &\\
$(\overline{\nu}^C_{R} \nu_{R})_{\textbf{2}} (\phi^* \phi_s)_{\textbf{2}},
(\overline{\nu}^C_{R} \nu_{R})_{\textbf{3}} (\phi^* \phi_s)_{\textbf{3}},
(\overline{\nu}^C_{R} \nu_{R})_{\textbf{3}^'} (\phi^* \phi_s)_{\textbf{3}^'}, (\overline{\nu}^C_{R} \nu_{R})_{\textbf{1}} (\varphi \phi^*_s)_{\textbf{1}},$ &\\
$(\overline{\nu}^C_{R} \nu_{R})_{\textbf{2}} (\varphi \phi^*_s)_{\textbf{2}},
(\overline{\nu}^C_{R} \nu_{R})_{\textbf{3}} (\varphi \phi^*_s)_{\textbf{3}},
(\overline{\nu}^C_{R} \nu_{R})_{\textbf{3}^'} (\varphi \phi^*_s)_{\textbf{3}^'},
(\overline{\nu}^C_{R} \nu_{R})_{\textbf{1}} (\rho \phi^*_s)_{\textbf{1}}, (\overline{\nu}^C_{s} \nu_{R})_{\textbf{3}} \varphi,$&\\
$ (\overline{\nu}^C_{s} \nu_{R})_{\textbf{3}} (\phi \chi^*)_{\textbf{3}}, (\overline{\nu}^C_{s} \nu_{s})_{\textbf{1}} (\phi^2)_{\textbf{1}},
(\overline{\nu}^C_{s} \nu_{s})_{\textbf{1}} (\phi^{*2}_s)_{\textbf{1}}, (\overline{\nu}^C_{s} \nu_{s})_{\textbf{1}} (\phi^* \varphi^*)_{\textbf{1}}, (\overline{\nu}^C_{s} \nu_{s})_{\textbf{1}} (\varphi \phi_s)_{\textbf{1}}$.&\\ \hline
$(\overline{\psi}_{L} l_{1R})_{\mathbf{3}} (H^'\phi)_{\mathbf{3}^'}, (\overline{\psi}_{L} \nu_{R})_{\mathbf{1}} (\widetilde{H}^'\rho)_{\mathbf{1}^'}. $&$S_4$  \\ \hline
$(\overline{\psi}_{L} l_{1R})_{\mathbf{3}} (H\varphi)_{\mathbf{3}}, (\overline{\psi}_{L} l_{1R})_{\mathbf{3}} (H\phi_s)_{\mathbf{3}}, (\overline{\psi}_{L} l_{\al R})_{\mathbf{3}} (H\varphi)_{\mathbf{3}}, (\overline{\psi}_{L} l_{\al R})_{\mathbf{3}} (H\phi_s)_{\mathbf{3}}, $&\multirow{3}{2 cm}{\hspace{0.8 cm}$Z_3$}  \\
$(\overline{\psi}_{L} l_{\al R})_{\mathbf{3}} (H^'\varphi)_{\mathbf{3}^'}, (\overline{\psi}_{L} l_{\al R})_{\mathbf{3}} (H^'\phi_s)_{\mathbf{3}^'},
(\overline{\psi}_{L} \nu_{R})_{\mathbf{3}}(\widetilde{H}\phi)_{\mathbf{3}},
(\overline{\psi}_{L} \nu_{R})_{\mathbf{3}}(\widetilde{H}\phi_s)_{\mathbf{3}},$ &\\
$(\overline{\psi}_{L} \nu_{R})_{\mathbf{3}^'} (\widetilde{H}^'\phi)_{\mathbf{3}^'}, (\overline{\psi}_{L} \nu_{R})_{\mathbf{3}^'} (\widetilde{H}^'\phi_s)_{\mathbf{3}^'}, (\overline{\nu}^C_{s} \nu_{R})_{\textbf{3}} (\phi \chi)_{\textbf{3}}, (\overline{\nu}^C_{s} \nu_{R})_{\textbf{3}} (\varphi \chi)_{\textbf{3}}.$ &\\ \hline

$(\overline{\psi}_{L} l_{\al R})_{\mathbf{3}} (H^'\phi^*_s)_{\mathbf{3}^'}, (\overline{\psi}_{L} \nu_{R})_{\mathbf{1}}\widetilde{H},
(\overline{\psi}_{L} \nu_{R})_{\mathbf{3}}(\widetilde{H}\varphi^*)_{\mathbf{3}}, (\overline{\psi}_{L} \nu_{R})_{\mathbf{1}}(\widetilde{H}\rho^*)_{\mathbf{1}}, $&\multirow{3}{2 cm}{\hspace{0.8 cm}$Z_4$}  \\
$(\overline{\psi}_{L} \nu_{R})_{\mathbf{3}^'} (\widetilde{H}^'\varphi^*)_{\mathbf{3}^'}, (\overline{\psi}_{L} \nu_{s})_{\textbf{3}} (\widetilde{H}\phi^*)_{\textbf{3}}, (\overline{\psi}_{L} \nu_{s})_{\textbf{3}} (\widetilde{H}\phi_s)_{\textbf{3}}, (\overline{\nu}^C_{R} \nu_{R})_{\textbf{3}} (\varphi \chi)_{\textbf{3}}, $ &\\
$(\overline{\nu}^C_{R} \nu_{R})_{\textbf{3}} (\varphi^* \chi)_{\textbf{3}}, (\overline{\nu}^C_{R} \nu_{R})_{\textbf{1}} (\rho\chi)_{\textbf{1}},
(\overline{\nu}^C_{R} \nu_{R})_{\textbf{1}} (\rho^*\chi)_{\textbf{1}}, (\overline{\nu}^C_{s} \nu_{R})_{\textbf{3}} (\phi^* \chi)_{\textbf{3}}.$ &\\ \hline
\end{tabular}
\end{center}
\end{table}}
\section{\label{Higgspotential} Higgs potential invariant under $\mathbf{\Gamma}$ symmetry}
\begin{center}
\begin{figure}[ht]
\begin{center}
\vspace*{-0.75cm}
\hspace*{-1.0cm}\includegraphics[width=1.0\textwidth]{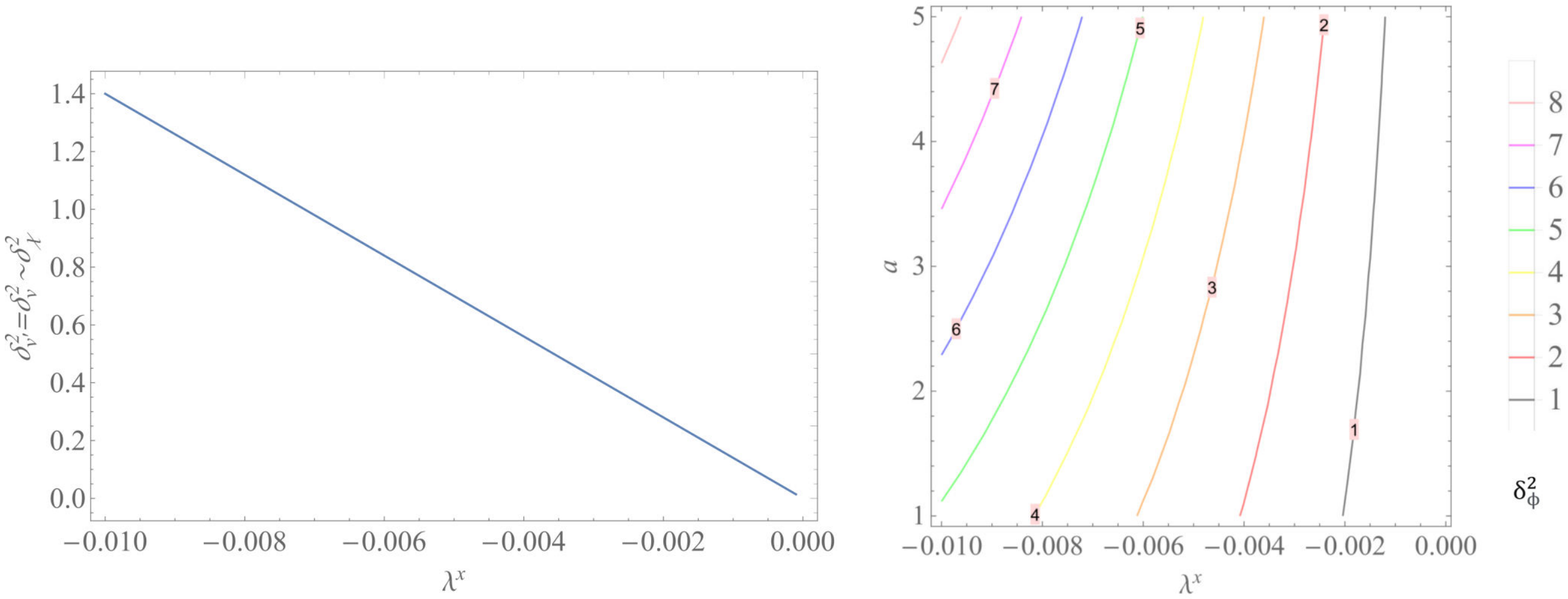}\hspace*{-1.0cm}
\vspace*{-5.35 cm}
\caption[$\delta^2_{v}=\delta^2_{v^'}\simeq \delta^2_{\chi}$ versus $\lambda^{x}$ (left panel), and $\delta^2_{\phi}$ versus $\lambda^{x}$ and $a$ (right panel) with $\lambda^{x}\in (-10^{-2}, -10^{-4})$ and $a\in (1.0, 5.0)$]{$\delta^2_{v}=\delta^2_{v^'}\simeq \delta^2_{\chi}$ versus $\lambda^{x}$ (in the left panel), and $\delta^2_{\phi}$ versus $\lambda^{x}$ and $a$ (in the right panel) with $\lambda^{x}\in (-10^{-2}, -10^{-4})$ and $a\in (1.0, 5.0)$}\label{del1263sq}
\vspace*{-0.75 cm}
\end{center}
\end{figure}
\end{center}

\begin{center}
\begin{figure}[ht]
\begin{center}
\vspace*{-1.0cm}
\hspace*{-1.0cm}\includegraphics[width=0.675\textwidth]{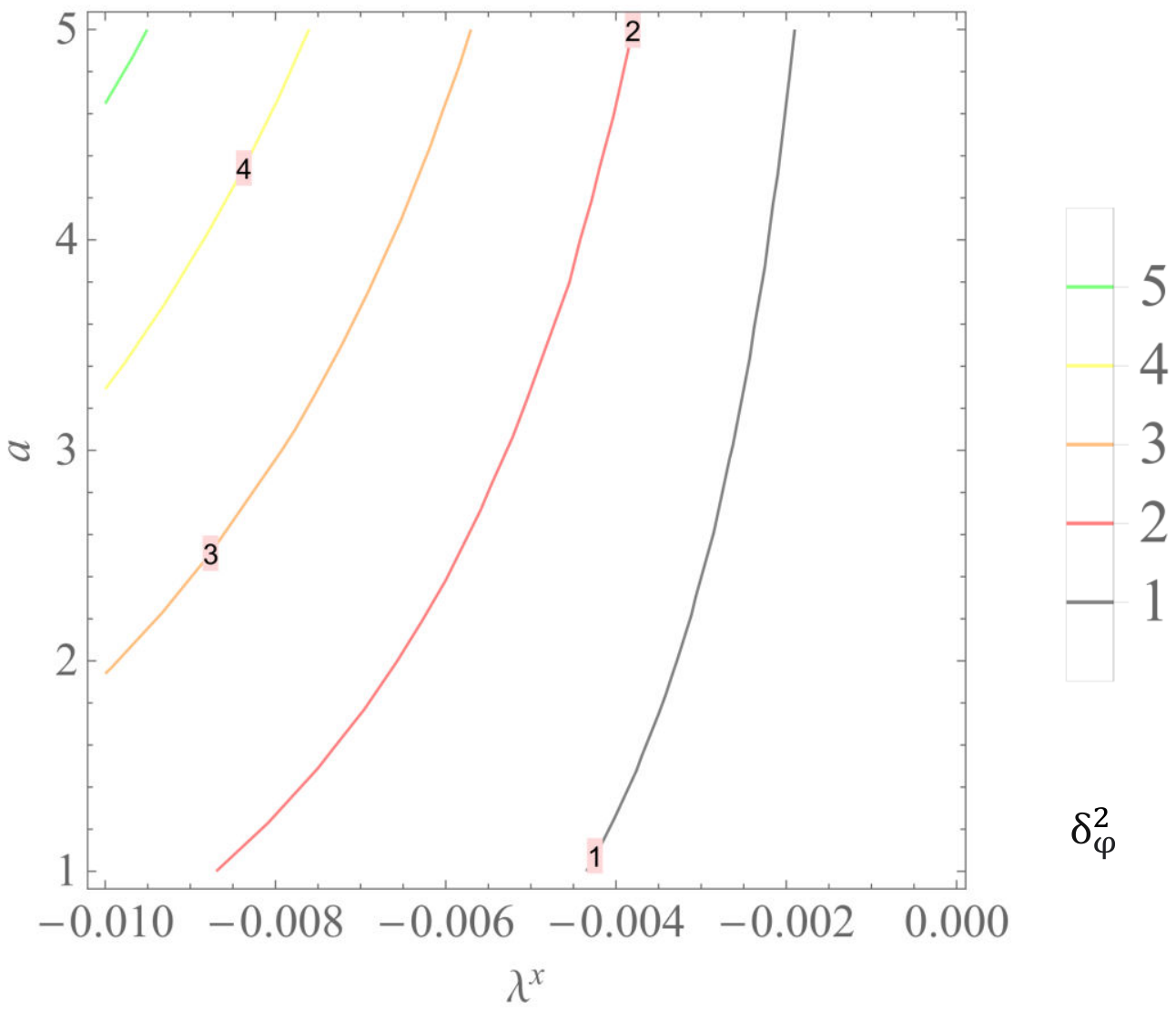}\hspace*{-3.75cm}
\includegraphics[width=0.675\textwidth]{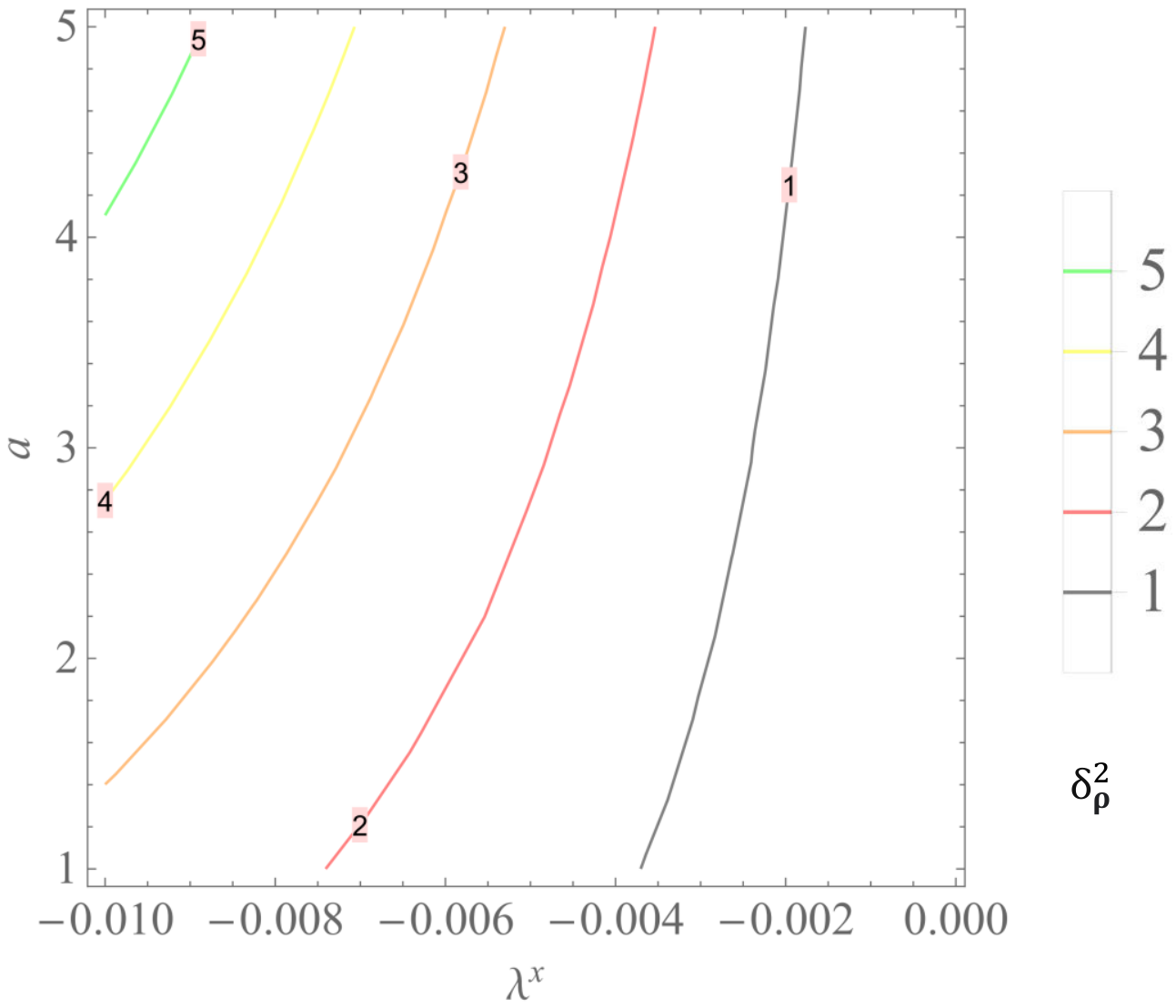}\\
\vspace*{-9.1cm}
\hspace*{-1.0cm}\includegraphics[width=0.675\textwidth]{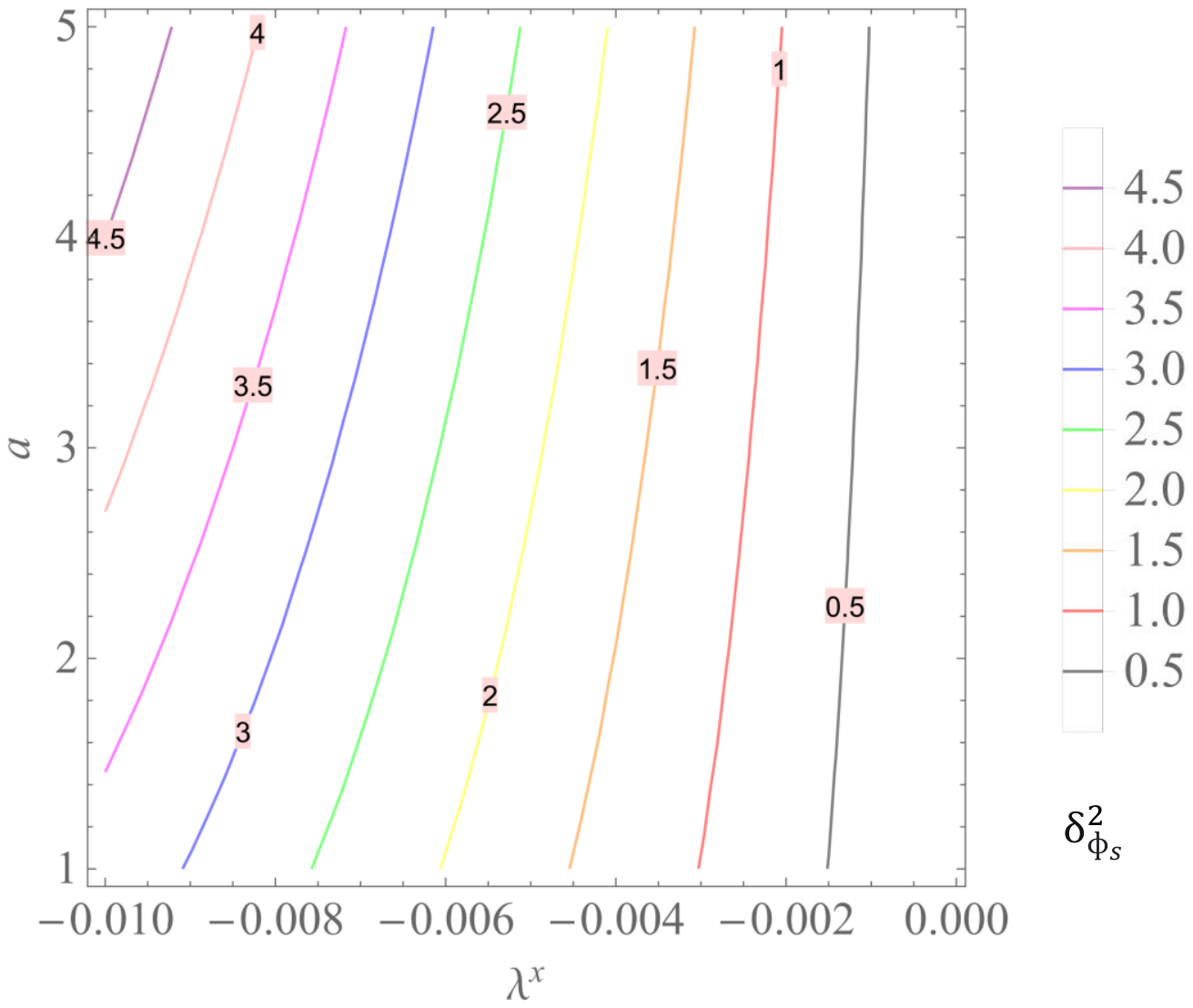}\hspace*{-3.75cm}
\includegraphics[width=0.675\textwidth]{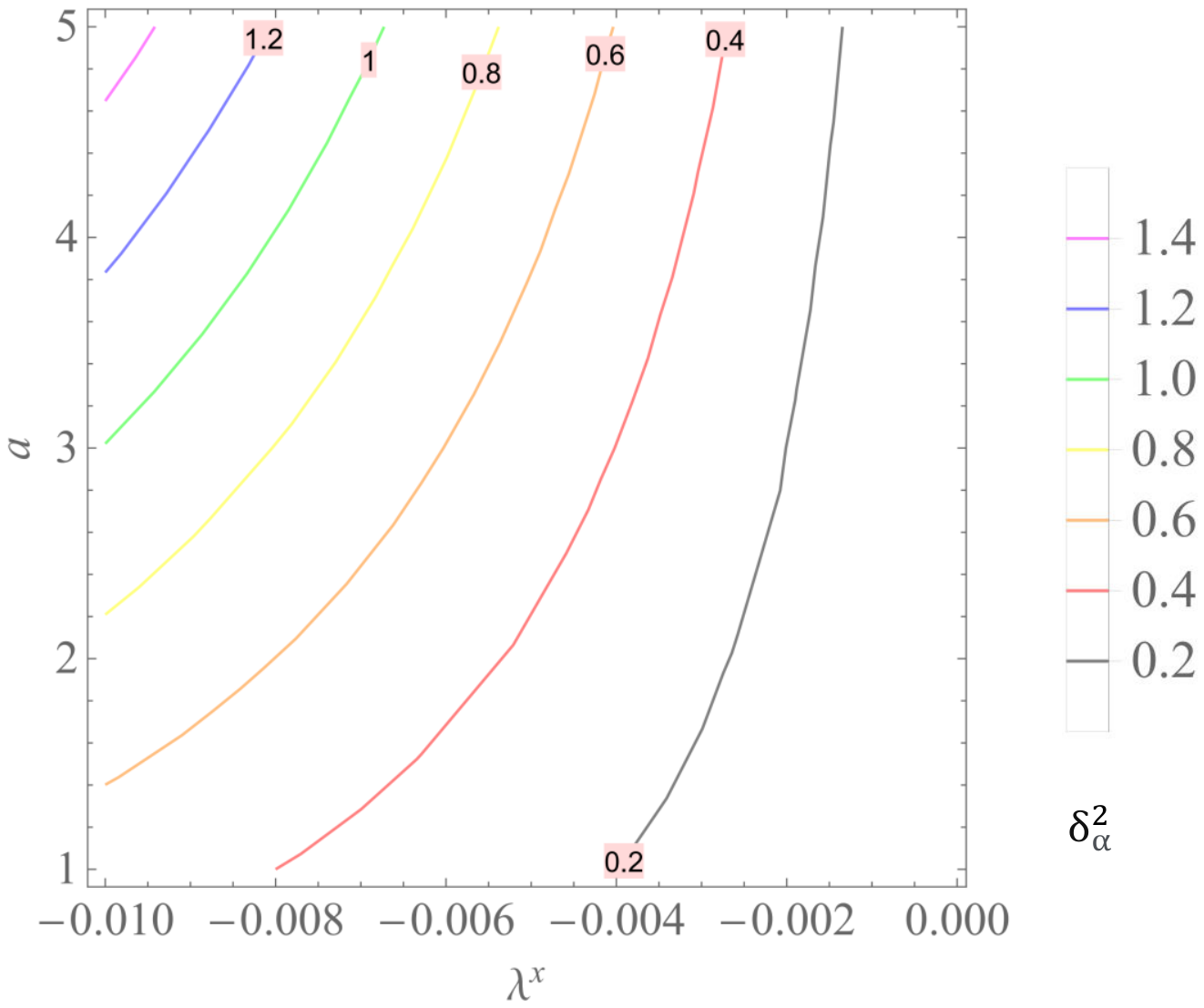}
\vspace*{-9.5cm}
\caption[$\delta^2_{\varphi}$ (upper left), $\delta^2_{\rho}$ (upper right), $\delta^2_{\phi_s}$ (bottom left) and $\delta^2_{\alpha}$ (bottom right) versus $\lambda^{x}$ and $a$ with $\lambda^{x}\in (-10^{-2}, -10^{-4})$ and $a\in (1.0, 5.0)$]{$\delta^2_{\varphi}$ (upper left), $\delta^2_{\rho}$ (upper right), $\delta^2_{\phi_s}$ (bottom left) and $\delta^2_{\alpha}$ (bottom right) versus $\lambda^{x}$ and $a$ with $\lambda^{x}\in (-10^{-2}, -10^{-4})$ and $a\in (1.0, 5.0)$}\label{delphisalphasq}
\vspace*{-1.5cm}
\end{center}
\end{figure}
\end{center}

The total renormalizable scalar potential 
invariant under $\mathrm{\Gamma}$ symmetry 
is given by\footnote{Here, $V(a_1 \rightarrow a_2, b_1 \rightarrow b_2,\cdots) \equiv V(a_1, b_1,\cdots)\!\!\!\mid_{\{a_1=a_2,\, b_1=b_2,\cdots \}}$.}:
\bea
V_{\mathrm{scal}}&=& V(H) +V(H^')+V(\phi)+V(\varphi)+V(\rho)+V(\chi)+V(\phi_s)+V(HH^')+V(H\phi)\crn
&+&V(H\varphi)+ V(H\rho)+V(H\chi)+V(H\phi_s)+ V(H^'\phi)+V(H^'\varphi)+V(H^'\rho)+V(H^'\chi)\crn
&+&V(H^'\phi_s)+ V(\phi\varphi)+V(\phi\rho)+V(\phi\chi)+V(\phi\phi_s)+V(\varphi\rho)+V(\varphi\chi)+V(\varphi\phi_s)\crn
&+&V(\rho\chi)+V(\rho\phi_s)+V(\chi\phi_s) +V_{\mathrm{trip}}+V_{\mathrm{quart}}, \label{Vtotal}
\eea
where
\bea
&&V(H)=\mu^2_H H^\+H +\lambda^H (H^\+H)^2,\hs V(H^')=V(H, H\rightarrow H^'), \crn
&&V(\phi)=\mu^2_\phi \phi^* \phi +\lambda^\phi_1 (\phi^* \phi)_1 (\phi^* \phi)_1
+\lambda^\phi_2 (\phi^* \phi)_{2}(\phi^* \phi)_{2}+\lambda^\phi_3 (\phi^* \phi)_{3}(\phi^* \phi)_{3} +\lambda^\phi_4 (\phi^* \phi)_{3^{'}}(\phi^* \phi)_{3^{'}},\crn
&&V(\varphi)=V(\phi\rightarrow \varphi), \, V(\rho)=\mu^2_\rho \rho^*\rho +\lambda^\rho (\rho^* \rho)_1 (\rho^* \rho)_1,\, V(\chi)=V(\rho\rightarrow \chi),\, V(\phi_s)=V(\phi\rightarrow \phi_s), \crn
&&V(H,H^')=\lambda^{HH^'}_1 (H^\+ H)_1(H^{'\+} H^')_1+\lambda^{HH^'}_2 (H^\+ H^')_{1^'}(H^{'\+} H)_{1^'},\, V(H,\phi)=\lambda^{H\phi}_1 (H^\+ H)_1(\phi^* \phi)_1\crn
&&\hspace{1.65 cm}+\,\lambda^{H\phi}_2 (H^\+\phi)_{3}(\phi^* H)_{3},\hs
V(H,\varphi)=V(H,\phi\rightarrow \varphi), \hs V(H,\rho)=\lambda^{H\rho}_1 (H^\+ H)_1(\rho^* \rho)_1\crn
&&\hspace{1.65 cm}+\,\lambda^{H\rho}_2 (H^\+\rho)_{1}(\rho^* H)_{1},\hs V(H,\chi)=V(H,\rho\rightarrow \chi), \hs\,\, V(H,\phi_s)=V(H,\phi\rightarrow \phi_s), \nonumber \eea
\bea
&&V(H,H^')=\lambda^{HH^'}_1 (H^\+ H)_1(H^{'\+} H^')_1+\lambda^{HH^'}_2 (H^\+ H^')_{1^'}(H^{'\+} H)_{1^'},\, V(H,\phi)=\lambda^{H\phi}_1 (H^\+ H)_1(\phi^* \phi)_1\crn
&&\hspace{1.65 cm}+\,\lambda^{H\phi}_2 (H^\+\phi)_{3}(\phi^* H)_{3},\hs
V(H,\varphi)=V(H,\phi\rightarrow \varphi), \hs V(H,\rho)=\lambda^{H\rho}_1 (H^\+ H)_1(\rho^* \rho)_1\crn
&&\hspace{1.65 cm}+\,\lambda^{H\rho}_2 (H^\+\rho)_{1}(\rho^* H)_{1},\hs V(H,\chi)=V(H,\rho\rightarrow \chi), \hs\,\, V(H,\phi_s)=V(H,\phi\rightarrow \phi_s), \crn
&&V(H^',\phi)=\lambda^{H^'\phi}_1 (H^{'\+} H^')_1(\phi^* \phi)_1+\lambda^{H^'\phi}_2 (H^{'\+}\phi)_{3^'}(\phi^* H^')_{3^'}, \hs\, V(H^',\varphi)=V(H^',\phi\rightarrow \varphi), \crn
&&V(H^',\rho)=\lambda^{H^'\rho}_1 (H^{'\+} H^')_1(\rho^* \rho)_1+\lambda^{H^'\rho}_2 (H^{'\+}\rho)_{1^'}(\rho^* H^')_{1^'},\hs\, V(H^',\chi)=V(H^',\rho \rightarrow \chi), \crn
&&V(H^',\phi_s)=V(H^',\phi \rightarrow \phi_s),\hs V(\phi,\rho)=\lambda^{\phi\rho}_1 (\phi^{*} \phi)_1(\rho^{*} \rho)_1+\lambda^{\phi\rho}_2 (\phi^{*} \rho)_{3}(\rho^{*} \phi)_{3},\crn
&&V(\phi,\chi)=V(\phi,\rho \rightarrow \chi),\hs V(\phi,\varphi)=\lambda^{\phi\varphi}_1 (\phi^{*} \phi)_1(\varphi^{*} \varphi)_1+\lambda^{\phi\varphi}_2 (\phi^{*} \phi)_2(\varphi^{*} \varphi)_2 +\lambda^{\phi\varphi}_3 (\phi^{*} \phi)_3(\varphi^{*} \varphi)_3\crn
&&\hspace{1.35 cm}+\,\lambda^{\phi\varphi}_4 (\phi^{*} \phi)_{3^'}(\varphi^{*} \varphi)_{3^'}+\lambda^{\phi\varphi}_5 (\phi^{*} \varphi)_1(\varphi^{*} \phi)_1 +\lambda^{\phi\varphi}_6 (\phi^{*} \varphi)_2(\varphi^{*} \phi)_2
+\lambda^{\phi\varphi}_7(\phi^{*} \varphi)_3(\varphi^{*} \phi)_3\crn
&&\hspace{1.35 cm}+\,\lambda^{\phi\varphi}_8 (\phi^{*} \varphi)_{3^'}(\varphi^{*} \phi)_{3^'}, \hs V(\phi,\phi_s)=V(\phi,\varphi\rightarrow \phi_s), \hs V(\varphi,\chi)=V(\phi \rightarrow \varphi,\chi), \nonumber\\
&&V(\varphi,\phi_s)=V(\phi\rightarrow \varphi,\phi_s),\hs V(\rho,\chi)=\lambda^{\rho\chi}_1 (\rho^{*} \rho)_1(\chi^{*} \chi)_1+\lambda^{\rho\chi}_2 (\rho^{*} \chi)_{1}(\chi^{*} \rho)_{1}, \crn
&& V(\rho,\phi_s)=V(H\rightarrow \rho,\phi \rightarrow \phi_s), \hs V(\chi,\phi_s)=V(\rho\rightarrow \chi,\phi_s), \crn
&&V_{\mathrm{trip}}=\lambda^{\phi\varphi\rho}_1 (\phi^*\phi)_3 (\varphi^*\rho)_3+\lambda^{\phi\varphi\rho}_2 (\phi^*\phi)_3 (\varphi\rho^*)_3
+\lambda^{\phi\varphi\rho}_3 (\phi^*\rho)_3 (\varphi^*\phi)_3+\lambda^{\phi\varphi\rho}_4 (\phi^*\varphi)_3 (\phi\rho^*)_3, \crn
&&V_{\mathrm{quart}}=\lambda^{\phi\varphi\rho\phi_s}_1 (\phi\varphi)_3 (\rho\phi_s)_3
+\lambda^{\phi\varphi\rho\phi_s}_2 (\phi\phi_s)_3 (\rho\varphi)_3
+\lambda^{\phi\varphi\rho\phi_s}_3 (\phi_s\varphi)_3 (\rho \phi)_3\crn
&&\hspace{1.05 cm}+\, \lambda^{\phi\varphi\rho\phi_s}_4 (\phi^*\varphi^*)_3 (\rho^*\phi^*_s)_3
+\lambda^{\phi\varphi\rho\phi_s}_5 (\phi^*\phi^*_s)_3 (\rho^*\varphi^*)_3
+\lambda^{\phi\varphi\rho\phi_s}_6 (\phi^*_s\varphi^*)_3 (\rho^*\phi^*)_3. \label{Vn}
\eea
Now we show that the VEV alignments in Eq. (\ref{scalarvev}) satisfy the
minimization condition of the scalar potential $V_{\mathrm{scal}}$ of the model whose explicit expression is given in Eqs. (\ref{Vtotal})-(\ref{Vn}). For this purpose we suppose that the VEVs of the scalars $H, H^', \phi, \rho, \chi$ and $\phi_s$ are real while that of $\varphi$ is complex, i.e., $v^*=v,\, v^{'*}=v^',\, v^*_\phi=v_\phi,\, v^*_\rho=v_\rho,\, v^*_\chi=v_\chi, v^*_{\phi_s}=v_{\phi_s}$ and $v_{\varphi}=v_{0} e^{i\alpha}$. The minimization condition of $V_{\mathrm{scal}}$ reads
\bea
\frac{\partial V_{\mathrm{scal}}}{\partial v_\lambda} &=&0,\hs
\delta^2_{\lambda}\sim \frac{\partial^2 V_{\mathrm{scal}}}{\partial v^2_\lambda} >0 \hs\hs (v_\lambda=v,\, v^',\, v_\phi,\, v_\rho,\, v_\chi,\, v_{\phi_s},\, v_{0},\, \alpha). \label{conditionv}
\eea
For simplicity we will work with the following benchmark points:
\bea
&&\lambda^{H\varphi}_1 = \lambda^{H\varphi}_2 =
\lambda^{H^'\varphi}_1 = \lambda^{H^'\varphi}_2 = 
\lambda^{H\chi}_1 = \lambda^{H\chi}_2 =
\lambda^{H^'\chi}_1 = \lambda^{H^'\chi}_2 
=\lambda^{HH^'}_1 = \lambda^{HH^'}_2 
\crn
&&\hspace{0.8 cm}=\lambda^{H\phi}_1 = \lambda^{H\phi}_2 =
\lambda^{H^'\phi}_1 = \lambda^{H^'\phi}_2 =
\lambda^{H\rho}_1 = \lambda^{H\rho}_2 =
\lambda^{H^'\rho}_1 = \lambda^{H^'\rho}_2 =
\lambda^{H\phi_s}_1 = \lambda^{H\phi_s}_2\crn
&&\hspace{0.8 cm} 
= \lambda^{H^'\phi_s}_1 = \lambda^{H^'\phi_s}_2 =
\lambda^{\phi\chi}_1 = \lambda^{\phi\chi}_2 =
\lambda^{\phi\varphi}_1 =
  \lambda^{\phi\varphi}_5 =
   \lambda^{\phi\varphi}_6 = \lambda^{\phi\varphi}_7 = \lambda^{\phi\varphi}_8=\lambda^{H^'\phi}_1 = \lambda^{H^'\phi}_2
   \crn
&&\hspace{0.8 cm} =
\lambda^{\phi\rho}_1 = \lambda^{\phi\rho}_2 = 
\lambda^{\phi\phi_s}_1 =
  \lambda^{\phi\phi_s}_3 =
   \lambda^{\phi\phi_s}_5 =
    \lambda^{\phi\phi_s}_6 = \lambda^{\phi\phi_s}_7 =\lambda^{\phi\phi_s}_8 = 
\lambda^{\varphi\chi}_1 = \lambda^{\varphi\chi}_2=\lambda^{\varphi\rho}_1 \crn
&&\hspace{0.8 cm} = \lambda^{\varphi\rho}_2 =
\lambda^{\rho\chi}_1 = \lambda^{\rho\chi}_2 =
\lambda^{\rho\phi_s}_1 = \lambda^{\rho\phi_s}_2 =
\lambda^{H\chi}_1 = \lambda^{H\chi}_2 =
\lambda^{\chi\phi_s}_1 = \lambda^{\chi\phi_s}_2 =
\lambda^{\varphi\phi_s}_1 =
 \lambda^{\varphi\phi_s}_2 \crn
 &&\hspace{0.8 cm }=
   \lambda^{\varphi\phi_s}_7 = \lambda^{\varphi\phi_s}_8 = 
   \lambda^{\phi\varphi\rho\phi_s}_3 =\lambda^{\phi\varphi\rho\phi_s}_2=\lambda^{\phi\varphi\rho\phi_s}_1 =\lambda^{\phi\varphi\rho}_3=\lambda^{\phi\varphi\rho}_1=\lambda^{x},
   \\
&&\lambda^{\phi\varphi\rho\phi_s}_6=\lambda^{\phi\varphi\rho\phi_s}_5 =
\lambda^{\phi\varphi\rho\phi_s}_4= a \lambda^{\phi\varphi\rho\phi_s}_1=a \lambda^{x},  \,\,
\lambda^{\phi\varphi\rho}_4 =\lambda^{\phi\varphi\rho}_2 = 
a \lambda^{x}, \\
&&\lambda^{\phi}_1 = \lambda^{\phi}_3 = \lambda^{\phi}, \hs
\lambda^{\varphi}_1 = \lambda^{\varphi}_2 = \lambda^{\varphi}, \hs \lambda^{\phi_s}_1 =\lambda^{\phi_s}_2= \lambda^{\phi_s}_3 =\lambda^{\phi_s}.
\eea
As a consequence, the condition (\ref{conditionv})  becomes
\bea
\mu_H^2 + 2 \lambda^{H} v^2 +
 2 \lambda^{x} \left(v_\varphi^2 + v_\chi^2 + v^{'2} + 3 v_\phi^2 + v_\rho^2 + 2 v_s^2\right) =0,\hs &&\label{eq1v} \\
\mu_{H^'}^2 + 2 \lambda^{H^'} v^{'2} +
 2 \lambda^{x} \left(v^2 + v_\varphi^2 + v_\chi^2 + 3 v_\phi^2 + v_\rho^2 + 2 v_s^2\right)=0,\hs &&\label{eq2v} \\
6 \mu_\phi^2 v_\phi + 84 \lambda^{\phi} v_\phi^3 +
 2 \lambda^{x} \left[6 v_\phi (v^2 + v_\varphi^2 + v_\chi^2 + v^{'2} + v_\rho^2) + 20 v_\phi v_s^2 +   e^{i\al} v_\varphi v_\rho (4 a v_\phi + 3 v_s) \right.\crn
 \left.  +\, e^{-i \al} v_\varphi v_\rho (4 v_\phi + 3 a v_s)\right]=0, &&\label{eq3v}\\
\mu_\varphi^2 v_\varphi + 6 \lambda^{\varphi} v_\varphi^3 +
 \lambda^{x} \left[2 v_\varphi (v^2 + v_\chi^2 + v^{'2} + 3 v_\phi^2 + v_\rho^2) +
    e^{i \al} v_\phi v_\rho (2 a v_\phi + 3 v_s) \right.\crn
 \left.  +\,     e^{-i \al} v_\phi v_\rho (2 v_\phi + 3 a v_s)\right]=0,&&\label{eq4v}\\
v_\rho \left[\mu_\rho^2 + 2 \lambda^{\rho} v_\rho^2 +
    2 \lambda^{x} (v^2 + v_\varphi^2 + v_\chi^2 + v^{'2} + 3 v_\phi^2 + 2 v_s^2)\right]+e^{i \al} \lambda^{x} v_\varphi v_\phi (2 a v_\phi + 3 v_s) &&\crn
+ e^{-i \al} \lambda^{x} v_\varphi v_\phi (2 v_\phi + 3 a v_s) =0, &&\label{eq5v}\\
\mu_\chi^2 + 2 \lambda^{\chi} v_\chi^2 +
 2 \lambda^{x} (v^2 + v_\varphi^2 + v^{'2} + 3 v_\phi^2 + v_\rho^2 + 2 v_s^2)=0, &&\label{eq6v}\\
3 \lambda^{x} v_\varphi v_\phi v_\rho\left(a e^{-i \al}  + e^{i \al}\right)+
 2 v_s \left[\mu_{\phi_s}^2 + 2 \lambda^{x} (v^2 + v_\chi^2 + v^{'2} + 5 v_\phi^2 + v_\rho^2) +
    10 \lambda^{\phi_s} v_s^2\right]=0, &&\label{eq7v}\\
e^{2 i \al} (2 a v_\phi + 3 v_s)-2 v_\phi - 3 a v_s =0, &&\label{eq8v}\\
\mu_H^2 + 6 \lambda^{H} v^2 +
 2 \lambda^{x} (v_\varphi^2 + v_\chi^2 + v^{'2} + 3 v_\phi^2 + v_\rho^2 + 2 v_s^2) > 0, &&\label{ieq1}\\
\mu_{H^'}^2 + 6 \lambda^{H^'} v^{'2} +
 2 \lambda^{x} (v_\varphi^2 + v_\chi^2 + v^{2} + 3 v_\phi^2 + v_\rho^2 + 2 v_s^2)>0,&&\label{ieq2}\\
3 \mu_\phi^2 + 126 \lambda^{\phi} v_\phi^2 + 4 \lambda^{x} v_\varphi v_\rho\left(e^{-i \al} + b e^{i \al}\right) + 6 \lambda^{x} (v^2 + v_\varphi^2 + v_\chi^2 + v^{'2} + v_\rho^2) +
 20 \lambda^{x} v_s^2>0, &&\label{ieq3}\\
\mu_\varphi^2 + 18 \lambda^{\varphi} v_\varphi^2 +
 2 \lambda^{x} (v^2 + v_\chi^2 + v^{'2} + 3 v_\phi^2 + v_\rho^2)>0, &&\label{ieq4}\\
\mu_\rho^2 + 6 \lambda^{\rho} v_\rho^2 +
 2 \lambda^{x} (v^2 + v_\varphi^2 + v_\chi^2 + v^{'2} + 3 v_\phi^2 + 2 v_s^2)>0, &&\label{ieq5}\\
\mu_\chi^2 + 6 \lambda^{\chi} v_\chi^2 +
 2 \lambda^{x} (v^2 + v_\varphi^2 + v^{'2} + 3 v_\phi^2 + v_\rho^2 + 2 v_s^2)>0, &&\label{ieq6}\\
\mu_{\phi_s}^2 + 2 \lambda^{x} (v^2 + v_\chi^2 + v^{'2} + 5 v_\phi^2 + v_\rho^2) +
 30 \lambda^{\phi_s} v_s^2>0, &&\label{ieq7}\\
-2 v_\phi - 3 a v_s - e^{2 i \al} \left(2 a v_\phi + 3 v_s\right)>0. &&\label{ieq8}\eea
The system of Eqs. (\ref{eq1v})-(\ref{eq8v}) yields the following solution:
\bea
&&\lambda^H=-\frac{\mu_H^2 + 2 \lambda^{x} (v_\varphi^2 + v_\chi^2 + v^{'2} + 3 v_\phi^2 + v_\rho^2 + 2 v_s^2)}{2v^2},\crn
&&\lambda^{H^'}=-\frac{\mu_{H^'}^2 + 2 \lambda^{x} (v_\varphi^2 + v_\chi^2 + v^{2} + 3 v_\phi^2 + v_\rho^2 + 2 v_s^2)}{2v^{'2}}, \crn
&&\lambda^{\phi}=-\frac{1}{42 v_\phi^3} \left[(3 \mu_\phi^2 + 6 \lambda^{x} v^2 + 6 \lambda^{x} v_\varphi^2 +
   6 \lambda^{x} v_\chi^2 + 6 \lambda^{x} v^{'2} + 6 \lambda^{x} v_\rho^2 +
   20 \lambda^{x} v_s^2) v_\phi  \right. \crn
   &&\left. \hspace{0.5 cm}+\, \lambda^{x} v_\varphi v_\rho \left(\frac{(4 a v_\phi + 3 v_s) \sqrt{2 v_\phi + 3 a v_s}}{\sqrt{
   2 a v_\phi + 3 v_s}} + \frac{\sqrt{2 a v_\phi + 3 v_s} (4 v_\phi + 3 a v_s)}{\sqrt{
   2 v_\phi + 3 a v_s}}\right)\right], \nonumber \eea
   \bea
&&\lambda^{\varphi}=-\frac{\mu_\varphi^2 v_\varphi +
 2 \lambda^{x} \left[v_\varphi (v^2 + v_\chi^2 + v^{'2} + 3 v_\phi^2 + v_\rho^2) +
    v_\phi v_\rho \sqrt{(2 a v_\phi + 3 v_s)(2 v_\phi + 3 a v_s)}\right]}{6 v_\varphi^3},\crn
    &&\lambda^{\rho}=-\frac{\mu_\rho^2 v_\rho +
 2 \lambda^{x} \left[v_\rho (v^2 + v_\varphi^2 + v_\chi^2 + v^{'2} + 3 v_\phi^2 + 2 v_s^2) + v_\varphi v_\phi \sqrt{(2 a v_\phi + 3 v_s)(2 v_\phi + 3 a v_s)}\right]}{2 v_\rho^3}, \crn
 &&\lambda^{\chi}=-\frac{\mu_\chi^2 + 2 \lambda^{x} (v^2 + v_\varphi^2 + v^{'2} + 3 v_\phi^2 + v_\rho^2 + 2 v_s^2)}{2 v^2_\chi}, \hs \al= i \log \left(\sqrt{\frac{2 a v_\phi + 3 v_s}{2 v_\phi + 3 a v_s}}\right),\crn
 &&\lambda^{\phi_s}=-\frac{\mu_{\phi_s}^2 v_s + 2 \lambda^{x} (v^2 + v_\chi^2 + v^{'2} + 5 v_\phi^2 + v_\rho^2) v_s }{10 v_s^3}
 -\frac{3\lambda^{x} v_\varphi v_\phi v_\rho (v_\phi + a^2 v_\phi + 3 a v_s)}{
 10 v_s^3 \sqrt{(2 a v_\phi + 3 v_s)(2 v_\phi + 3 a v_s)}}.\label{solutionminimaleq}
\eea
With the aid of the solution (\ref{solutionminimaleq}), expressions
(\ref{ieq1})-(\ref{ieq4}) become
\bea
\delta^2_{v} \sim -\mu_H^2-2 \lambda^{x} \left(v_\varphi^2+v_\chi^2+v^{'2}+3v_\phi^2+v_\rho^2+2 v_s^2\right)>0, \label{ineq1n}&&\\
\delta^2_{v^'} \sim -\mu_{H^'}^2-2 \lambda^{x} \left(v_\varphi^2+v_\chi^2+v^{2}+3v_\phi^2+v_\rho^2+2 v_s^2\right)>0, \label{ineq1n}&&\\
\delta^2_\phi \sim -3 \mu_\phi^2 - \lambda^{x}\left[ 6 v^2 + 6 v_\varphi^2 + 6 (v_\chi^2 + v^{'2} + v_\rho^2) + 20 v_s^2
\hspace{0.7 cm} \right.&&\crn
\left. +\, \frac{v_{\varphi} v_{\rho} \left[a v_{\phi} (21 a v_{s}+16 v_{\phi})+3 v_{s} (9 a v_{s}+7 v_{\phi})\right]}{v_{\phi} \sqrt{3 a v_{s}+2 v_{\phi}} \sqrt{2 a v_{\phi}+3 v_{s}}} \right] >0,&&\\
\delta^2_\varphi \sim -\mu_\varphi^2 - 2 \lambda^{x} (v^2 + v_\chi^2 + v^{'2} + 3 v_\phi^2 + v_\rho^2)-\frac{3 \lambda^{x} v_\phi v_\rho \sqrt{(2 a v_\phi + 3 v_s)(2 v_\phi + 3 a v_s)}}{v_\varphi}>0,&&\\
\delta^2_\rho \sim -\mu_\rho^2 - 2 \lambda^{x} (v^2 + v_\chi^2 + v^{'2} + 3 v_\phi^2 + v_\varphi^2)-\frac{3 \lambda^{x} v_\phi v_\varphi \sqrt{(2 a v_\phi + 3 v_s)(2 v_\phi + 3 a v_s)}}{v_\rho}>0,&&\\
\delta^2_\chi =-\mu_\chi^2 - 2 \lambda^{x} (v^2 + v_\varphi^2 + v^{'2} + 3 v_\phi^2 + v_\rho^2 + 2 v_s^2)>0,&&\\
\delta^2_{\phi_s} \sim -2\mu_{\phi_s}^2 - 4 \lambda^{x} (v^2 + v_\chi^2 + v^{'2} + 5 v_\phi^2 + v_\rho^2)-\frac{9 \lambda^{x} v_\varphi v_\phi v_\rho (v_\phi + a^2 v_\phi + 3 a v_s)}{v_s \sqrt{(2 a v_\phi + 3 v_s)(2 v_\phi + 3 a v_s)}}>0,&&\crn
\delta^2_{\alpha}\sim -\lambda^{x} v_{\varphi} v_{\phi} v_{\rho} \sqrt{3 a v_{s}+2 v_{\phi}} \sqrt{2 a v_{\phi}+3 v_{s}}>0. \label{ineq8n} &&\eea
Assuming that $\mu^2_H,\, \mu^2_{H^'},\, \mu^2_\phi,\, \mu^2_\varphi,\, \mu^2_\rho,\, \mu^2_\chi$ and $\mu^2_{\phi_s}$ are negative and of the same order of magnitude and the same as that of the SM \cite{PDG2020}\footnote{In the SM \cite{PDG2020}, $|\mu|=88.4$ GeV. Here, we use $|\mu_H|\sim |\mu_\phi|\sim|\mu_l|\sim|\mu_\nu|=10^2$\, GeV for their scales.},
\bea
\mu^2_H\sim \mu^2_{H^'}\sim \mu^2_\phi \sim  \mu^2_\varphi \sim \mu^2_\rho \sim \mu^2_\chi \sim\mu^2_{\phi_s}\sim - 10^4\, \mathrm{GeV}. \label{assume}
\eea
Expressions Eqs. (\ref{SMBLscale}), (\ref{flavonscale}), (\ref{ineq1n})-(\ref{ineq8n}) and (\ref{assume}) tell us that $\delta^2_{v} = \delta^2_{v^'}$ and $\delta^2_\chi$ depend on one parameter $\lambda^{x}$ while $\delta^2_\phi, \delta^2_\varphi, \delta^2_\rho,\delta^2_{\phi_s}$ and $\delta^2_\alpha$ depend on two parameters $\lambda^{x}$ and $a$ which are respectively plotted in Figs. \ref{del1263sq} and \ref{delphisalphasq} with $\lambda^{x}\in (-10^{-2}, -10^{-4})$ and $a\in (1.0, 5.0)$. These figures imply that the expressions (\ref{ineq1n})--(\ref{ineq8n}) are always satisfied by the VEV alignments in Eq. (\ref{scalarvev}).
\newpage
\vspace*{2 cm}
\section{\label{UijNH} The dependence of $|U_{ij}|\, (i=1,2,3; j=1,3)$ on $s_{13}$ and $s_{23}$ for normal hierarchy}
\begin{center}
\begin{figure}[h]
\begin{center}
\vspace{-1.0 cm}
\hspace*{-2.2 cm}
\includegraphics[width=0.8\textwidth]{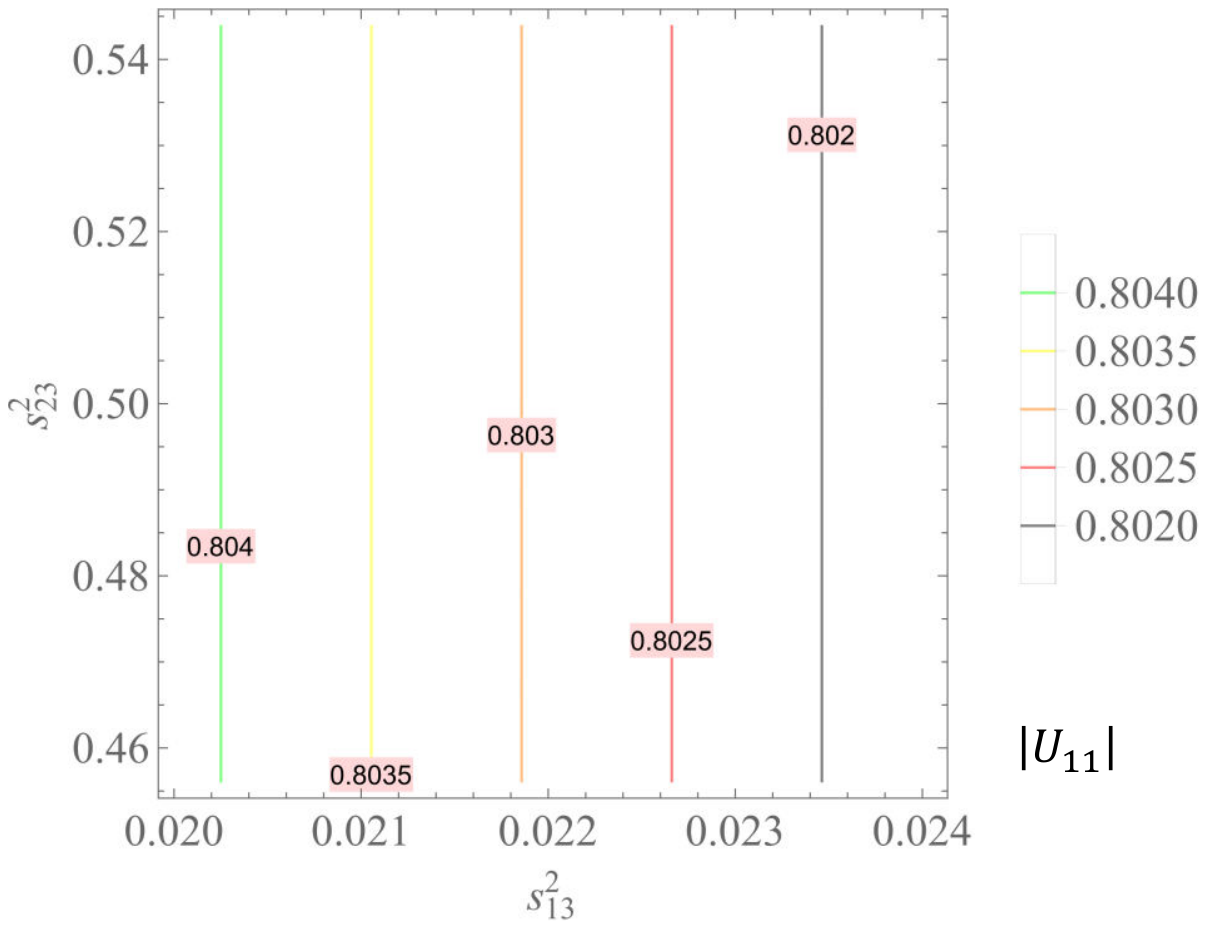}\hspace*{-5.25 cm}
\includegraphics[width=0.8\textwidth]{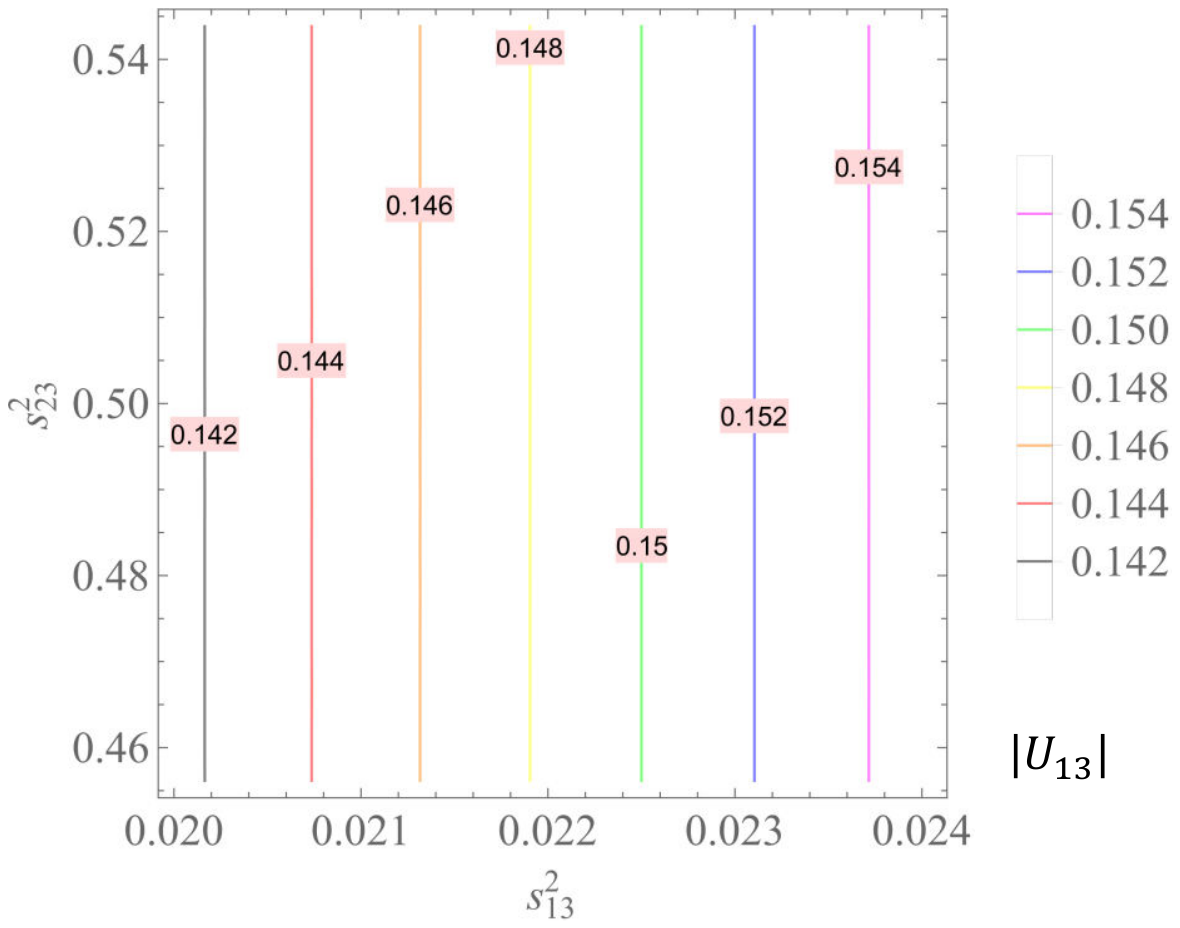}\hspace*{-1.0 cm}\\
\vspace{-12.15 cm}
\hspace*{-2.2 cm}
\includegraphics[width=0.8\textwidth]{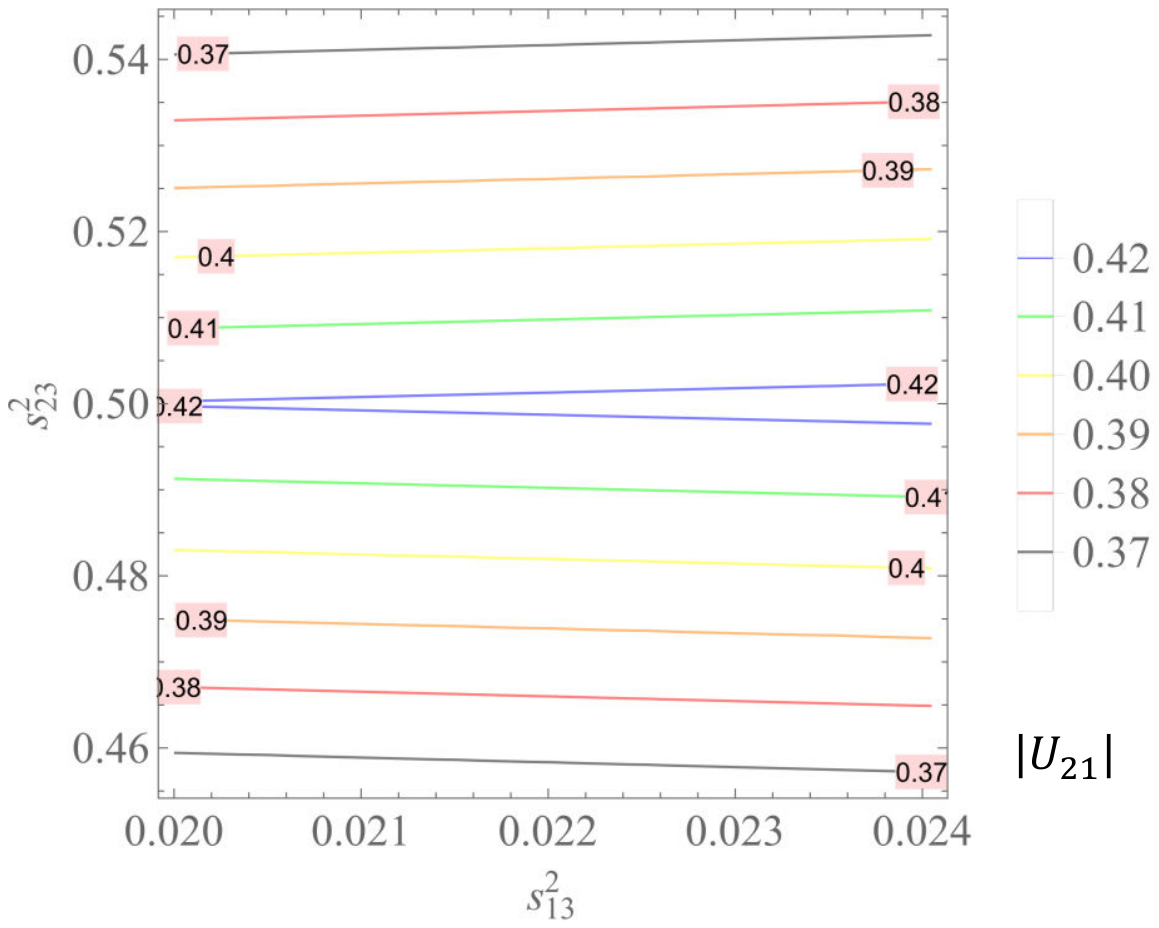}\hspace*{-5.25 cm}
\includegraphics[width=0.8\textwidth]{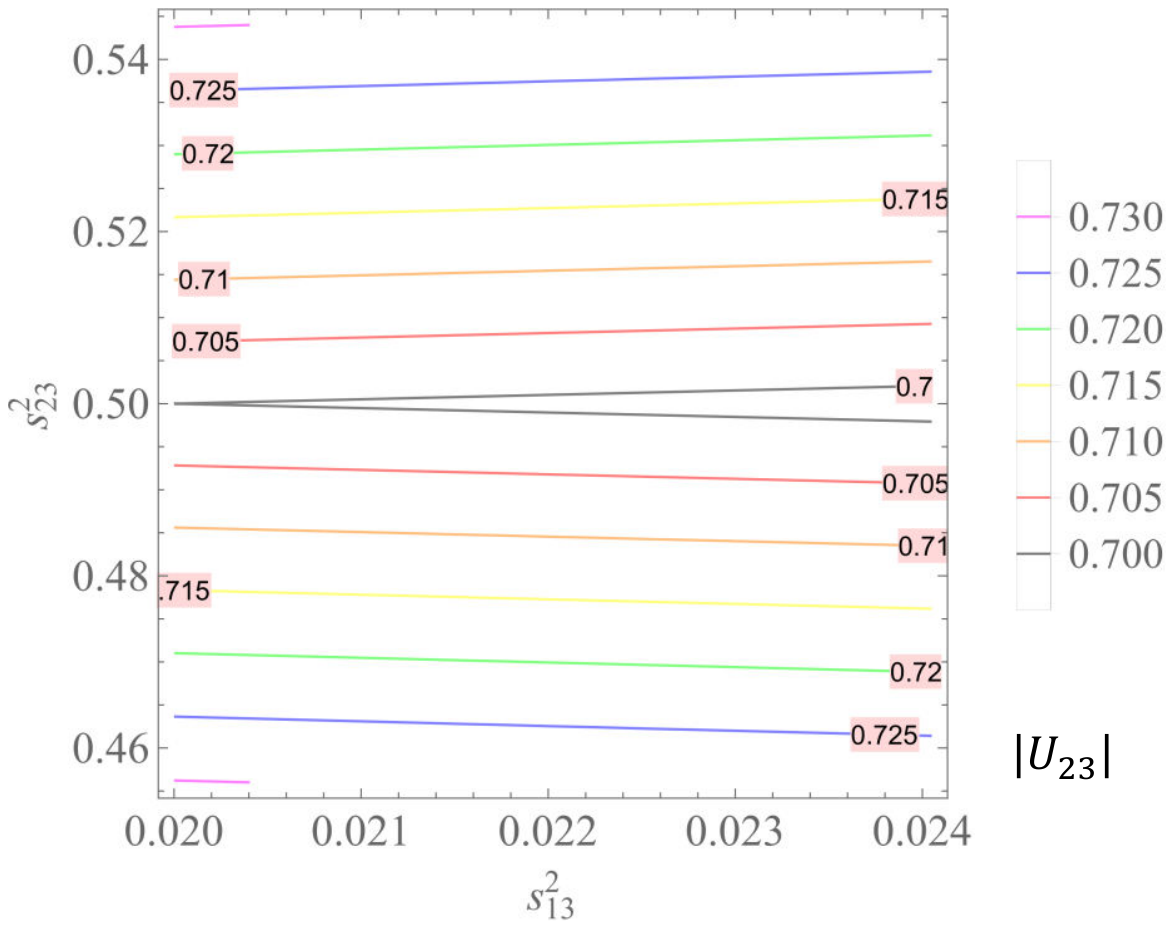}\hspace*{-1.0 cm}\\
\vspace{-12.15 cm}
\hspace*{-2.2 cm}
\includegraphics[width=0.8\textwidth]{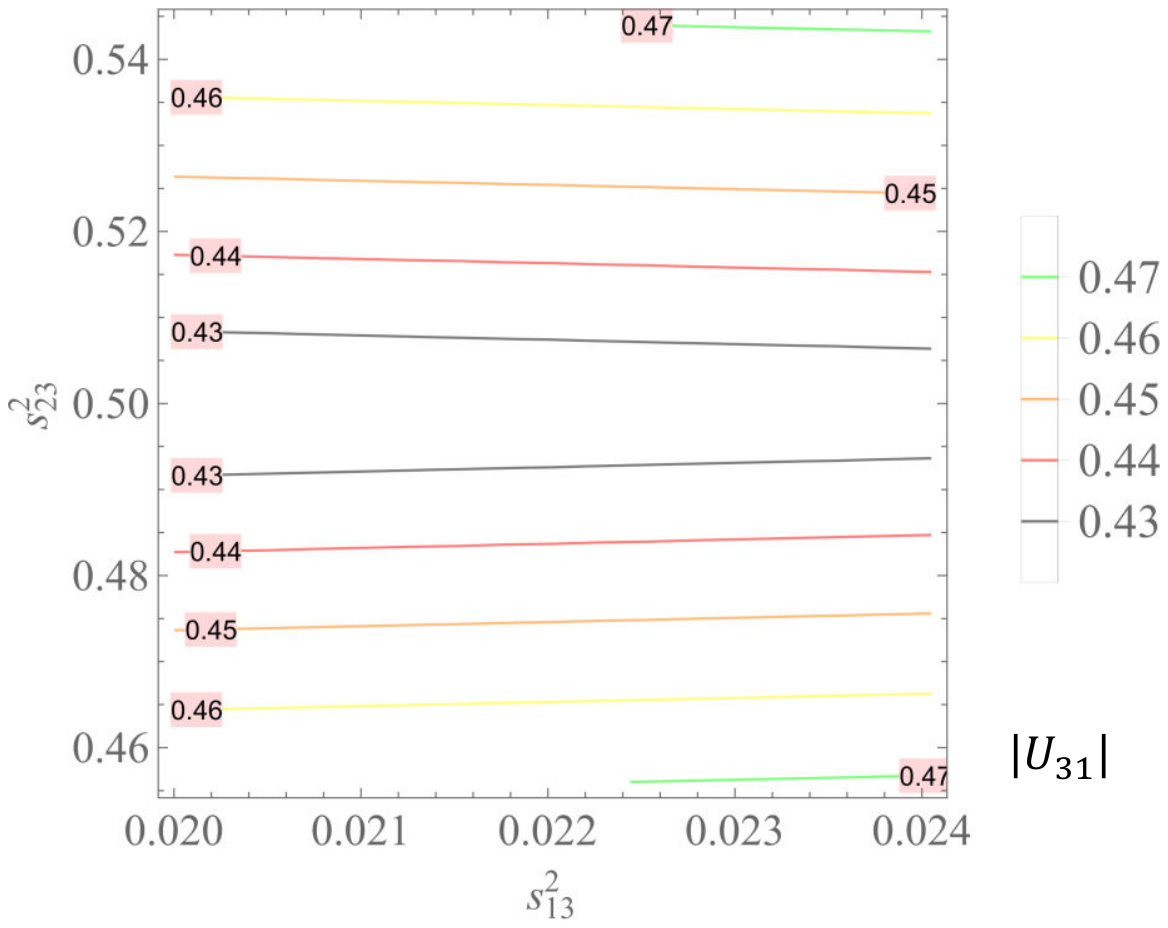}\hspace*{-5.25 cm}
\includegraphics[width=0.8\textwidth]{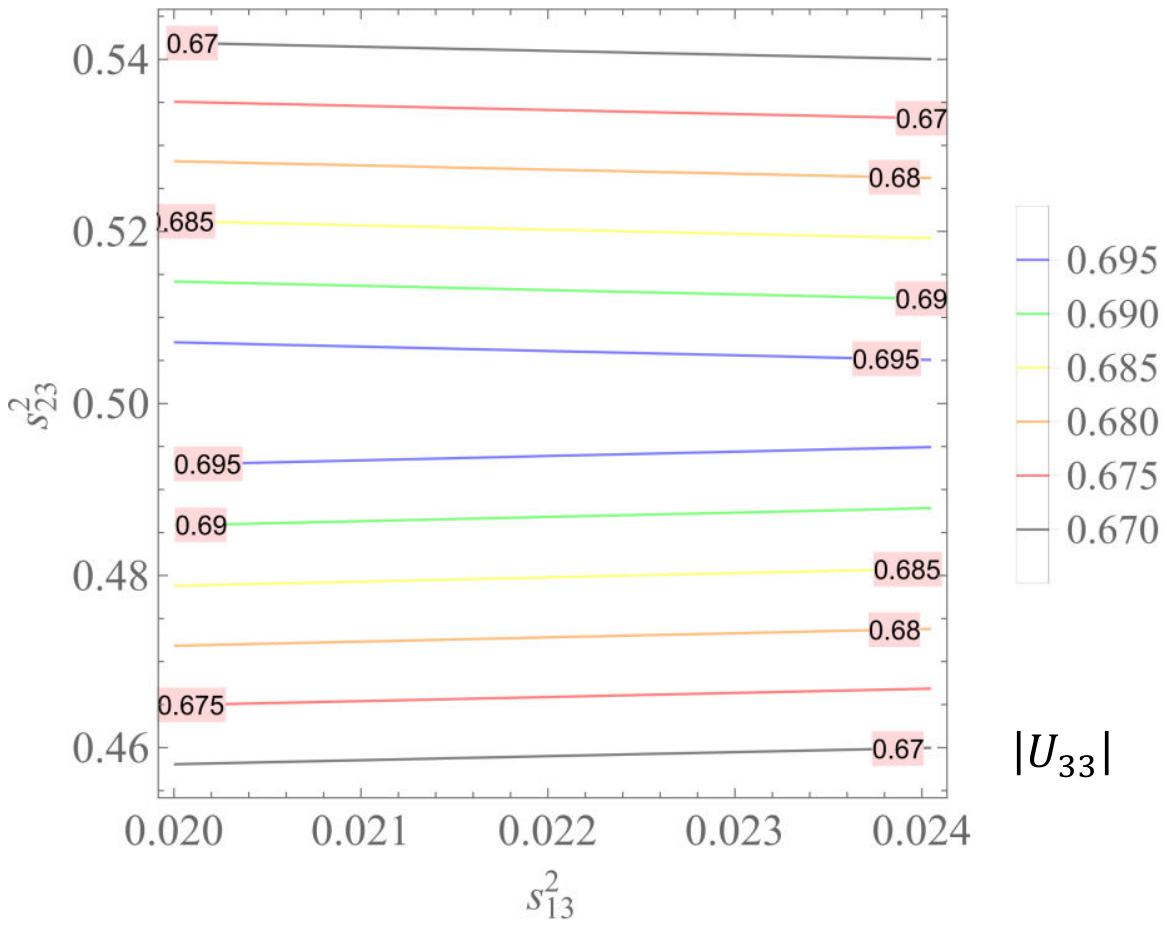}\hspace*{-1.0 cm}
\end{center}
\vspace{-12.0 cm}
\caption{$|U_{ij}|\, (i=1,2,3; j=1,3)$ as functions of $s^2_{23}$ and $s^2_{13}$ with $s^2_{23}\in (0.456, 0.544)$ and $s^2_{13}\in (2.00, 2.405)10^{-2}$ for NH}
\label{UijNF}
\end{figure}
\vspace{-2.0 cm}
\end{center}
\newpage
\section{\label{UijIH} The dependence of $|U_{ij}|\, (i=1,2,3; j=1,3)$ on $s_{13}$ and $s_{23}$ for inverted hierarchy}
\begin{center}
\begin{figure}[h]
\begin{center}
\vspace{-1.0 cm}
\hspace*{-2.2 cm}
\includegraphics[width=0.8\textwidth]{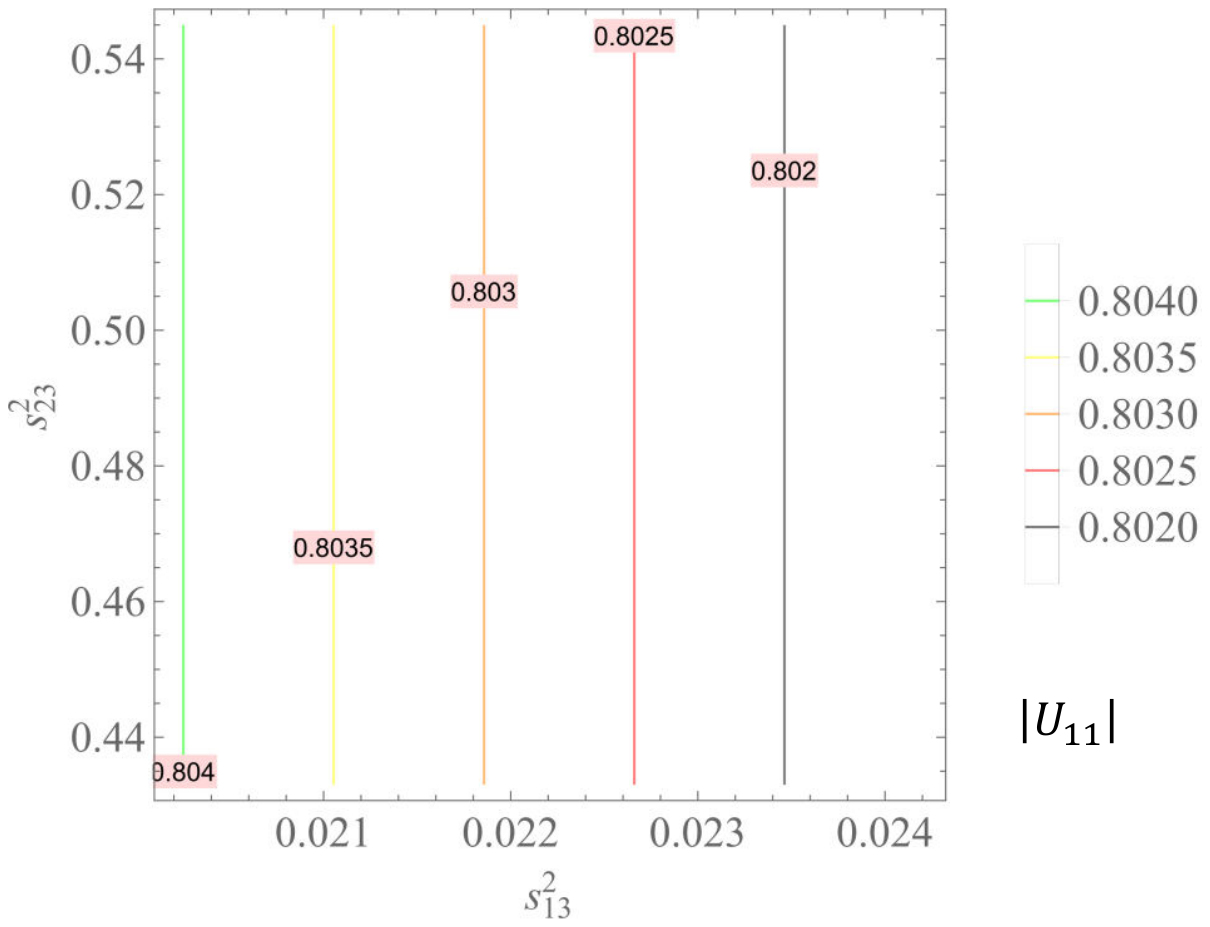}\hspace*{-5.25 cm}
\includegraphics[width=0.8\textwidth]{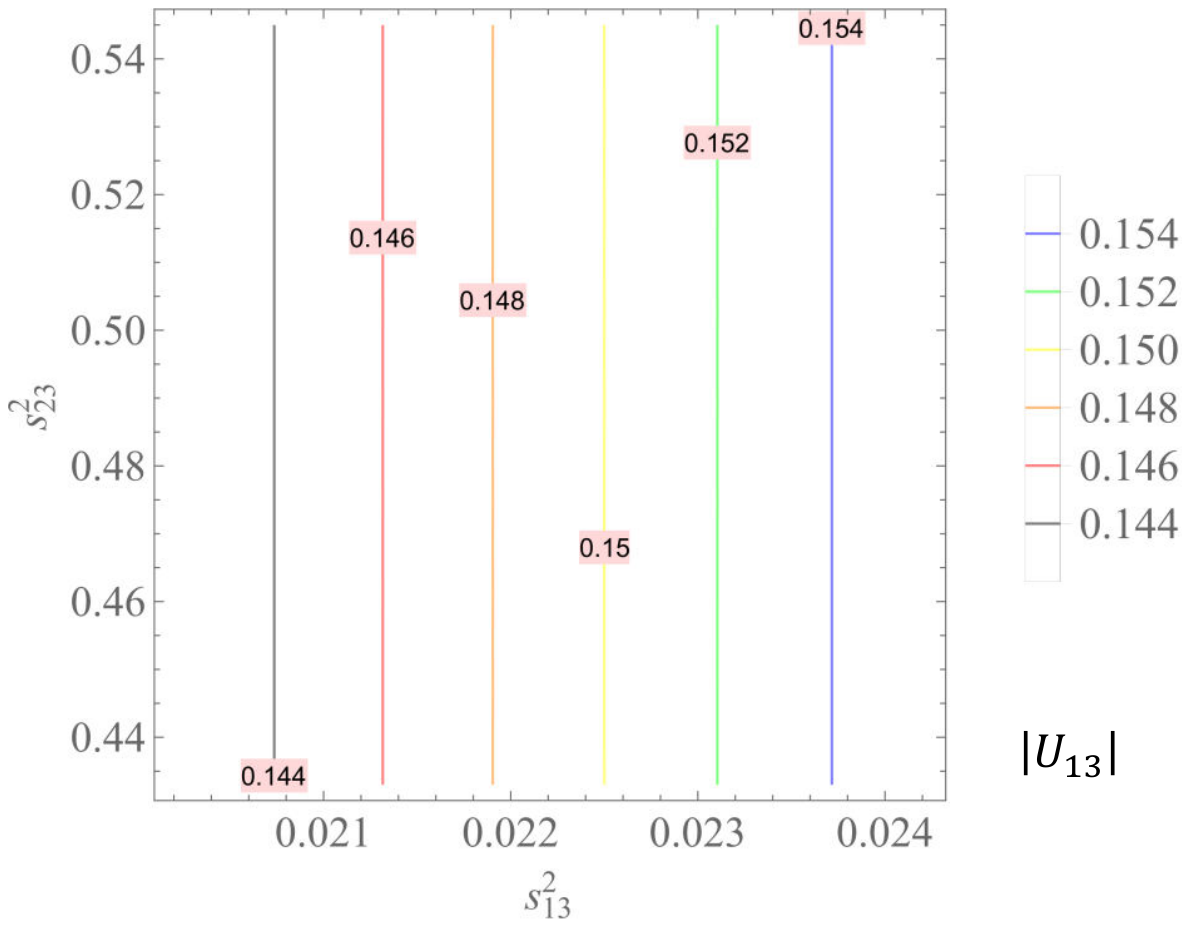}\hspace*{-1.0 cm}\\
\vspace{-12.15 cm}
\hspace*{-2.2 cm}
\includegraphics[width=0.8\textwidth]{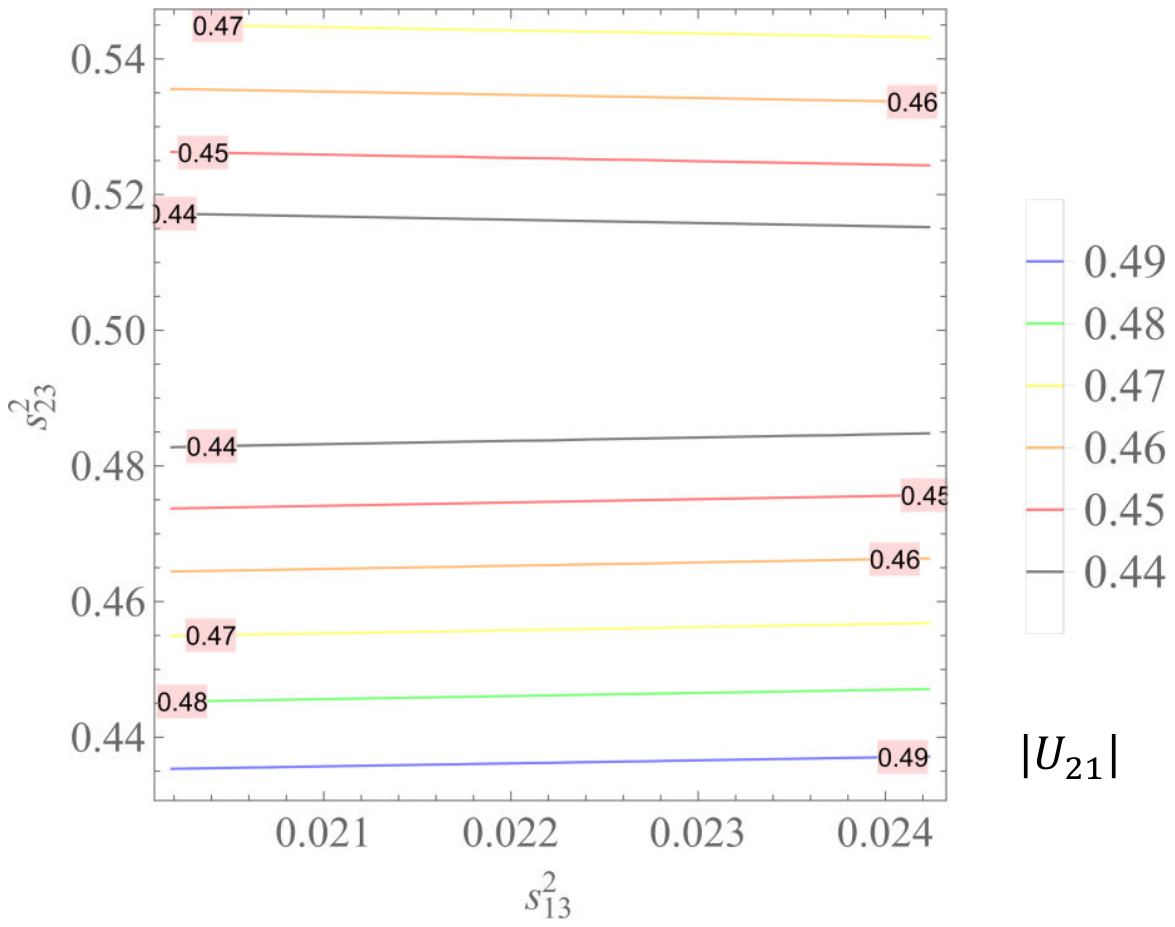}\hspace*{-5.25 cm}
\includegraphics[width=0.8\textwidth]{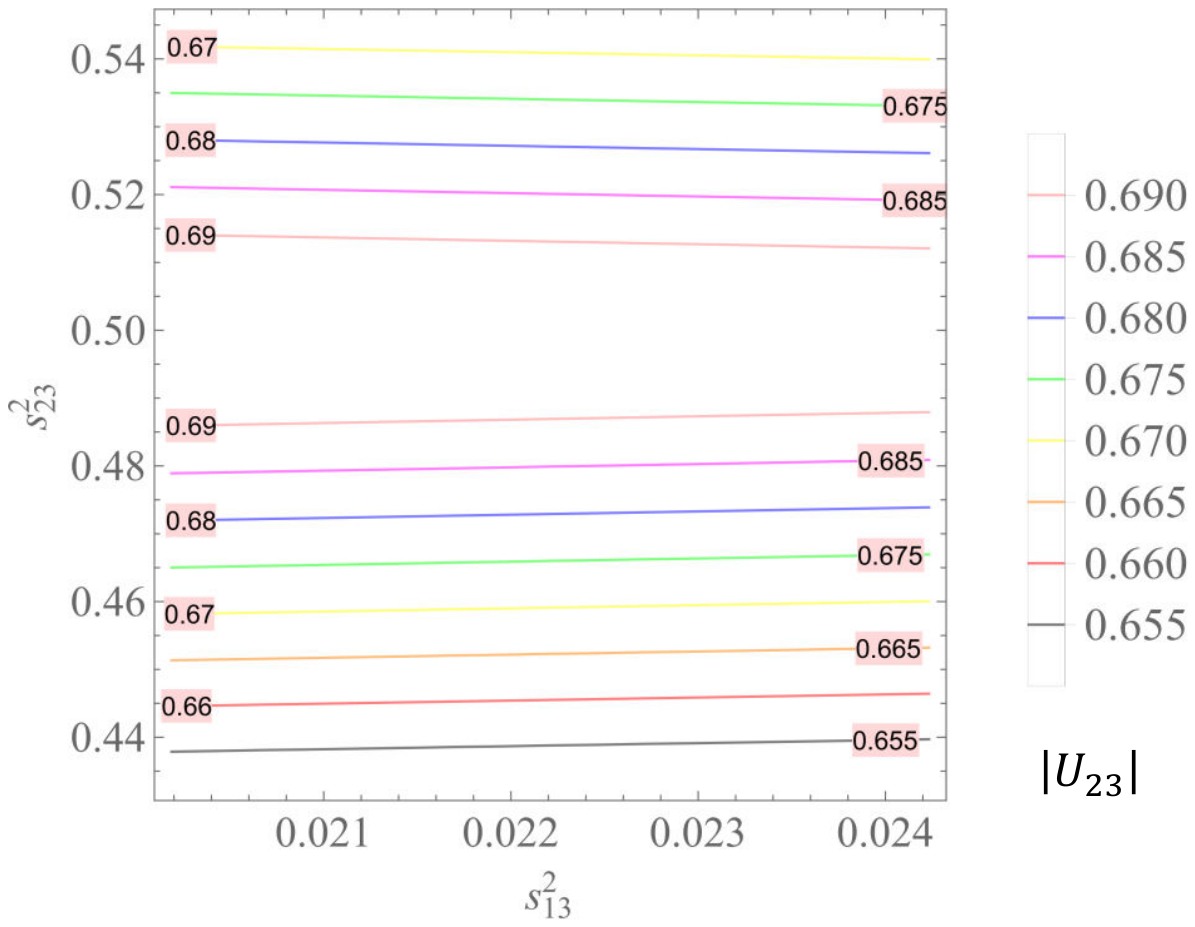}\hspace*{-1.0 cm}\\
\vspace{-12.15 cm}
\hspace*{-2.2 cm}
\includegraphics[width=0.8\textwidth]{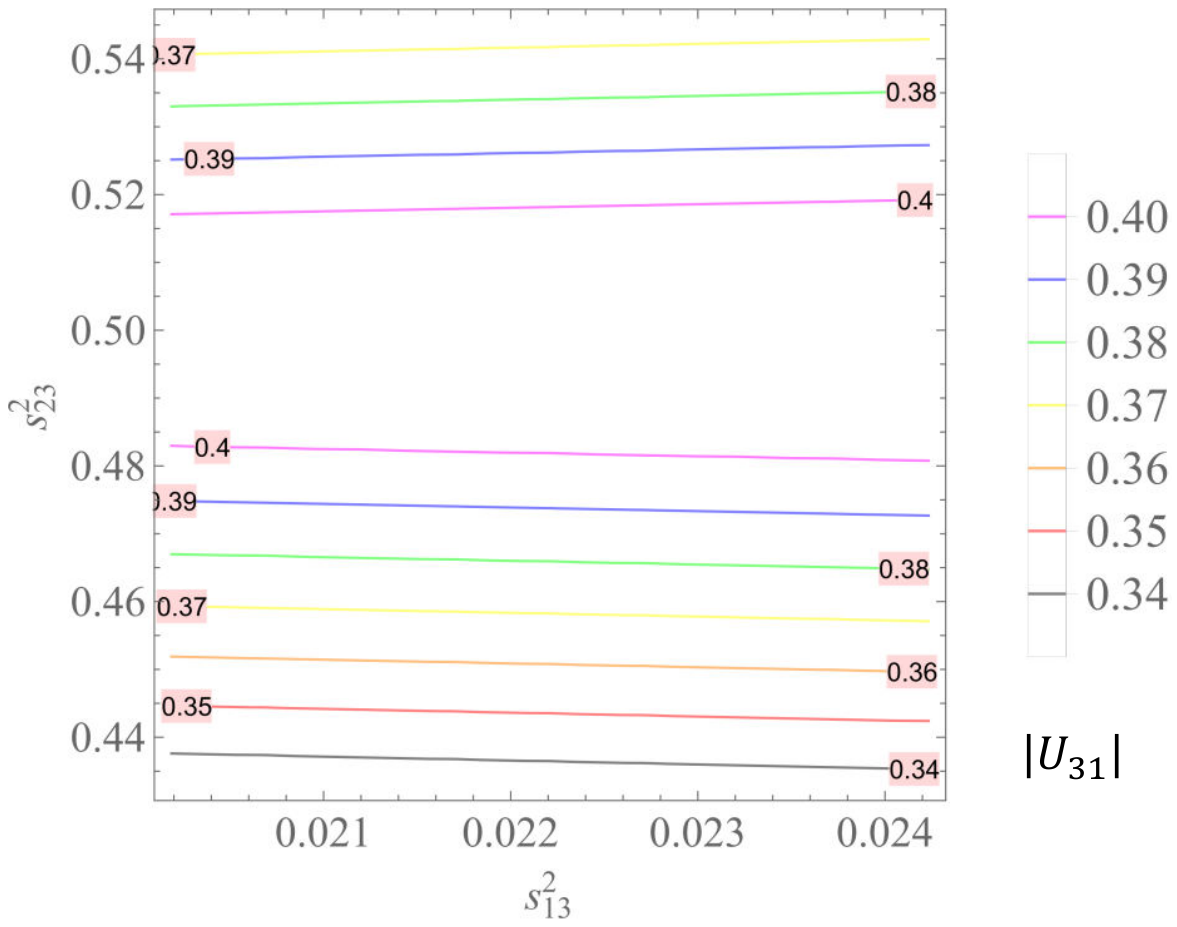}\hspace*{-5.25 cm}
\includegraphics[width=0.8\textwidth]{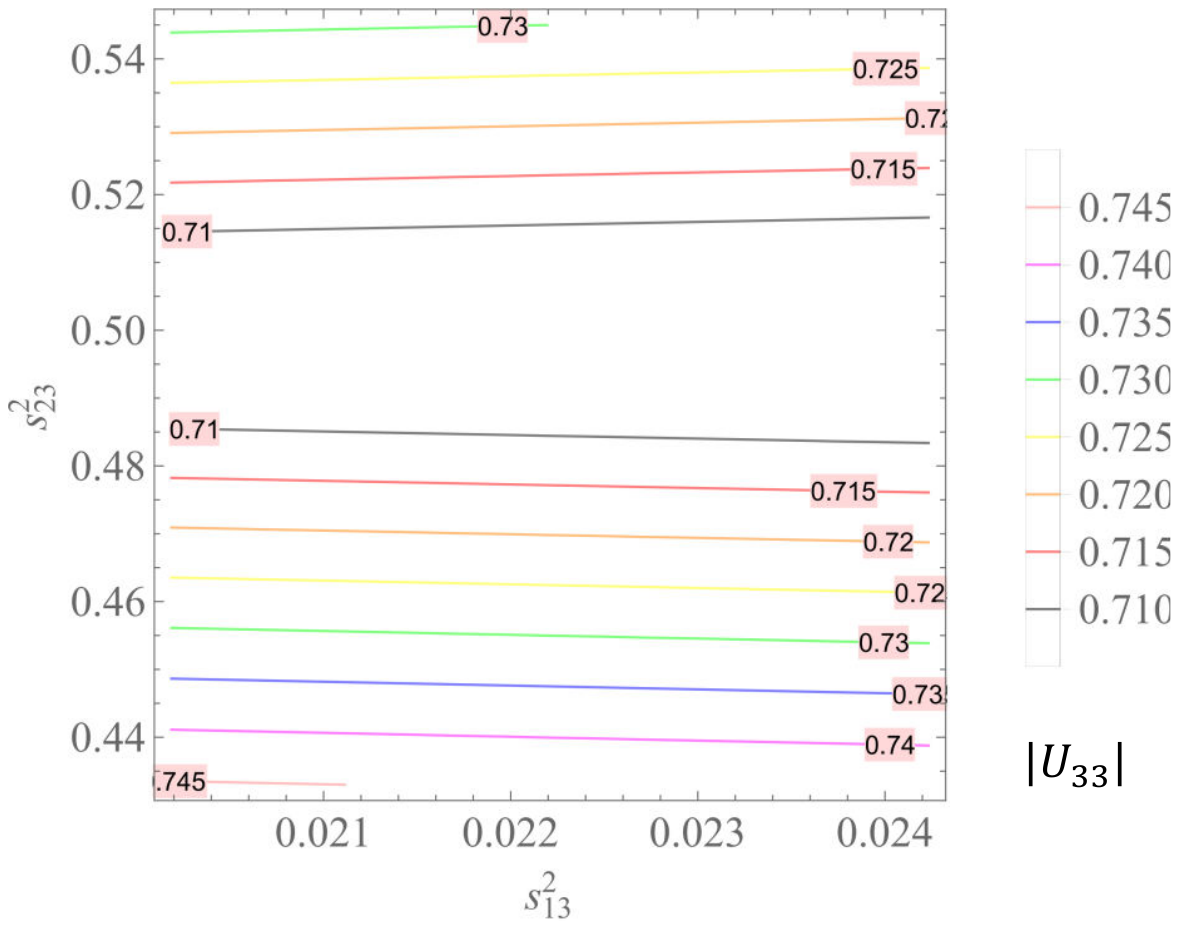}\hspace*{-1.0 cm}
\end{center}
\vspace{-12.0 cm}
\caption{$|U_{ij}|\, (i=1,2,3; j=1,3)$ as functions of $s^2_{23}$ and $s^2_{13}$ with $s^2_{23}\in (0.433, 0.545)$ and $s^2_{13}\in (2.018, 2.242)10^{-2}$ for IH}
\label{UijIF}
\end{figure}
\hspace*{-1.0 cm}
\end{center}
\newpage

\end{document}